\documentclass[fleqn,usenatbib]{mnras}

\usepackage{newtxtext,newtxmath}

\usepackage[T1]{fontenc}

\DeclareRobustCommand{\VAN}[3]{#2}
\let\VANthebibliography\thebibliography
\def\thebibliography{\DeclareRobustCommand{\VAN}[3]{##3}\VANthebibliography}

\usepackage{graphicx}	
\usepackage{amsmath}	

\title[FRB Polarization]{Polarization of Fast Radio Bursts: radiation mechanisms and propagation effects} 

\author[Y. Qu and B. Zhang]{
Yuanhong Qu$^{1,2}$\thanks{E-mail: yuanhong.qu@unlv.edu}
and
Bing Zhang$^{1,2}$\thanks{E-mail: bing.zhang@unlv.edu}
\\
$^{1}$Department of Physics and Astronomy, University of Nevada Las Vegas, Las Vegas, NV 89154, USA\\
$^{2}$Nevada Center for Astrophysics, University of Nevada, Las Vegas, NV 89154
}

\date{}

\pubyear{2023}

\begin{document}
\label{firstpage}
\pagerange{\pageref{firstpage}--\pageref{lastpage}}
\maketitle

\begin{abstract}
Fast radio bursts (FRBs) are observed to be highly polarized. Most have high linear polarization but a small fraction show significant circular polarization. We systematically investigate a variety of polarization mechanisms of FRBs within the magnetar theoretical framework considering two emission sites inside and outside the magnetosphere. For each site, we discuss both intrinsic radiation mechanisms and propagation effects. 
Inside the magnetosphere, we investigate the polarization properties of both coherent curvature radiation and inverse Compton scattering by charged bunches and conclude that both mechanisms produce 100\% linear polarization at an on-axis geometry but can produce circular polarization if the viewing angle is off axis. The lack of circular polarization for the majority of bursts requires that the bunches have a large transverse dimension size. Resonant cyclotron absorption within magnetosphere may produce high circular polarization if electrons and positrons have an asymmetric Lorentz factor distribution. 
Outside the magnetosphere, the synchrotron maser emission mechanism in general produces highly linearly polarized emission. Circular polarization would appear at off-beam angles but the flux is greatly degraded and such bursts are not detectable at cosmological distances. 
Synchrotron absorption in a nebula with ordered magnetic field may reduce the circular polarization degree. Cyclotron absorption in a strongly magnetized medium may generate significant circular polarization. Faraday conversion in a medium with field reversal can convert one polarization mode to another. The two absorption processes require stringent physical conditions. Significant Faraday conversion may be realized in a magnetized dense environment involving binary systems or supernova remnants.
\end{abstract}




\section{Introduction}
Fast Radio Bursts are bright radio bursts with extremely high brightness temperatures $\sim10^{36} \ \rm K$ \citep{Lorimer2007,Petroff2016}, implying the intrinsic emission mechanisms must be coherent. Coherent emission mechanisms can be generally divided into maser (including both vacuum maser and plasma instabilities or plasma maser) and the antenna mechanisms \citep{Ginzburg1969}. Within the FRB context, both types of models can operate either within or outside the magnetosphere of a the FRB source (e.g. a magnetar, see \citet{Lu&Kumar2018} and \citet{zhangRMP22} for an analysis of various coherent mechanisms).
The breakthrough discovery of a bright radio burst (FRB 200428) \citep{Bochenek2020,CHIME/FRB2020}  in association with a hard X-ray burst \citep{Mereghetti20,CKLi21,konus,AGILE} from the Galactic magnetar SGR 1935+2154 suggests that at least some FRBs are produced by magnetars born from the core collapse of massive stars. However, it is still unknown whether all FRBs, especially the active repeaters from cosmological distances, are powered by magnetars. 

Polarization properties carry important information about radiation mechanism and environment properties of FRB sources. Observationally, many non-repeaters and 9 repeaters with polarization properties have been reported, which show diverse polarization patterns. These include $\sim 100\%$ linearly polarized emission in most sources \citep{petroff2019} and most bursts for individual repeating sources \citep{Xu2021,Jiang22}, both constant \citep{Michilli2018} and varying \citep{Luo2020nature} polarization angles, and both secular \citep{Michilli2018} and short-term \citep{Xu2021,Anna2022} variations of the Faraday rotation measure (RM) for repeating sources.

The dominant feature of FRB emission is its high linear polarization. This can be generally generated in ordered magnetic fields in the emission region. Circular polarization, on the other hand, is usually not straightforwardly expected and it carries useful information about the intrinsic radiation mechanisms and propagation effects. Interestingly, current data show interesting but puzzling features regarding circular polarization, which we summarize below:
\begin{itemize}
\item  Strong circular polarization has been detected in a good fraction of non-repeating FRBs \citep{masui2015,petroff2015,Cho2020,Day2020}. 
\item It was suspected that circular polarization may be the characteristics of non-repeating FRBs that differentiate them from repeaters \citep{Dai21}.  Indeed early observations of repeating FRBs did not show significant circular polarization. Rather, they are mostly nearly 100\% linearly polarized. Some reported repeating FRB sources have not shown circular polarization yet. These include FRB 20180301A \citep{Price2019,Luo2020nature}, FRB 20180916B \citep{Nimmo2021}, FRB 20190417A \citep{Feng2022}, FRB 20190604A \citep{Feng2022}, FRB 20190711A \citep{Day2020,PKumar2021}, and a few other CHIME repeaters \citep{Fonseca2020}.
\item The situation changed after the intense observations of an active repeater FRB 20201124A. The source was detected to enter a period of increased activity in March 2021 by CHIME \citep{CHIME2021} and has been extensively studied by FAST \citep{Xu2021,ZhouDJ2022,ZhangYK2022,Jiang22,NiuJR2022} and other telescopes.
\cite{PKumar2021} reported a burst with significant circular polarized emission with circular polarization degree of $\Pi_V\sim47\%$ using the Ultra-wideband Low receiver at the Parkes radio telescope.  
Later, FAST detected many more cases of circularly polarized bursts. In an active episode 2021 April 1 to June 11, \cite{Xu2021} detected 1863 polarized bursts from FRB 20201124A in 54 days. 
In particular, some bursts (e.g. Bursts 779 and 926) show clear oscillating features as a function of wavelength, in linear and circular polarization degrees as well as the total intensity, showing evidence of possible Faraday conversion and/or synchrotron absorption \citep{Xu2021}. Later, in another 4-day active episode, more than 90\% bursts were detected with a total degree of polarization greater than 90\% \citep{Jiang22}. Some bursts have $\Pi_V > 50\%$ and even reaching $\Pi_V = 75\%$. Interestingly, those bursts with high $\Pi_V$ values usually have reduced $\Pi_L$ so that the total polarization degree remains close to $\sim 100\%$ \citep{Jiang22}.
\item A further scrutiny of archival data revealed circular polarization in two more active repeaters \citep{Feng2022b}. The first case is FRB 20121102A, the first repeater discovered by Arecibo Telescope \citep{Spitler2014,Spitler2016}.
Most of its bursts are $\sim$ 100\% linearly polarized as measured in the C-band with a very large RM in the source frame
\citep{Michilli2018}. However, the linear polarization degree decreases in the L-band and is not detectable with the Five-hundred-meter Aperture Spherical radio Telescope (FAST) \citep{LiD21}. This was interpreted as the large RM scatter due to the multi-path effect \citep{Feng2022,YXZ}. Recently, circular polarization was detected in a dozen of bursts out of nearly 2000 bursts \citep{Feng2022b}. 
\item Another case is FRB 20190520B, a FAST-discovered repeater as a close analogy of FRB 20121102A  \citep{Niu2022}. Most bursts are linear polarized, but a few bursts with circular polarization (as high as $\Pi_V\sim42\%$) have been detected \citep{Anna2022,Feng2022b}.
\end{itemize}

\begin{figure}
	\includegraphics[width=\columnwidth]{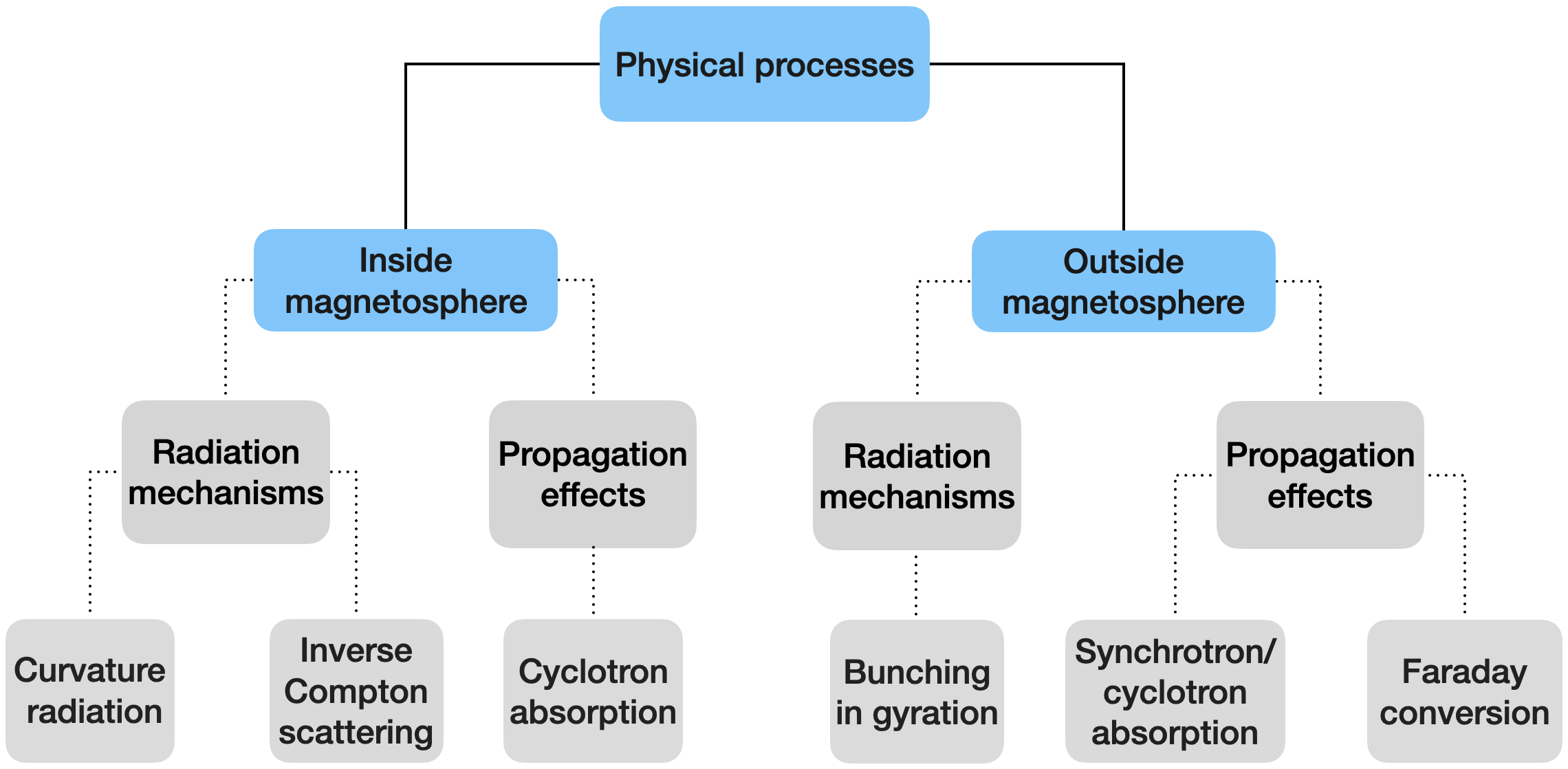}
    \caption{The physical processes to generate linear and circular polarization discussed in this paper. Two emission sites (inside and outside the magnetospheres) are considered. For each case, both intrinsic radiation mechanisms and propagation effects are discussed.}
    \label{fig:treemap}
\end{figure}

In general, linear and circular polarization can be generated both intrinsically via direct radiation mechanisms or extrinsically via propagation effects. These can happen both inside and outside the magnetospheres of the FRB sources (likely magnetars). The relevant processes that are discussed in this paper are summarized in Figure \ref{fig:treemap}.

For intrinsic radiation mechanism models, one can generally classify them to two classes based on the location of the coherent emission \citep{Zhang2020}: pulsar-like models that invoke emission processes inside or slightly outside the magnetospheres \citep[e.g.][]{Kumar2017,Yang&zhang2018,Yang&zhang2021,Wadiasingh19,kumar&Bosnjak2020,Lu20,Zhang22,QZK} 
and GRB-like models that invoke emission processes in relativistic shocks far from the magnetospheres 
\citep[e.g.][]{Lyubarsky2014,Beloborodov2017,Beloborodov2020,Plotnikov&Sironi2019,Metzger2019,Margalit2020}. The direct radiation mechanisms that may generate circular polarization include the three mechanisms as listed below. 
\begin{itemize}
\item Curvature radiation: bunched net charges moving in curved magnetic field lines have been discussed by many authors as the radiation mechanism for FRBs, with the requirement that a parallel electric field exists in the emission region to continuously supply energy to the emitting bunches \citep[e.g.][]{katz2014,Kumar2017,Yang&zhang2018,Lu20,Cooper2021,Wang2022,Wang2022b,QZK}. It has been known that linear polarization exists for an on-axis observation and circular polarization can be generated with this mechanism if the line of sight is off the emission beam \citep{Wangshort,Tong2022,Wang2022b}.This mechanism will be further analyzed in this paper with a physical understanding (Sect. \ref{sec:CR}). 
\item Inverse Compton scattering: For magnetar central engine models invoking crust cracking that sends Alfv\'en waves to the magnetosphere, the same oscillations would also emit low-frequency electromagnetic waves. As a result, bunched charges in charge starvation regions can also emit FRB emission via inverse Compton scattering (ICS) off the low-frequency waves \citep{Zhang22,QZK}. The polarization properties of such a process have been studied within the context of radio pulsars \citep{Qiao1998,Xu2000}. We will provide a fresh analysis of this problem and show that high circular polarization could be produced at an off-axis geometry in this paper (Sect \ref{sec:ICS}).
\item Synchrotron radiation: For models invoking relativistic shocks, synchrotron maser mechanism in an ordered magnetic field has been widely discussed as a mechanism to produce FRBs \citep{Lyubarsky2014,Metzger2019,Beloborodov2020,Plotnikov&Sironi2019}. The polarization properties of this mechanism have not been well studied in the literature, which will be analyzed in detail (Sec. \ref{sec:syn}).

\end{itemize}

\begin{figure}
	\includegraphics[width=\columnwidth]{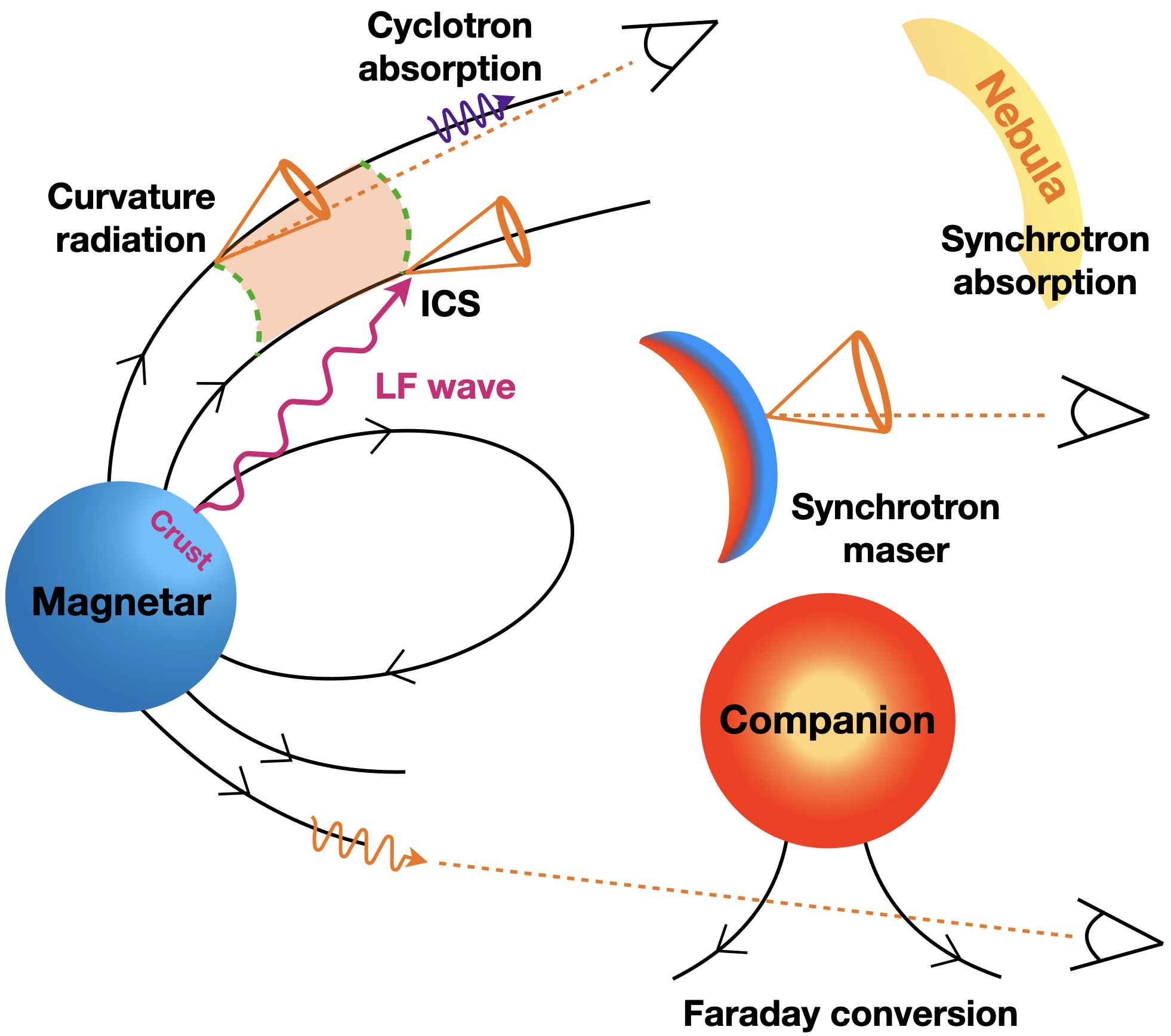}
    \caption{A cartoon picture of various physical processes discussed in this paper, which include radiation mechanisms (curvature radiation and ICS inside the magnetosphere and synchrotron radiation in relativistic, magnetized shocks outside the magnetosphere) and propagation effects (cyclotron/synchrotron absorption inside the magnetosphere and Faraday conversion / synchorotron absorption outside the magnetosphere). The orange circular cones and the orange wiggler denote FRB emission. Dashed orange lines are the directions of line of sight. The purple wiggler demotes low-frequency waves generated from magnetar surface, which are upscattered to produce FRB waves in the ICS model. The dark, purple wiggler denotes cyclotron motion of particles that potentially absorb the FRB waves.}
    \label{fig:cartoon}
\end{figure}

Various propagation effects of producing circular polarization for radio pulsars have been discussed in the literature. Here we summarize some relevant physical processes below, which will be investigated in detail within the context of FRBs in the paper:
\begin{itemize}
\item Cyclotron absorption: In the outer part of the pulsar magnetosphere, radio waves may undergo cyclotron absorption at cyclotron resonance under special conditions that invoke extremely asymmetric electron-positron plasmas \citep{Luo&Melrose2001,Petrova2006,Wang2010}. This may generate circular polarization for pulsar radio emission. The importance of this process for FRBs is investigated in Sect. \ref{sec:cyclotron-absorption}.
\item Generalized Faraday rotation (also called Faraday conversion): This effect can convert linearly polarized waves to partially circularly polarized waves under certain conditions \citep{Melrose2010,Vedantham&Ravi2019,Gruzinov&Levin2019}. 
The incident waves can be decomposed into the R-mode and L-mode in the quasi-parallel regime ($\vec{k}\parallel\vec{B}$), or into the X-mode and O-mode in the quasi-perpendicular regime\footnote{In the literature \citep{Booker1935,Melrose&McPhedran1991,Wang2010}, this is also called as quasi-tangential (QT).} ($\vec{k}\perp\vec{B}$). For each case, the dispersion relations of the two eigen-modes are different so that they propagate with different speeds. 
For the quasi-parallel regime case, different propagating speeds of the R-mode and L-mode 
would lead to rotation of the linear polarization angle with frequency, the so-called Faraday rotation, but the waves remain linearly polarized. 
For the quasi-perpendicular case (which may invoke a field line reversal), on the other hand, the phase difference between the X-mode and O-mode would make the superposed waves elliptically polarized, or linear polarization is partially converted to circular polarization. This is Faraday conversion. We investigate this process in detail in this paper under three possible scenarios (Sect \ref{sec:conversion}).
\item Selected synchrotron absorption: If the FRB source is surrounded by a synchrotron-radiating nebula, FRB waves may undergo synchrotron absorption. If the synchrotron nebula carries an ordered magnetic field, it is possible that the synchrotron absorption optical depth for two different polarization modes differ. If one mode has $\tau \gg 1$ while the other mode has $\tau \ll 1$, FRB wave modes can be selectively absorbed. This would change the final superposed polarization status (e.g.  enhancing the relative linear polarization degree for an elliptically polarized wave), causing apparent Faraday conversion. This is studied in Sect. \ref{sec:synchrotron-absorption}.
\end{itemize} 

In this paper, we generally discuss a variety of intrinsic radiation mechanisms and propagation effects of FRBs inside and outside magnetospheres, aiming to offer explanations to the polarization properties (especially circular polarization) of some FRBs. Figure \ref{fig:cartoon} is a cartoon picture for all the physical processes discussed in this paper. Our general guide line is to adequately investigate possible intrinsic radiation mechanisms and propagation effects within the FRB problem and judge whether each of them could contribute to the generation of circular polarization in FRBs. 
This paper is organized as follows. In section \ref{sec:polarization}, we introduce the basic polarization theory for a radiation field and the dispersion relations of FRBs in a cold plasma. In section \ref{sec:magnetosphere}, we discuss the possible linear/circular polarization processes within a magnetar magnetophere, including coherent curvature and ICS radiation by charged bunches and cyclotron resonance absorption.
In section \ref{sec:outside}, we discuss possible linear/circular polarization processes outside a magnetar magnetophere, including synchrotron maser radiation, Faraday conversion, and selected synchrotron absorption. 
The main conclusions and discussions are summarized in section \ref{sec:conclusions}. Throughout the paper, the convention $Q=10^n Q_n$ in cgs units is adopted.

\section{Polarization of a radiation field and radiation transfer of polarized emission}\label{sec:polarization}

The polarization properties of a quasi-monochromatic electromagnetic wave can be described by four Stokes parameters \citep{Rybicki&Lightman1979}
\begin{equation}\label{Stokes}
\begin{aligned}
&I=\frac{1}{2}(E_x^{*}E_x+E_y^{*}E_y)\\
&Q=\frac{1}{2}(E_x^{*}E_x-E_y^{*}E_y)\\
&U={\rm Re}(E_x^{*}E_y)\\
&V={\rm Im}(E_x^{*}E_y),
\end{aligned}
\end{equation}
where $E_x$ and $E_y$ are the electric vector amplitudes of two linearly polarized wave eigen-modes  perpendicular to the line of sight (LOS), the superscript $``*"$ denotes the conjugation of $E_{x/y}$, $I=\vert \Vec{E} \vert^2$ defines the total intensity, $Q$ and $U$ define linear polarization and its position angle, and $V$ describes  circular polarization. The linear, circular, and the overall degree of polarization are described by $\Pi_{L}=(Q^2+U^2)^{1/2}/I$, $\Pi_{V}=V/I$ and $\Pi_{P}=(Q^2+U^2+V^2)^{1/2}/I$, respectively, all of which are $\leq1$. The Faraday rotation angle is defined as $\Phi=\tan^{-1}(U/Q)$. 

One can define a four vector using the Stokes parameters, which can undergo generalized Faraday rotation and absorption. The generalized Faraday rotation can be described as a $4\times4$ matrix \citep{Melrose&McPhedran1991}
\begin{equation}
\rho_{AB}=\left( 
  \begin{array}{cccc}  
    0 & 0 & 0 & 0\\
    0 & 0 & -\rho_V & \rho_U\\
    0 & \rho_V & 0 & -\rho_Q\\
    0 & -\rho_U & \rho_Q & 0\\
  \end{array}
\right),
\end{equation}
where $\rho_V$ is Faraday rotation coefficient, and $\rho_Q$ and $\rho_U$ are Faraday conversion coefficients. One can also write an absorption matrix \citep{Melrose&McPhedran1991}
\begin{equation}
\eta_{AB}=\left( 
  \begin{array}{cccc}  
    \eta & \eta_Q & \eta_U & \eta_V\\
    \eta_Q & \eta & 0 & 0\\
    \eta_U & 0 & \eta & 0\\
    \eta_V & 0 & 0 & \eta\\
  \end{array}
\right),
\end{equation}
where $\eta$, $\eta_Q$, $\eta_U$ and $\eta_V$ are the absorption coefficients for $I$, $Q$, $U$ and $V$, respectively. Therefore, the general radiation transfer equation can be written as\footnote{The third matrix describing the generalized Faraday rotation and absorption is obtained by $\rho_{AB}-\eta_{\rm AB}$.} \citep{Sazonov1969}
\begin{equation}\label{general}
\frac{d}{ds}\left(
  \begin{array}{ccc}  
    I\\
    Q\\
    U\\
    V\\
  \end{array}
\right)=\left(
  \begin{array}{ccc}
    \epsilon_I\\
    \epsilon_Q\\
    \epsilon_U\\
    \epsilon_V\\
  \end{array}
\right)-\left( 
  \begin{array}{cccc}  
    \eta & \eta_Q & \eta_U & \eta_V\\
    \eta_Q & \eta & \rho_V & -\rho_U\\
    \eta_U & -\rho_V & \eta & \rho_Q\\
    \eta_V & \rho_U & -\rho_Q & \eta\\
  \end{array}
\right)\left( 
  \begin{array}{ccc}  
    I\\
    Q\\
    U\\
    V\\
  \end{array}
\right).
\end{equation}
{where $\epsilon_I$, $\epsilon_Q$, $\epsilon_U$ and $\epsilon_V$ are the spontaneous emission coefficients.}
Without loss of generality, one can define a coordinate system to make $\eta_U=0$, so that $\eta$, $\eta_Q$ and $\eta_V$ can be considered as isotropic, linear and circular absorption coefficients, respectively\footnote{For comparison, the linear absorption coefficient $\eta_{Q}$ defined here is $\eta_{L}$ in \citep{Xu2021}.}. Note that the $\eta$ values are defined as positive for absorption. Under special conditions, they can be negative, which denote the inverse emission process. For such a case, the waves would undergo maser amplifications\footnote{This would correspond to certain FRB generation mechanisms. Since the proposed processes discussed in the literature \citep[e.g.][]{waxman17} does not predict polarized radiation, we do not discuss these models in this paper.}. 

In principle, one can integrate the radiation transfer equation along the wave path from the source to the observer once the parameters of the medium and initial polarized conditions are known. The initial four Stokes parameters are determined by incident orthogonal modes. The absorption coefficients and the Faraday coefficients can be determined by waves dispersion relations in the plasma. 

We consider the background magnetic field $\vec B$ is along the $z$-axis and the plasma is cold, non-relativistic, and uniformly distributed in space. We describe the plasma dispersion relations in the linear regime below.
\begin{itemize}
\item  For the cold electron-ion plasma case: In general, the refractive index of a wave propagating through a cold electron-ion plasma can be written as \citep{Stix1992}
\begin{equation}
n^2=1-\frac{2\omega_p^2(\omega^2-\omega_p^2)/\omega^2}{2(\omega^2-\omega_p^2)-\Omega_e^2\sin^2\theta\pm\Omega_e\Delta},
\end{equation}
where $\omega_p=\sqrt{4\pi q^2n/m_e}$ is the plasma frequency, {$\theta$ is the angle between the wave vector and the background magnetic field direction,} $n$ is the lepton number density, $\Delta=[\Omega_e^2\sin^4\theta+4(\omega^2-\omega_p^2)\cos^2\theta/\omega^2]^{1/2}$ and $\Omega_e=-eB/(m_ec)=-\omega_B$ is the electron cyclotron frequency  (which is the Larmor frequency $\omega_B$ with a negative sign).
The conditions of quasi-perpendicular and quasi-parallel regimes can be written as $\omega_B^2\sin^4\theta\gg4(\omega^2-\omega_p^2)^2\cos^2\theta/\omega^2$
and $\omega_B^2\sin^4\theta\ll4(\omega^2-\omega_p^2)^2\cos^2\theta/\omega^2$, respectively. 

Under the quasi-perpendicular condition, the dispersion relations of the X-mode and O-mode can be expressed as (see the Appendix \ref{B} for a derivation)
\begin{equation}\label{Xmodeion}
n^2_{\rm X}=\frac{(\omega^2-\omega_p^2)^2-\omega^2\Omega_e^2\sin^2\theta}{\omega^2(\omega^2-\omega_p^2)-\omega^2\Omega_e^2\sin^2\theta}\simeq1-\frac{\omega_p^2(\omega^2-\omega_p^2)}{\omega^4-\omega^2(\omega_p^2+\omega_B^2)},
\end{equation}
and
\begin{equation}\label{Omodeion}
n_{\rm O}^2=\frac{\omega^2-\omega_p^2}{\omega^2-\omega_p^2\cos^2\theta}\simeq1-\frac{\omega_p^2}{\omega^2},
\end{equation}
where we have applied $\theta=\pi/2$ in the last step to denote the case when the wave vector is perpendicular to background magnetic field. 

Under the quasi-parallel condition, on the other hand, 
the dispersion relations can be written as
\begin{equation}\label{Rmode}
n_{\rm R}^2=1-\frac{\omega_p^2}{\omega(\omega-\omega_B\cos\theta)}\simeq1-\frac{\omega_p^2}{\omega(\omega-\omega_B)}
\end{equation}
and
\begin{equation}\label{Lmode}
n_{\rm L}^2=1-\frac{\omega_p^2}{\omega(\omega+\omega_B\cos\theta)}\simeq1-\frac{\omega_p^2}{\omega(\omega+\omega_B)},
\end{equation}
where the approximation $\theta\rightarrow 0$ is applied in the last step to denote the case when the wave vector is parallel to background magnetic field.

\item  For the cold pair (electron-positron) plasma case: The dispersion relations in the quasi-perpendicular case can be written as (see Appendix \ref{B} for a derivation)
\begin{equation}
n_{\rm X}^2=1-\frac{\omega_p^2}{\omega^2-\omega_B^2\sin^2\theta}\simeq1-\frac{\omega_p^2}{\omega^2-\omega_B^2},
\end{equation}
and
\begin{equation}\label{pairO}
n_{\rm O}^2=\frac{(\omega^2-\omega_p^2)(\omega^2-\omega_p^2-\omega_B^2)}{\omega^2(\omega^2-\omega_p^2)+\omega_B^2(\omega_p^2-\omega^2+\omega_p^2\cos2\theta)}\simeq1-\frac{\omega_p^2}{\omega^2},
\end{equation}
where we apply $\theta=\pi/2$ in the last step. For the quasi-parallel case, the dispersion relations  
can be written as
\begin{equation}
n_{\rm R}^2=n_{\rm L}^2=1-\frac{\omega_p^2}{(\omega-\omega_B\cos\theta)(\omega+\omega_B\cos\theta)}\simeq1-\frac{\omega_p^2}{\omega^2-\omega_B^2},
\end{equation}
where we apply $\theta=0$ in the last step.
\end{itemize}

One can see that the dispersion relations for two eigen-modes are usually different (the only exception is the pair plasma in the parallel case). The different propagation speeds of the two eigen-modes would result in Faraday rotation (quasi-parallel) or Faraday conversion (quasi-perpendicular). In the following, we will discuss the processes within a magnetosphere (with the plasma being an electron-positron plasma) and outside a magnetosphere (with the medium plasma being an electron-ion plasma) separately.

\section{Inside magnetospheres}\label{sec:magnetosphere}
In this section, we discuss two coherent emission mechanisms (bunched curvature and inverse Compton scattering) and one possible propagation effect inside magnetospheres. 

In order to quantitatively calculate the dispersion relations of a pair plasma, one needs to calculate the characteristic values of the plasma frequency and Larmor frequencies. We consider a magnetospheric plasma in the open field line region at a typical altitude $\hat{r} = r/R_* \sim10-100$, where $R_{\star}=10^6 \ \rm cm$ is the radius of magnetar. At this location, the plasma frequency can be estimated as
\begin{equation}
\begin{aligned}
\omega_p&=\sqrt{\frac{4\pi e^2\xi n_{\rm GJ}}{m_e}}\simeq\sqrt{\frac{4\pi e\xi B_\star}{Pm_ec}}\left(\frac{r}{R_\star}\right)^{-3/2}\\
&\simeq(4.7\times10^9 \ {\rm rad \ s^{-1}}) \ \xi_2^{1/2}B_{\star,15}^{1/2}P^{-1/2} \hat r_2^{-3/2},
\end{aligned}
\end{equation}
where $\xi$ is the pair multiplicity factor with respect to the Goldreich Julian density, and $B_{\star}$ is the surface magnetic field of the magnetar. The cyclotron frequency is
\begin{equation}
\omega_B=\frac{eB}{m_ec}\simeq\frac{eB_\star}{m_ec}\left(\frac{r}{R_\star}\right)^{-3}\simeq(1.8\times10^{16} \ {\rm rad \ s^{-1}}) \ B_{\star,15} \hat r_2^{-3}.
\end{equation}
The ratio between the plasma frequency and the cyclotron frequency can be calculated as 
\begin{equation}
\begin{aligned}
\frac{\omega_p}{\omega_B}&=\frac{m_ec}{eB}\sqrt{\frac{4\pi e^2\xi n_{\rm GJ}}{m_e}}\simeq\sqrt{\frac{4\pi\xi m_ec}{B_\star Pe}}\left(\frac{r}{R_\star}\right)^{3/2}\\
&\simeq2.7\times10^{-7} \ \xi_2^{1/2}B_{\star,15}^{-1/2}P^{-1/2} \hat r_2^{3/2}
\ll 1.
\end{aligned}
\end{equation}

\subsection{Emission mechanisms}
In this section, we discuss coherent curvature and ICS radiation by bunches, respectively. For simplicity, we consider a point-like bunch so that the emission beaming angle is $1/\gamma$, where $\gamma$ is the Lorentz factor of the bunch. 
Hereafter, we define the on-beam case is within the $1/\gamma$ cone and off-beam case is outside the cone. In reality, the bunch shape can be more complicated (a pancake shape) so that the bunch opening angle can be larger \citep[e.g.][]{zhangRMP22}. Emission properties for such a more complicated geometry has been studied by \cite{Wang2022b} within the framework of bunched curvature radiation. 

When studying emission mechanisms, what matters is the emission properties at the emission region. For the convenience of treatment, hereafter we replace the electric field in the detected waves to the amplitude of emission in the source region, i.e.  $\vec{A}=(c/4\pi)^{1/2}(R\vec{E})_{\rm rec}$, where the power radiated per unit solid angle can be written as ${dP}/{d\Omega}=|\vec A|^2$ \citep{Jackson1998}, and $R$ is the distance from the radiation source to the field point (observer) at the retarded time. Therefore, based on Eq. (\ref{Stokes}), the degree of linear and circular polarization in the Fourier space can be re-written as
\begin{equation}\label{degree}
\begin{aligned}
&\Pi_L=\frac{\sqrt{(A_\parallel^2+A_\perp^2)(A_\parallel^{*2}+A_\perp^{*2})}}{A_\parallel A_\parallel^{*}+A_\perp A_\perp^{*}},\\
&\Pi_V=-i\frac{A_\parallel A_\perp^{*}-A_\perp A_\parallel^{*}}{A_\parallel A_\parallel^{*}+A_\perp A_\perp^{*}}.
\end{aligned}
\end{equation} 

\begin{figure*}
\begin{center}
\begin{tabular}{ll}
\resizebox{80mm}{!}{\includegraphics[]{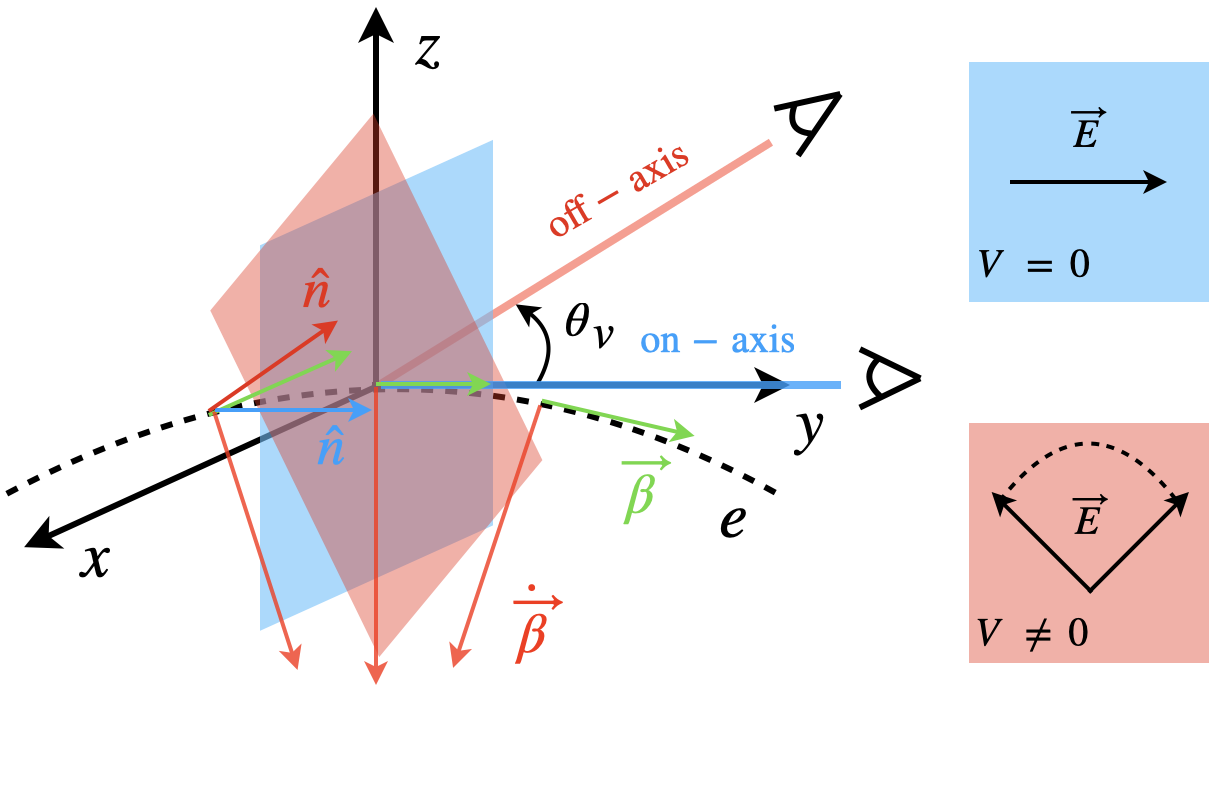}}&
\resizebox{80mm}{!}{\includegraphics[]{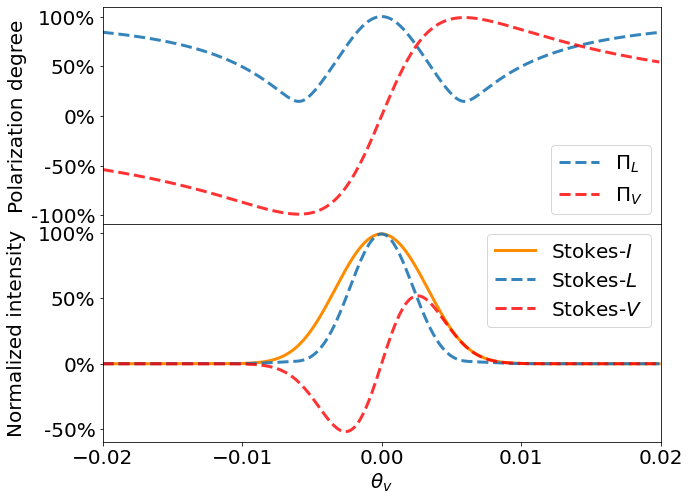}}
\end{tabular}
\caption{Left panel: A direct image way to understand how circular polarization is detected in off-beam case. The dashed curve is the electron's trajectory and background magnetic field in $x-y$ plane. LOS ($\hat n$) along on-axis and off-axis cases are blue  and red line, respectively. The green arrow is electron velocity ($\vec\beta$). The curvature acceleration $\dot{\vec\beta}$ is perpendicular to trajectory in $x-y$ plane. Blue and red planes are perpendicular to LOS along on-axis and off-axis directions, respectively. Right panel: Simulated circular ($V$), linear ($L$) and total intensity ($I$) polarization fractions of coherent bunches curvature emission as a function of  angle $\phi$ between LOS and the trajectory plane from -0.02 to 0.02. The Lorentz factor of the bunch is $\gamma=100$, inclination angle between rotation axis and magnetic axis is $30^\circ$ and emission frequency equal to critical curvature radiation frequency, i.e. $\omega=\omega_c$ are applied.}
\label{fig:cur}
\end{center}
\end{figure*}

\subsubsection{Coherent curvature radiation}\label{sec:CR}

A commonly discussed FRB emission mechanism inside a magnetar magnetosphere is coherent curvature radiation by bunches. Within this model, relativistic bunch of particles, typically at a radius of 10s to 100s times of neutron star radius, is believed to radiate coherently to power the bright FRB emission. To sustain high emission power, the bunches need to be continuously accelerated by an electric field ($E_\parallel$) parallel to the local magnetic field, possibly produced when crust-oscillation-driven Alf\'ven waves propagate to the charge starvation region \citep{Lu20,kumar&Bosnjak2020,kumar22}.
\cite{Yang&zhang2018} have calculated the spectrum of coherent curvature radiation in different geometric conditions of bunches. 
The polarization properties of curvature radiation have been studied within the framework of pulsars \citep[e.g.][]{Gil1990} and FRBs \citep{Gangadhara2010,Wang2022,Wang2022b,Tong2022,Liu22} by various authors.

Consider a bunch with charge $Q = N_e e$ and Lorentz factor $\gamma$ moving along the local magnetic field with curvature radius $\rho$. The condition of coherence 
requires that the longitudinal size is smaller than the wavelength of FRB emission $\lambda\simeq(30 \ {\rm cm}) \ \nu_9^{-1}$.
The net charge number density in the magnetosphere in terms of GJ density \citep{GJ1969} can be written as
\begin{equation}
n=\zeta n_{\rm{GJ}}\simeq\frac{\xi B_{\star}\Omega}{2\pi qc}\left(\frac{r}{R_{\star}}\right)^{-3}\simeq(7\times10^{9} \ {\rm cm^{-3}}) \ \zeta_{2}B_{\star,15}P^{-1}\hat{r}_{2}^{-3},
\end{equation}
where $\zeta$ is the net charge factor normalized to $10^2$.
Thus the total number of net charges in one bunch can be estimated as $N_e=nAl_\parallel\simeq10^{20} \ \zeta_2B_{\star,15}P^{-1}\hat{r}_2^{-3}\nu_9^{-3}$, where $A \sim \pi (\gamma \lambda)^2$ is the cross section of the bunch. To produce $\sim$ 1 $\rm GHz$ curvature radiation, the required Lorentz factor can be written as
\begin{equation}
\gamma\simeq241(\rho_8\nu_9)^{1/3}.
\end{equation}
To maintain the bunch in the radiation-reaction-limited regime to power the bright FRB emission, one requires a parallel electric field \citep{Kumar2017}
\begin{equation}
E_{\parallel}\sim(1.1\times10^{4} \ {\rm{esu}}) \ N_{e,20} \rho_8^{-2/3}\nu_9^{4/3}.
\end{equation}

In the following, we discuss the curvature radiation polarization properties of the bunch. The radiation properties of a macro charge bunch is similar to those of a single electron, which is well described in textbooks  \citep[e.g.][]{Jackson1998}. For convenience, in the following we consider one electron when the polarization properties of a certain radiation mechanism is considered.  

The radiation electric field of a single moving electron is given by \citep{Jackson1998}
\begin{equation}
\vec E_{\rm rad}(\vec r,t)=\frac{e}{c}\left[\frac{\hat{n}\times\{(\hat{n}-\vec \beta)\times\Dot{\Vec{\beta}}\}}{(1-\hat{n}\cdot\Vec{\beta})R}\right]_{\rm rec},
\end{equation}
where $\hat{n}$ is the unit vector along the wave propagation direction. The polarization state of curvature radiation depends on the relative direction of relativistic beaming and the line of sight (LOS) significantly. We present a cartoon picture in the left panel of Fig. \ref{fig:cur} and explain how circular polarization can be generated in an off-beam case below. We consider the electron moves along a background strong magnetic field, i.e. the black dashed curve. A linearly polarized wave means that the electric field vector is a straight line projected in the plane perpendicular to the LOS. When an observer is along the blue solid line, i.e. the on-axis case, the direction of $\vec{E}_{\rm rad}$ is along the vector $\hat{n}\times{(\hat{n}-\vec \beta)\times\Dot{\Vec{\beta}}}$, which is the projection of $\dot{\vec{\beta}}$ on the blue plane perpendicular to the LOS. The acceleration vector $\dot{\vec\beta}$ may change its amplitude but it is always in $x-y$ plane, thus the projection of $\dot{\vec{\beta}}$ is always a straight line, i.e. the observer always observes linearly polarized radiation. For the off-beam case (the red line direction), on the other hand, one should look at the the projection of vector $\dot{\vec{\beta}}$ on red plane. Carefully inspecting the projected direction of $\dot{\vec{\beta}}$, one can see that it tracks a curve as a function of time, i.e. the observer could see a circularly polarized radiation.

We now calculate the polarization properties of curvature radiation from the first principle.
The energy radiated per unit solid angle per unit frequency interval for a single charged particle is given by \citep{Rybicki&Lightman1979,Jackson1998}
\begin{equation}
\begin{aligned}
\frac{d^2I}{d\omega d\Omega}&=\frac{e^2\omega^2}{4\pi^2c}\left\vert \int_{-\infty}^{\infty}\vec n\times(\vec n\times\vec \beta)e^{i\omega(t-\Vec{n}\cdot\Vec{r}(t)/c)}dt\right\vert^2\\
&=\frac{e^2\omega^2}{4\pi^2c}\left\vert -\Vec{\epsilon}_{\parallel}A_{\parallel}(\omega)+\Vec{\epsilon}_{\perp}A_{\perp}(\omega)\right\vert^2,
\end{aligned}
\end{equation}
where the unit vector $\vec\epsilon_\parallel$ is along the direction of the instantaneous curvature radius in the orbital plane and $\vec \epsilon_\perp=\vec n\times\vec\epsilon_\parallel$ is the orthogonal polarization vector.
The amplitudes of electric fields can be written as
\begin{equation}
A_{\parallel}(\omega)\simeq\frac{\rho}{c}\left(\frac{1}{\gamma^2}+\theta_v^2\right)\int_{-\infty}^{\infty}xe^{\left[i\frac{3}{2}\varsigma\left(x+\frac{1}{3}x^3\right)\right]}dx,
\end{equation}
and
\begin{equation}
A_{\perp}(\omega)\simeq\frac{\rho}{c}\theta_v\left(\frac{1}{\gamma^2}+\theta_v^2\right)^{1/2}\int_{-\infty}^{\infty}e^{\left[i\frac{3}{2}\varsigma\left(x+\frac{1}{3}x^3\right)\right]}dx,
\end{equation}
where $x=ct/[\rho(1/\gamma^2+\theta^2)^{1/2}]$ is a replacement variable, $\varsigma=\omega\rho(1/\gamma^2+\theta_v^2)^{3/2}/3c$ and $\theta_v$ is the angle between the LOS and the trajectory plane. The integrals in the above equation can be written as 
\begin{equation}
\int_{-\infty}^{\infty}x\sin\left[{\frac{3}{2}\varsigma\left(x+\frac{1}{3}x^3\right)}\right]=\frac{2}{\sqrt{3}}K_{2/3}(\varsigma),
\end{equation}
and
\begin{equation}
\int_{-\infty}^{\infty}\cos\left[{\frac{3}{2}\varsigma\left(x+\frac{1}{3}x^3\right)}\right]=\frac{2}{\sqrt{3}}K_{1/3}(\varsigma),
\end{equation}
where $K_{\nu}(\varsigma)$ is the modified Bessel function. Without loss generality, we consider an electron moving along the magnetic field line in the direction with an angle $\chi$ with respect to the initial trajectory at the retarded time $t=0$, the amplitudes of emission can be written as
\begin{equation}
\begin{aligned}
A_{\parallel}(\omega)&\simeq\frac{i2\rho}{\sqrt{3}c}\left(\frac{1}{\gamma^2}+\theta_v^2+\chi^2\right)K_{2/3}(\varsigma)\\
&+\frac{2\rho}{\sqrt{3}c}\chi\left(\frac{1}{\gamma^2}+\theta_v^2+\chi^2\right)^{1/2}K_{1/3}(\varsigma),
\end{aligned}
\end{equation}
and
\begin{equation}
A_{\perp}(\omega)\simeq\frac{2\rho}{\sqrt{3}c}\theta_v\left(\frac{1}{\gamma^2}+\theta_v^2+\chi^2\right)^{1/2}K_{1/3}(\varsigma),
\end{equation}
where $x=ct/[\rho(1/\gamma^2+\theta^2+\chi^2)^{1/2}]$ and $\varsigma=\omega\rho(1/\gamma^2+\theta_v^2+\chi^2)^{3/2}/3c$ are modified.
We integrate the two components of the electric field amplitudes and insert them into Eq. (\ref{Stokes}) to obtain the evolution of the Stokes parameters. The linear and circular polarization degrees ($\Pi_L$ and $\Pi_V$) as well as the $I$, $L$ and $V$ parameters as a function of viewing angle are presented in the right panel of Fig. \ref{fig:cur}. One can see $\Pi-L$ is $100$\% at $\theta_v=0$, i.e. $A_{\perp}(\omega)=0$. As $\theta_v$ increases, circular polarization starts to appear. The $V$ parameter reaches the peak at $\theta_v \sim 0.0026$ and $\Pi_V$ reaches the peak at $\theta_v \sim 0.006$ and stays high at $\theta_v > 1/\gamma$ where the emission flux is much degraded. Since within the $1/\gamma$ cone the emission power is comparable, the probability of detecting high circular polarization for point-like bunches is quite high. Observationally, since circular polarization only appears in a small fraction of bursts, one needs to invoke more complicated geometry, e.g. the emission beam solid angle is much larger than $1/\gamma$, to account for observations. In such a geometry, within the broader emission beam there are always on-beam bunches that produce $\sim 100\%$ linear polarization \citep{Wang2022b}.

\subsubsection{Coherent inverse Compton scattering}\label{sec:ICS}

The coherent inverse Compton scattering (ICS) process by charged bunches off low frequency electromagnetic waves produced by inner gap sparking and the corresponding polarization properties have been discussed by \cite{Qiao1998} and \cite{Xu2000} within the context of radio pulsars. Within the context of FRBs, \cite{Zhang22} proposed the coherent ICS model invoking low-frequency electromagnetic waves generated by near-surface oscillating charges due to crust shaking that also drives Alfv\'en waves: Particles accelerated in the charge starvation region at a high altitude (e.g. due to charge depletion as Alfv\'en waves propagate to a critical radius \citep{kumar&Bosnjak2020}) would coherently upscatter the incoming electromagnetic waves with angular frequency $\omega_{i}$ and power the observed $\sim$ GHz FRB emission. 
With a general incident angle $\theta_i$ between the low-frequency waves and charge motion direction, one can generally define two eigen-modes:
Mode (1) or "X"-mode: the wave electric field $\vec E$ is perpendicular to the $(\vec k, \vec B)$ plane, and Mode (2) or "O"-mode: $\vec E$ is parallel to the $(\vec k, \vec B)$ plane. It should be pointed out that the terminology so called X-mode and O-mode is more relevant in the quasi-perpendicular case, i.e. $\vec k$ and $\vec B$ are nearly perpendicular to each other. In the quasi-parallel case ($\vec k$ and $\vec B$ are nearly parallel), both Modes (1) and (2) are nearly extraordinary and the so-called O-mode (Mode (2)) is similar to the X-mode and can propagate in a magnetar magnetosphere \citep{QZK}. 
Different from curvature emission, coherent ICS by bunches discussed in this model only requires merely charge density fluctuations in an relativistic particle outflow to satisfy the observed luminosity of FRBs \citep{Zhang22}. In order to produce 1-GHz radio waves through the ICS mechanism, one requires {the Lorentz factor of the bunch to be} $\Gamma_{\rm ICS}\simeq316 \ \nu_{\rm 9}^{1/2}\nu_{i,4}^{-1/2}(1-\beta\cos\theta_{i})^{-1/2}$, {where the frequency of the incident wave is normalized to $\nu_i=10$ kHz}. The rapid radiative cooling of the bunch requires that a parallel electric field $E_{\parallel,\rm ICS}$ along the magnetic field lines must be produced. Thus the balance between acceleration and ICS radiation cooling requires \citep{Zhang22}
\begin{equation}
\begin{aligned}
E_{\parallel,\rm ICS}&\simeq(8.6\times10^{11} \ {\rm esu}) \ \zeta_2\nu_9^{-3}\Gamma_{\rm ICS,2.5}^{2}B_{\star,15}P^{-1}f(\theta_{i})\delta B_{0,6}^2\hat{r}_2^{-5}\\
&\simeq(8.6\times10^{6} \ {\rm esu}) \ \zeta_2\nu_9^{-3}\Gamma_{\rm ICS,2.5}^{2}B_{\star,15}P^{-1}f(\theta_{i})\delta B_{0,6}^2\hat{r}_3^{-5},
\end{aligned}
\end{equation}
{where $f(\theta_i)=\sin^2\theta_i/(1-\beta\cos\theta_i)$ is defined to describe the cross section and $\delta B$ is the magnetic field strength of the low frequency waves near the magnetar surface region, which we normalize to a relatively small value $\delta B=10^6$ G.}
{One can see that the parallel electric field strength} is so large in the radius range ($10^8-10^9$ cm) that the surrounding plasma could be separated since $E_{\parallel,\rm ICS}$ can overcome the plasma Coulomb potential and the plasma suppression effect can be ignored \citep{QZK}. In this subsection, our plan as follows. First, we will discuss ICS by a single non-relativistic electron and study its polarization properties. 
Then, we will present a detail calculation on ICS by a charged bunch and the generation of circular polarization. Finally, we will discuss the main differences of polarization properties between coherent curvature radiation and ICS radiation by bunches. Hereafter, subscripts $i$ and $s$ denote the incident and scattered waves, respectively.

\begin{figure*}
\begin{center}
\begin{tabular}{ll}
\resizebox{80mm}{!}{\includegraphics[]{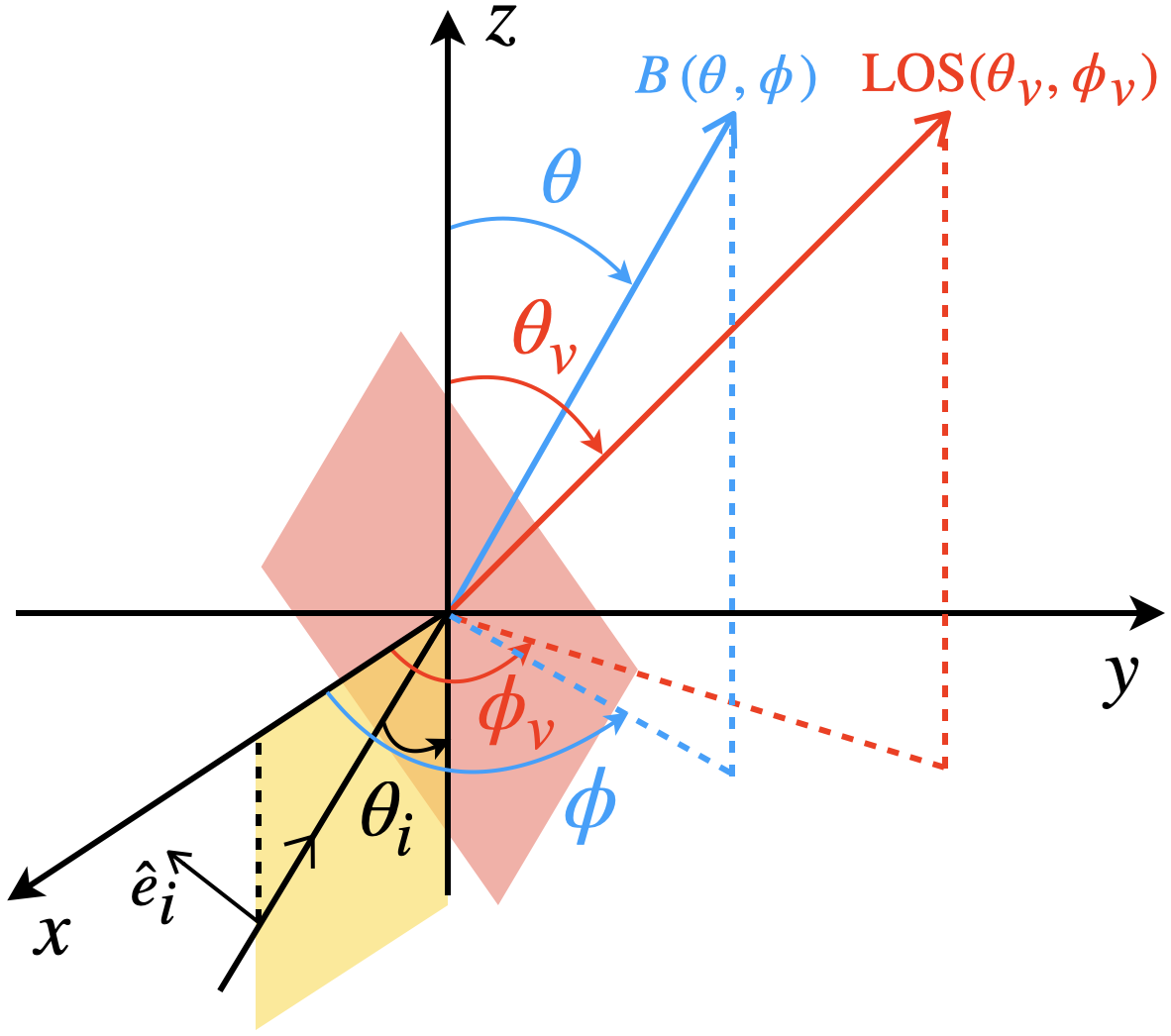}}&
\resizebox{80mm}{!}{\includegraphics[]{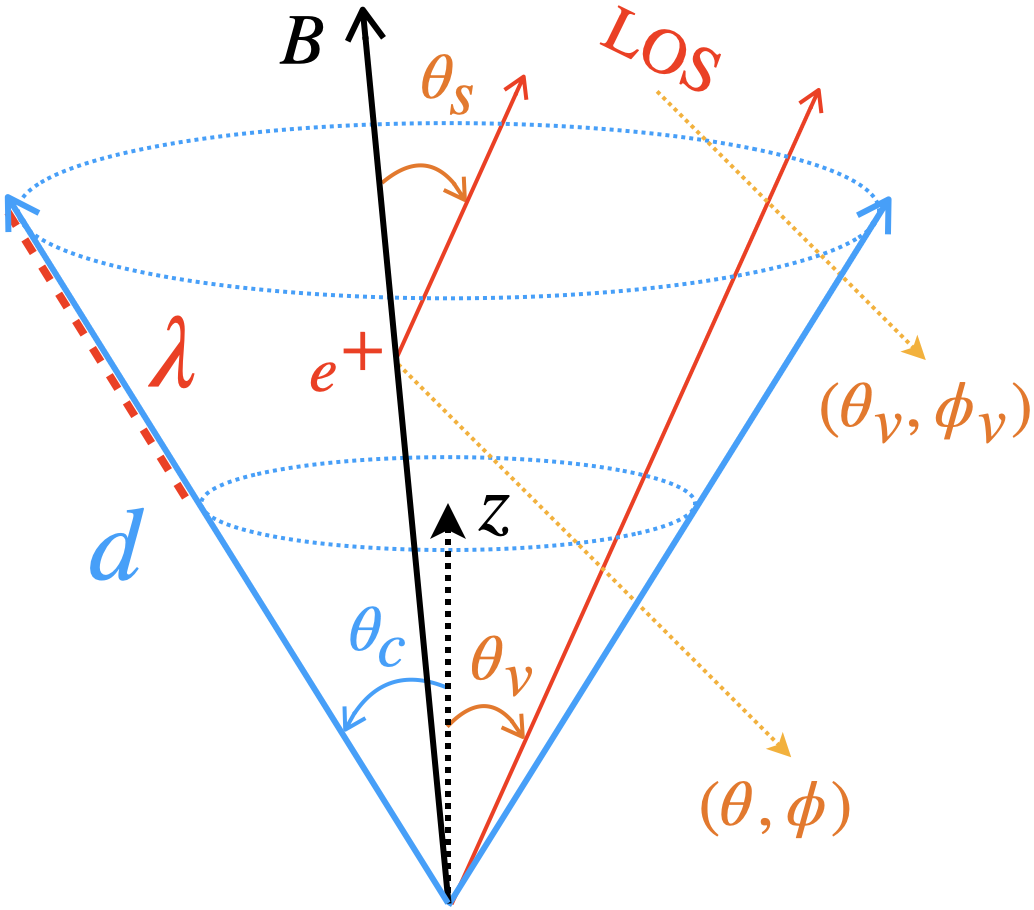}}
\end{tabular}
\caption{Left panel: The geometry of inverse Compton scattering for one relativistic electron at the origin off a low-frequency photon with an incident angle $\theta_i$. The incident waves have a unit vector of electric field $\hat{e}_i$ in the $x-z$ plane (yellow). The background magnetic field is along the direction ($\theta,\phi$). The LOS (red line) is along an arbitrary direction ($\theta_v,\phi_v$) and the red plane is perpendicular to the LOS. Right panel: The geometry of an ICS bunch. The distance $d$ is from the origin to the bunch boundary and the wavelength $\lambda$ is the longitudinal length of the bunch. The black dashed line (the $z$ axis) is the symmetric axis of the bunch and the angle $\theta_v$ is between the $z$-axis and the LOS (red lines). The angle $\theta_s$ is between the LOS and arbitrary magnetic field (black line) in the cone. The angle $\theta_c$ is between the cone boundary and the $z$-axis. Following parameters are adopted: bunch longitudinal size $\lambda=30$ cm, $d=10^5$ cm, $\theta_c=1/\gamma=0.01$ and Lorentz factor $\gamma$=100.}
\label{fig:ICSfig}
\end{center}
\end{figure*}

(i) ICS radiation by a single electron: The ICS geometry of a single electron is presented in the left panel of Fig. \ref{fig:ICSfig}.
Generally, we consider an arbitrary strong background magnetic field $\vec B$ along $(\theta,\phi)$ (blue line) and the electron at origin can only move along the magnetic field. The unit vector of LOS (red line) is chosen as $\vec{n}=(\sin\theta_v\cos\phi_v,\sin\theta_v\sin\phi_v,\cos\theta_v)$ in the lab frame (hereafter, all quantities in co-moving frame are denoted with a prime ($'$)). The incident low-frequency electromagnetic wave vector is in the $x-z$ plane with an incident angle $\theta_{i}$ between the wave vector and $z$-axis. Assuming that the incident waves are linearly polarized with the electric field vector in the $x-z$ plane, one has the electric field unit vector $\hat{e}_i=(\cos\theta_{i},0,\sin\theta_i)$ and the three components of the incident electric field as $E_i=(E_{i,0}\cos\theta_{i},0,E_{i,0}\sin\theta_i)$.

The initial rest electron motion equation in the lab frame can be written as
\begin{equation}
m\frac{d^2\vec{r}}{dt^2}=e\Vec{E}+\frac{e}{c}\left(\frac{d\vec{r}}{dt}\times\Vec{B}\right),
\end{equation}
where $\vec{r}$ is the position vector.
The scattered electric field can be calculated by applying Larmor’s formula as \citep{Rybicki&Lightman1979}
\begin{equation}
\begin{aligned}
\vec E_s&=\frac{e}{c}\left[\frac{\vec{n}\times(\vec{n}\times\Dot{\vec{{\beta}}})}{R}\right]_{\rm rec}.
\end{aligned}
\end{equation}
The direction of $\vec E_s$ is determined by $\vec{n}\times(\vec{n}\times\Dot{\vec{{\beta}}})$ and this vector is $\Dot{\vec{{\beta}}}$ projected in the plane perpendicular to LOS. In the non-relativistic case, the direction of $\Dot{\vec{{\beta}}}$ is the same as that of the incident electric field $\vec E_i$, i.e. we can project $\vec E_i$ onto the red plane in Fig. \ref{fig:ICSfig}.
One can see that the electric field perpendicular to the LOS only has one direction and the observer can always detect linear polarized waves. This is consistent with classical result that the upscattered photons in ICS are always linearly polarized.

In order to obtain the scattered electric field of a single relativistic electron through ICS, we take the following steps 
(see Appendix \ref{A} for detailed derivations): In the first step, we solve the scattered electric field $E_s'$ in the co-moving frame of one electron by applying Larmor's formula written in the $xyz$-frame as
\begin{equation}
\begin{aligned}
E_{s,x}'=&-\frac{e\omega_i'^2}{c^2D'}(-{r_x'}\sin^2{\phi_v'} \sin ^2{\theta_v'}+{r_y'}\sin{\phi_v'}\cos{\phi_v'}\sin^2{\theta_v'}\\
&+{r_z'}\cos{\phi_v'}\sin{\theta_v'} \cos {\theta_v'}-{r_x'}\cos^2{\theta_v'}).
\end{aligned}
\end{equation}
\begin{equation}
\begin{aligned}
E_{s,y}'=&-\frac{e\omega_i'^2}{c^2D'}({r_x'} \sin{\phi_v'} \cos {\phi_v'} \sin ^2{\theta_v'}-{r_y'} \cos^2{\phi_v'}\sin^2{\theta_v'}\\
&+{r_z'}\sin {\phi_v'}\sin{\theta_v'}\cos{\theta_v'}-{r_y'}\cos^2{\theta_v'}).
\end{aligned}
\end{equation}
\begin{equation}
\begin{aligned}
E_{s,z}'=&-\frac{e\omega_i'^2}{c^2D'}({r_x'} \cos{\phi_v'} \sin {\theta_v'} \cos{\theta_v'}+{r_y'} \sin{\phi_v'} \sin{\theta_v'}\cos{\theta_v'}\\
&-{r_z'} \sin^2{\phi_v'} \sin ^2{\theta_v'}-{r_z'} \cos^2{\phi_v'} \sin^2{\theta_v'}),
\end{aligned}
\end{equation}
where $r_{x/y/z}'$ are given by Eqs. (\ref{Eq:r_x'}), (\ref{Eq:r_y'}) and (\ref{Eq:r_z'}) and $D'$ is the distance between the retarded emission point to the observer.
Then we decompose the $\Vec{E}_s'$ into the parallel and perpendicular directions which are with respect to the electron's moving direction ($E_{s,z,B'}'$ and $E_{s,x/y,B'}'$). We then perform the relativistic transformation of $\vec E_s'$ to obtain $\vec E_s$ in the lab frame\footnote{As explained below (and also in Appendix \ref{A}), there exist some unjustified omissions in the derivation of the scattered waves in \cite{Xu2000}. They solved the scattered electric field in the lab frame by considering that the electron is at rest initially ($\gamma=1$) and then they performed the Doppler transformation of co-moving frame scattered electric field to the lab frame. The transformation of the electric field amplitude is more general and only reduces to the Doppler factor under very special conditions.}
\begin{equation}
\Vec{E}_{\parallel,B}=\vec E_{s,z,B'}'
\end{equation}
and
\begin{equation}
\begin{aligned}
\Vec{E}_{\perp,B}&={\gamma}\left({\Vec{E}_{\perp}'}-\frac{\Vec{v}}{c}\times\Vec{B}_w'\right),
\end{aligned}
\end{equation}
where $\vec v$ is the electron's velocity.
Finally, we project the scattered electric field vector $\vec E_s$ back to the $xyz$-frame as
\begin{equation}
E_{\perp,i}=E_{\perp,B,i}\cos\phi-E_{\perp,B,j}\sin\phi.
\end{equation}
\begin{equation}
E_{\perp,j}=E_{\perp,B,i}\sin\phi\cos\theta+E_{\perp,B,j}\cos\phi\cos\theta+E_{\parallel,B,k}\sin\theta.
\end{equation}
\begin{equation}
E_{\parallel,k}=-E_{\perp,B,i}\sin\phi\sin\theta-E_{\perp,B,j}\cos\phi\sin\theta+E_{\parallel,B,k}\cos\theta.
\end{equation}
One notices that $E_{\perp,i}$ and $E_{\perp,j}$ have the same phase. Thus, we conclude that ICS produced by a single electron is always 100\% linearly polarized in any scattered direction. 

\begin{figure*}
\begin{center}
\begin{tabular}{ll}
\resizebox{80mm}{!}{\includegraphics[]{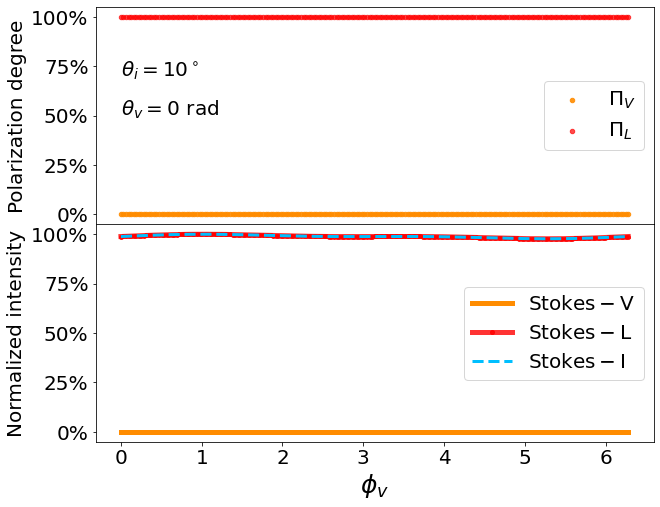}}&
\resizebox{80mm}{!}{\includegraphics[]{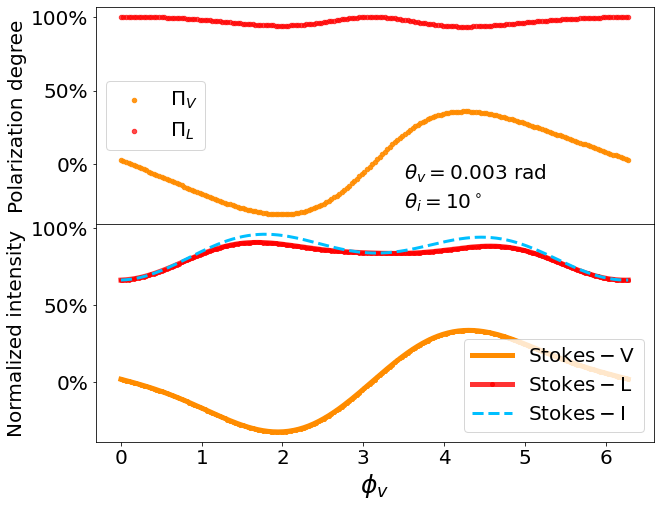}}\\
\resizebox{80mm}{!}{\includegraphics[]{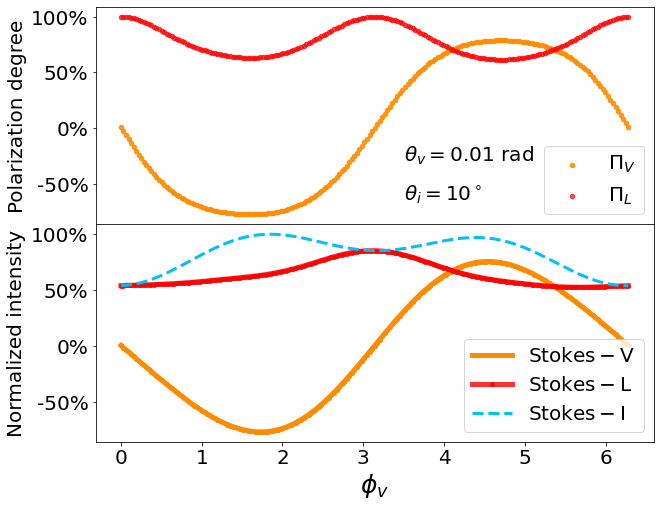}}&
\resizebox{80mm}{!}{\includegraphics[]{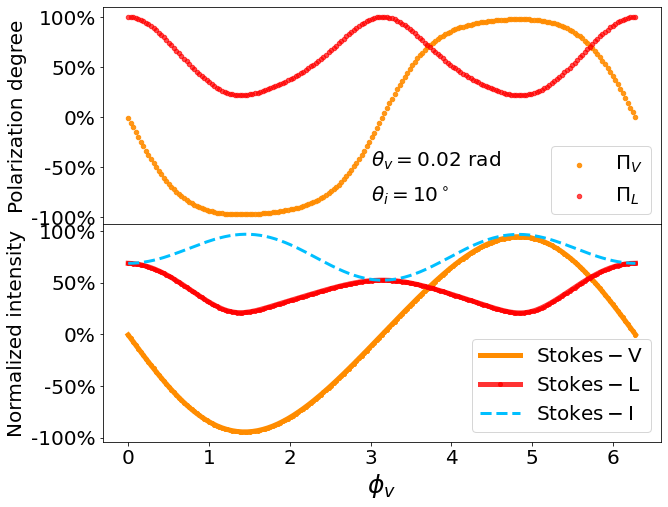}}
\end{tabular}
\caption{Circular and linear polarization degrees as a function of azimuthal angle $\phi_v$ for different viewing angle $\theta_v$. Following parameters are adopted: Incident angle $\theta_i=10^\circ$ is fixed, the bunch Lorentz factor is $\gamma=100$, the incident wave frequency is $\nu_i=10^4$ Hz, the scattered wave frequency is $\nu_{s}=\nu_{\rm frb}=10^9$ Hz, the bunch size is $\lambda=30$ cm and the distance from the source is $d=10^5$ cm.}
\label{fig:ICS10}
\end{center}
\end{figure*}

\begin{figure*}
\begin{center}
\begin{tabular}{ll}
\resizebox{80mm}{!}{\includegraphics[]{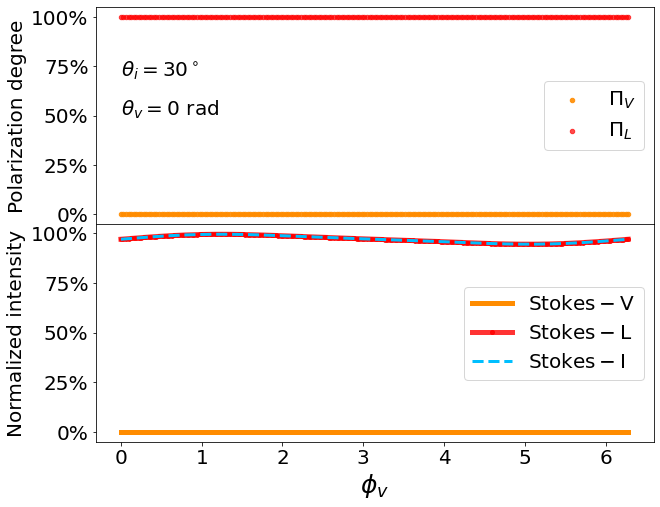}}&
\resizebox{80mm}{!}{\includegraphics[]{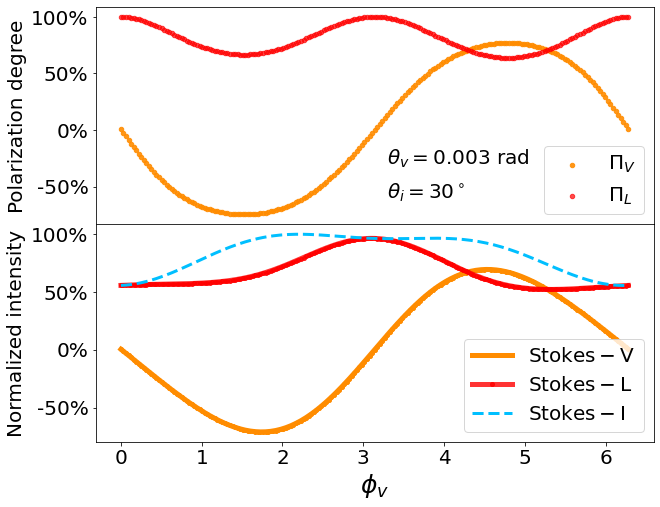}}\\
\resizebox{80mm}{!}{\includegraphics[]{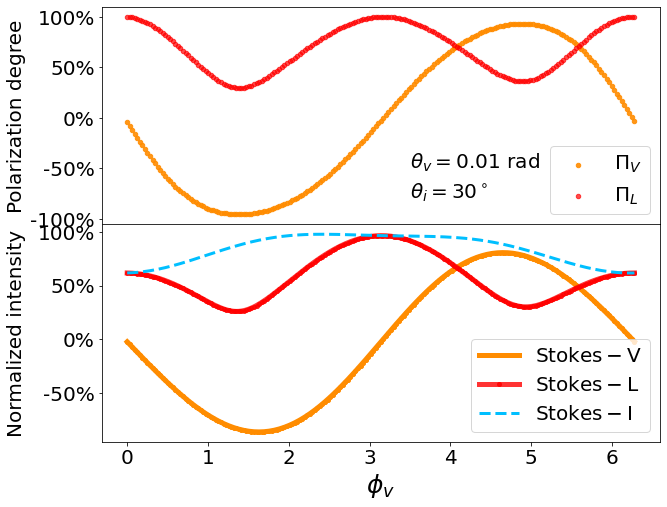}}&
\resizebox{80mm}{!}{\includegraphics[]{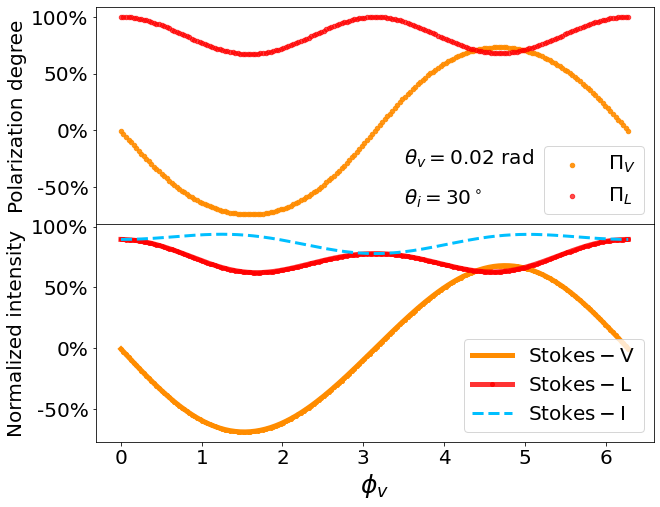}}
\end{tabular}
\caption{Same as Fig. \ref{fig:ICS10} but for $\theta_i=30^\circ$.}
\label{fig:ICS30}
\end{center}
\end{figure*}

(ii) Polarization properties of coherent ICS radiation by a bunch\footnote{In our following treatment, for simplicity we have assumed that the low-frequency electromagnetic waves that seed the ICS process are in the weak wave (low amplitude) regime. In principle, it could in the strong-wave regime, which requires more complicated treatments \citep[e.g.][]{QKZ}.}:
We consider a charged bunch that consists positive charged electrons and the scattered radiation is required to be coherent. 
In order to calculate the the scattered electric field of relativistic electrons, based on the general electric field expression we solved,
we project them onto the plane perpendicular to LOS and write the amplitudes of the two orthogonal components as
\begin{equation}
{A}_{\parallel}=E_{\perp,i}\cos\phi_v+E_{\perp,j}\sin\phi_v,
\end{equation}
and
\begin{equation}
{A}_{\perp}=-E_{\perp,i}\cos\theta_v\sin\phi_v+E_{\perp,j}\cos\theta_v\cos\phi_v-E_{\parallel,k}\sin\theta_v.
\end{equation}
The total scattered electric field perpendicular to the line of sight can be calculated as
\begin{equation}
\begin{aligned}
{A}_{\rm tot,\parallel}=\frac{1}{f_V}\int_0^{\theta_c}d\theta\int_0^{2\pi}d\phi\int_{r_{\rm min}}^{r_{\rm max}}r^2\sin\theta dr[A_\parallel],
\end{aligned}
\end{equation}
and
\begin{equation}
\begin{aligned}
{A}_{\rm tot,\perp}=\frac{1}{f_V}\int_0^{\theta_c}d\theta\int_0^{2\pi}d\phi\int_{r_{\rm min}}^{r_{\rm max}}r^2\sin\theta dr[A_\perp].
\end{aligned}
\end{equation}
where {$f_V$ is defined to describe the volume of the bunch,} $r_{\rm min}=(d-\lambda)\cos\theta_c/\cos\theta$ and $r_{\rm max}=d\cos\theta_c/\cos\theta$. Here again for simplicity, we consider the cone angle $\theta_{c}$ is equal to $1/\gamma$ of one bunch, but in reality it can be larger. The geometry of the bunch is presented in the right panel of Fig. \ref{fig:ICSfig}. The bunch longitudinal size is chosen as the typical FRB wavelength $\lambda=30$ cm and the transverse size is $\sim d\theta_{c}=10^3$ cm. Each particle is moving along a different local magnetic field line with the same Lorentz factor and radiates photons towards the LOS. We choose the minimum transverse size for the bunch as $\gamma\lambda\simeq \theta_c d$ for self-consistency.

We numerically integrate the scattered electric fields and calculate the degree of linear and circular polarization presented in Fig. \ref{fig:ICS10} and \ref{fig:ICS30}.
For the on-axis case, i.e. $\theta_{v}=0$, the ICS radiation is always 100\% linearly polarized\footnote{Strictly speaking, the incident wave cannot interact with every electron in one bunch, thus it should generate circular polarization even in $\theta_v=0$. The reason why $\Pi_V=0$ is that the incident wave is low frequency wave and wave vector value is extremely small $k_i=2\pi\nu_i/c\simeq(2\times10^{-6} \ {\rm cm}) \ \nu_{4}$ compared with bunch size. The phase differences of every electron are dominated by scattered wave. Thus the different phases contributed by incident wave can be ignored and every electron can be considered to upscatter the incident wave simultaneously. Otherwise, the minimum transverse size $\sim\gamma\lambda$ is much large than the longitudinal size and the highest circular polarization degree is mainly influenced by the transverse size of the bunch, i.e. for an off-beam case, there exists large phase difference between left and right side in the bunch.} ($\Pi_L=100\%$) and no circular polarized waves ($\Pi_V=0$) are produced (see Eq.(\ref{Stokes}) and Eq.(\ref{degree}) for definitions). We present the numerical results of circular and linear polarization degree as a function of $\phi_v$ for different incident angles in Figs. \ref{fig:ICS10} and \ref{fig:ICS30} for $\theta_i=10^\circ$ and $30^\circ$, respectively. Red line and orange line denote the degree of linear and circular polarization, respectively.
When $\theta_v\neq0$, one can see circular polarization shows up. the larger the $\theta_v$, the larger the circular polarization degree at a specific $\phi_v$. When $\theta_i=10^\circ$, the circular polarization degree could reach $\sim30^\circ$ for $\theta_v=0.003$ rad (on-beam case), and $\sim60\%$ for $\theta_v=0.01$ rad exactly at the cone edge, and $\sim90\%$ for $\theta_v=0.02$ rad outside the cone. When $\theta_i=30^\circ$, the circular polarization degree could reach $\sim60\%$ for $\theta_v=0.01$ rad exactly at the cone edge. One can also see that the polarization profile has a rotational symmetry with respect to $\phi_v=\pi$ since the incident wave is in the $x-z$ plane. Similar to curvature radiation for a bunch of opening angle of $1/\gamma$, the ICS bunch can also produce circular polarization in a wide solid angle of viewing angle. In order to accommodate the data that show a small fraction of circular polarization, one also needs to introduce a bunch with a much wider cross section so that within a large solid angle within the bunch cone, one can roughly have a symmetric scattering geometry so that the upscattered photons mostly carry linear polarization.

(iii) The polarization properties and main differences between coherent curvature and ICS radiation can be summarized as follows:
\begin{itemize}
\item For emission of a single electron, curvature radiation can produce both linear and circular polarization depending on the viewing angle, whereas ICS can only make linearly polarized upscattered waves. This can be most directly visualized from the geometric plots as shown in Figs. \ref{fig:cur} and \ref{fig:ICSfig}
by noticing the projection of the acceleration vector on the plane perpendicular to the LOS. 
For ICS, consider one electron moving in an incident low frequency electromagnetic wave. The acceleration unit vector is the same as the incident electric field of the wave and can be projected on the plane perpendicular to an arbitrary LOS, one can always see a straight line. Thus ICS by a single electron can only make 100\% linear polarization in any viewing angles. However, for the single electron curvature radiation off-beam case, the projection of the acceleration vector on such a plane is not always a straight line. Circular polarization can therefore be generated in the off-axis configuration. 
\item For the case of curvature radiation produced by a bunch, high degree of circular polarization can be observed in the off-axis case, similar to the single electron case. This is because in the off-beam case
the two orthogonal electric field components in the emitted waves have  different phases.
When the radiation cone ($\sim 1/\gamma$) is considered, the polarization properties of the bunch are similar to those of a point charge. 
\item For ICS emitted by a charged bunch, circular polarization could be generated because of the different phases of the electric fields of the scattered waves due to the spatial distribution of electrons within the bunch. The generation of circular polarization is not from the intrinsic ICS itself. In order to produce circular polarization, different phases and different polarization angles of the electric fields of the scattered waves are both needed. Both factors are influenced by the curved magnetic field geometry. If the magnetic field lines are straight lines towards one direction, no net circular polarization can be produced from a charged bunch. 
\item For both radiation mechanisms, the observed flux decreases rapidly once viewing angle is outside the radiation cone, which is assumed to be $1/\gamma$ in our calculations but could be in principle larger. In the off-beam geometry, even if high polarization can be generated, it is  unlikely to be observed because of the much lower flux. 

\end{itemize}

\subsection{Cyclotron absorption}\label{sec:cyclotron-absorption}
In view of the radiation mechanisms discussed, we consider propagation of the FRB waves across the open field line regions\footnote{Radio waves propagating across field lines (effectively across closed field line region) has been studied in \cite{Lu2019}.}.
Within the magnetosphere, one possible 
propagation effect to generate circular polarization is cyclotron absorption, which has been discussed within the context of radio pulsars \citep{Wang2010}.  We reinvestigate this mechanism within the context of FRBs in this sub-section. If FRB waves are generated in the inner magnetosphere of a magnetar along magnetic field lines, the wave vector is quasi-parallel\footnote{The relation between radius and the $\left<\vec{k}, \vec{B}\right>$ angle in a dipole magnetic field is calculated in \cite{QKZ}.} to the background magnetic field lines within the light cylinder and two eigen-orthogonal modes can be considered as the two circularly polarized modes: R-mode and L-mode. If the two modes undergo different levels of absorption, circular polarization components could be produced. 
Cyclotron emission/absorption can be considered as the energy level transition processes. We consider the incident FRB waves interact with background pair plasma and the electrons/positrons absorb the photon energy to jump to a higher energy level.
In a strong magnetic field, the allowed energy levels for electrons are the Landau energy levels. For simplicity, we consider a dipolar magnetic field 
with the surface field strength $B_{\star}$, which decreases rapidly with radius as $B=B_{\star}(r/R_{\star})^{-3}$ for $r>R_{\star}$ and $r < R_{\rm LC}$. The minimum energy of Landau level can be written as
\begin{equation}
E_{\rm min}=\hbar\omega_B\simeq \hbar\frac{eB_\star}{m_ec}\left(\frac{r}{R_\star}\right)^{-3},
\end{equation}
where $\hbar$ is the Planck constant.

In order to excite cyclotron resonance, electrons/positrons should be allowed to jump between two energy levels with an interval defined by the typical frequency of the FRB emission in the rest frame of electrons, i.e. 
\begin{equation}
\omega'=\gamma_{\pm}\omega_{\rm frb}(1-\beta\cos\theta_B)=\omega_B,
\end{equation}
where $\omega'$ is the FRB wave circular frequency in the rest frame of lepton, $\gamma_+$ and $\gamma_-$ are the Lorentz factors of positrons and electrons, respectively, $\theta_B$ is the angle between the wave vector and the magnetic field at the resonance radius. When the absorption optical depth is different for R-mode and L-mode, which requires different properties of positrons and electrons, one mode may be selectively absorbed so that net circular polarization could be generated from linear polarization. In order to detect FRBs at all, this mechanism should allow at least one mode escape freely without having the FRB being completely absorbed.

The resonance condition gives the cyclotron resonance absorption radius, which can be written as
\begin{equation}
\begin{aligned}
r_{\pm}&=\left[\frac{eB_\star R_\star^3}{m_ec\gamma_{\pm}\omega_{\rm FRB}(1-\beta\cos\theta_B)}\right]^{1/3}.
\end{aligned}
\end{equation}
\begin{figure*}
\begin{center}
\begin{tabular}{ll}
\resizebox{80mm}{!}{\includegraphics[]{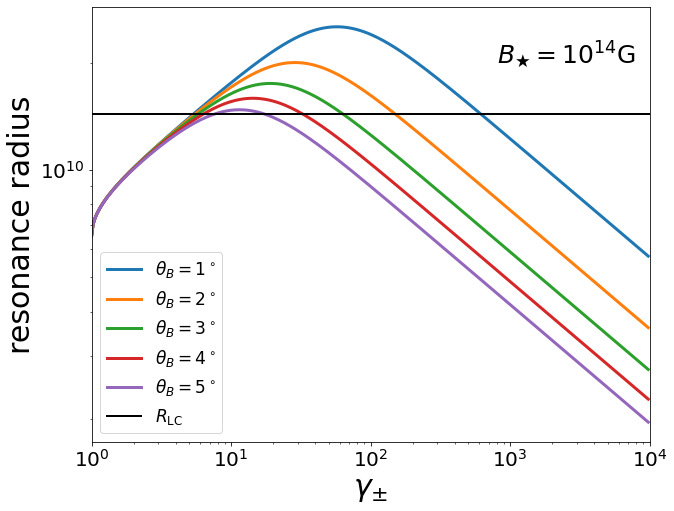}}&
\resizebox{80mm}{!}{\includegraphics[]{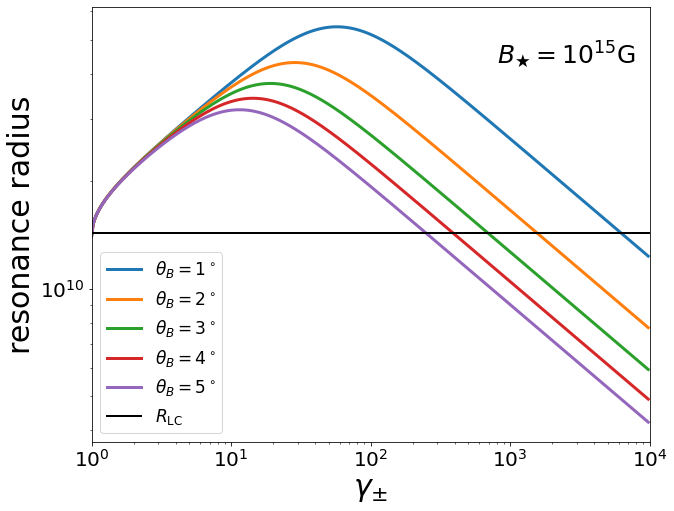}}
\end{tabular}
\caption{The cyclotron resonance radius as a function of pair plasma Lorentz factor for different angles $\theta_B$ between the wave vector and the background magnetic field from $1^\circ$ to $5^\circ$ and for two surface magnetic fields $B_\star=10^{14}$ G (left panel) and $B_\star=10^{15}$ G (right panel), respectively. The horizontal black solid line is the radius of light cylinder. Following parameters are adopted: magnetar radius $R_\star=10^6$ cm, FRB waves frequency $\nu_{\rm frb}=10^9$ Hz, and magnetar period $P=3$ s. One can see that the resonance condition can be satisfied inside the light cylinder if $\gamma_\pm $ is large enough.}
\label{fig:resonance radius}
\end{center}
\end{figure*}
We present a calculation of the cyclotron resonance radius as a function of $\gamma_\pm$ in Fig. \ref{fig:resonance radius}.
One can see the resonance radius can be within the magnetosphere if $\gamma_\pm$ is large enough. A lower $B_*$ and a longer period would allow 
a larger parameter space for cyclotron resonance.  
In general, the Lorentz factors of the leptons are required to be larger than a few times $10^2$ for most angles in order to satisfy the cyclotron resonance absorption condition. On the other hand, such as high Lorentz factor is naturally expected in the open field line region of a magnetar, as has been argued by \cite{QKZ}.

The total intensity for both incident circular polarization modes (L-mode and R-mode) may be considered to have the same strength, i.e. $I_{i,L}=I_{i,R}$. After the cyclotron resonance absorption, the intensities of the left-hand and right-hand polarized circular waves can be expressed as $I_{s,L}=I_{i,L}e^{-\tau_+}$ and $I_{s,R}=I_{i,R}e^{-\tau_-}$, respectively. In this calculation, we have ignored the emissivity of the plasma itself because the FRB is much brighter than the emission of the background plasma.
Thus, the Stokes-V that measures the intensity difference between the right-hand and left-hand polarized circular waves can be written as $V=I_{s,L}-I_{s,R}$, and the circular polarization degree can be calculated as
\begin{equation}
\Pi_V=\frac{|e^{-\tau_{+}}-e^{-\tau_{-}}|}{e^{-\tau_{+}}+e^{-\tau_{-}}},
\end{equation}
where $\tau_{\pm}$ corresponds to optical depths of positrons and electrons, respectively.
For the case of a symmetric pair plasma, i.e the two species of leptons have exactly the same Lorentz factor and number density, one can see $\Pi_V=0$, which means that the radio waves remain 100\% linearly polarized. When one of the two modes (e.g. R-mode) is completely absorbed 
(e.g. $\tau_{+}\gg1$) whereas the other mode (e.g. L-mode) is barely absorbed (e.g. $\tau_{-} \ll 1$),  
then the outgoing waves would be nearly 100\% circularly polarized.

We now estimate the optical depth for cyclotron absorption. The cross section for the electron/positron cyclotron resonance absorption in the co-moving frame of electron is given by \citep{Herold1979,Daugherty1978,Dermer1990}
\begin{equation}
\sigma_{\rm cyc,\pm}'=\frac{1}{2}\pi r_0c(1+\cos^2\theta')\phi(\nu_{\rm frb}'-\nu_B'),
\end{equation}
where $\nu'_B=\nu_B=\omega_B/(2\pi)$ and $\phi(\nu_{\rm frb}'-\nu_B')$ describes the line profile which obeys the Lorentz
profile as \citep{Rybicki&Lightman1979}
\begin{equation}
\phi(\nu'-\nu_B')=\frac{\Gamma_{lu}/(4\pi^2)}{(\nu_{\rm frb}'-\nu_B')^2+[\Gamma_{lu}/(4\pi)]^2},
\end{equation}
where the parameter $\Gamma_{lu}$ describes the transition from a lower Landau level to an upper Landau level. For transition from the ground state to the first Landau state, it can be written as
\begin{equation}
\Gamma_{lu}=\frac{4e^2\omega_B^2}{3m_ec^3}.
\end{equation}
The electron/positron number density can be estimated as $n_{\pm}=\xi_{\pm} n_{\rm GJ}$ and the optical depths for both species of leptons in the lab frame can be written as \citep{QKZ}
\begin{equation}
\begin{aligned}
\tau_{\pm}&=\int_{r_{\rm min}}^{r_{\rm max}}\xi_{\pm} n_{\rm GJ}\sigma_{\rm cyc,\pm}'\frac{(1-\beta_{\pm}\cos\theta_{B,\pm})^2}{\cos\theta_{B,\pm}}dr\\
&=\int_{r_{\rm min}}^{r_{\rm max}}\xi_{\pm} n_{\rm GJ}\pi r_0c\frac{(1-\beta_{\pm}\cos\theta_{B,\pm})^2(1+\cos\theta_{B,\pm}^2)}{2\cos\theta_{B,\pm}}\\
&\times\phi(\nu'_{\rm frb}-\nu'_B)dr.
\label{eq:cycoptical}
\end{aligned}
\end{equation}
Note that in order to have net cyclotron resonance absorption, $\beta_+/\beta_-$ and $\theta_{B,+} / \theta_{B,-}$ could be different, even though $\xi_+ \simeq \xi_-$ is expected to keep global neutrality of the generated pairs. 
When the incident FRB wave frequency is exactly equal to the Larmor frequency in the co-moving frame of a lepton, the cyclotron absorption cross section of reaches the maximum value at resonance
\begin{equation}
\sigma_{\rm cyc,max}\simeq3\pi\left(\frac{c}{\omega_B}\right)^2\simeq214.5 \ {\rm cm^2}.
\end{equation}

\begin{figure}
	\includegraphics[width=\columnwidth]{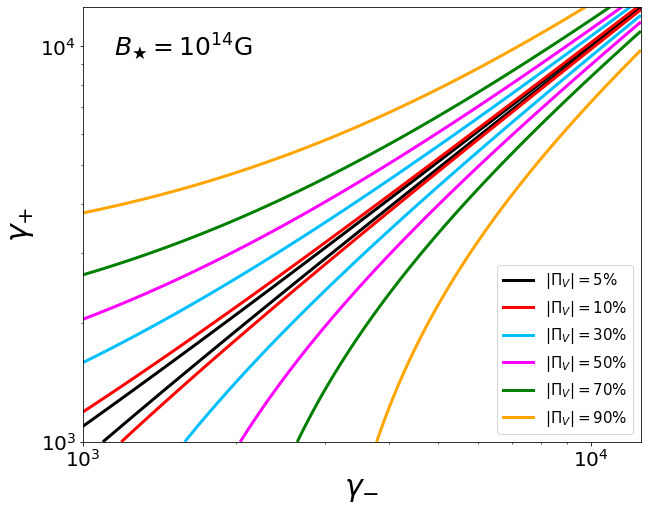}
    \caption{The circular polarization degree as a function of electron and positron Lorentz factors ($\gamma_+$ and $\gamma_-$) through cyclotron resonance absorption. Following parameters are adopted: surface magnetic field $B_\star=10^{14}$ G, FRB wave frequency $\nu_{\rm frb}=10^9$ Hz, magnetar period $P=3$ s and multiplicity factors $\xi_+=\xi_-=100$.}
    \label{fig:insidecyclotron}
\end{figure}

The profile of cyclotron resonance cross section as a function of radius has a sharp peak around the resonance radius. The cross section drops quickly when the radius changes slightly. We consider a characteristic length scale as the range where the cross section drops up to $10\%$ of $\sigma_{\rm cyc,max}$ and assume $\sigma_{\rm cyc,max}$ is valid for all points within such a length scale. One can then estimate the optical depths for both electrons and positrons and then calculate the net circular polarization degree assuming $\gamma_+$ and $\gamma_-$ are different (which means $\theta_{B,+}$ and $\theta_{B,-}$ could be different). 
We present the numerical results of the circular polarization degree as a function of $\gamma_{+}$ and $\gamma_-$ in Fig.\ref{fig:insidecyclotron}. One can see that when $\gamma_-=\gamma_+$, there is no circular polarization since the R-mode and L-mode of incident waves undergo the same absorption so that $\tau_-=\tau_+$. When $\gamma_-\neq\gamma_+$, net circular polarization can be generated through different cyclotron resonance absorption degrees for R-mode and L-mode. One can see that at higher Lorentz factors, a small relative Lorentz factor difference would generate a large circular polarization degree. The reason is that a higher Lorentz factor corresponds to a smaller resonance radius, where the pair plasma number densities are higher.

We summarize the main conclusions of this subsection as follows. 
For FRBs produced in the open field line region of a magnetar\footnote{\cite{Wangshort} proposed that the cyclotron absorption could be generated in the closed field line region. However, FRBs are unlikely generated from such regions because a parallel electric field required to power FRB emission cannot be developed because of the large plasma density. 
Furthermore, the large-amplitude wave effect likely chokes the FRB propagation in the closed field line region \citep{Beloborodov2021}.} the wave vector is quasi-parallel to background magnetic field. Thus the orthogonal eigne-modes of the incident waves can be conveniently set as R-mode and L-mode. An asymmetric distribution of the lepton Lorentz factors is needed to generate a relatively high circular polarization degree.

\section{Outside magnetosphere}\label{sec:outside}

In this section, we discuss the emission processes and propagation effects for FRBs generated far outside the magnetosphere of a magnetar. For the emission mechanism, we discuss the synchrotron maser model invoking highly ordered magnetic fields which can produce highly linearly polarized emission and conclude that highly circular polarized FRB waves can be rarely generated in such a scenario. 
We further discuss various propagation effects, including synchrotron/cyclotron absorption and Faraday conversion via magnetic field reversals within  various astrophysical scenarios. For the convenience of later discussion, we first list the characteristic frequencies in both a magnetar wind and an ambient interstellar medium (ISM).

In the magnetar wind region out side the light cylinder, the characteristic value of plasma frequency can be calculated as
\begin{equation}
\begin{aligned}
\omega_{p,\rm wind}&=\sqrt{\frac{4\pi e^2n_e}{m_e}}\simeq\sqrt{\frac{ e^2\xi\dot N_{\rm GJ}}{m_ecr^2}}\\
&\simeq(4.8\times10^3 \ {\rm rad \ s^{-1}}) \ \xi_2^{1/2}r_{13}^{-1}B_{\star,15}^{1/2}R_{\star,6}^{3/2}P^{-1},
\end{aligned}
\end{equation}
where $\dot N_{\rm GJ}=2cA_{\rm cap}n_{\rm pole}$ is the Goldreich-Julian particle ejection rate from the polar cap, $A_{\rm cap}\simeq\pi R_\star^3/R_{\rm LC}$ is the area of the polar cap and $n_{\rm pole}=B_\star/(Pec)$ is the Goldreich-Julian density at the magnetar surface, $B_\star$ is the surface magnetic field strength of the magnetar at the pole, and $\xi$ is pair multiplicity. The Larmor frequency can be calculated as
\begin{equation}
\begin{aligned}
\omega_{B,\rm wind}&=\frac{eB}{m_ec}\simeq\frac{eB_\star}{2m_ec}\left(\frac{R_{\rm LC}}{R_\star}\right)^{-3}\left(\frac{r}{R_{\rm LC}}\right)^{-1}\\
&\simeq(3.9\times10^7 \ {\rm rad \ s^{-1}}) \ B_{\star,15}P^{-2}R_{\star,6}^{-3}r_{13}^{-1}.
\end{aligned}
\end{equation}

The ratio between the plasma frequency and the Larmor (cyclotron) frequency in the magnetar wind region can be estimated as
\begin{equation}
\begin{aligned}
\frac{\omega_{p,\rm wind}}{\omega_{B,\rm wind}}
&\simeq1.2\times10^{-4} \ \xi_2^{1/2}B_{\star,15}^{-1/2}PR_{\star,6}^{-3/2}\ll 1,
\end{aligned}
\end{equation}
which is independent of distance $r$.
In the ISM region, the typical magnetic field strength is $B\sim10^{-6}$ G and electron number density is $n_e\sim1 \ \rm cm^{-3}$. We then have the plasma frequency
\begin{equation}
\omega_{p,\rm ISM}=\sqrt{\frac{4\pi e^2n_e}{m_e}}\simeq(5.6\times10^4 \ {\rm rad \ s^{-1}}) \ n_e^{1/2},
\end{equation}
and the Larmor frequency 
\begin{equation}
\omega_{B,\rm ISM}=\frac{eB}{m_ec}\simeq(17.6 \ {\rm rad \ s^{-1}}) \ B_{-6},
\end{equation}
and the ratio between the two can be estimated as
\begin{equation}
\frac{\omega_{p,\rm ISM}}{\omega_{B,\rm ISM}}\simeq3.2\times10^3 \ n_e^{1/2}B_{-6}^{-1} \gg 1.
\end{equation}

\subsection{Emission mechanism: synchrotron/cyclotron maser}\label{sec:syn}

\begin{figure*}
\begin{center}
\begin{tabular}{ll}
\resizebox{80mm}{!}{\includegraphics[]{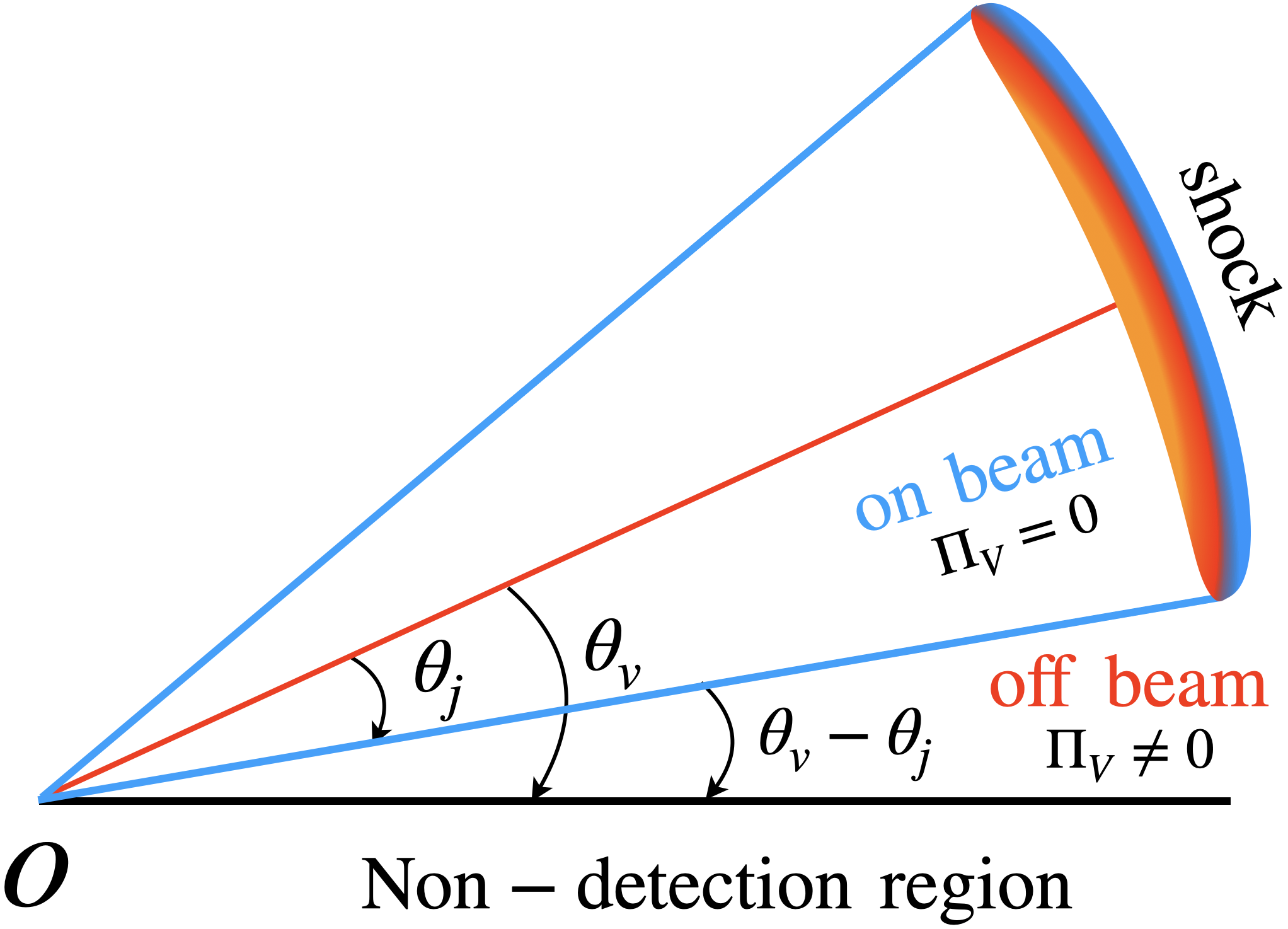}}&
\resizebox{80mm}{!}{\includegraphics[]{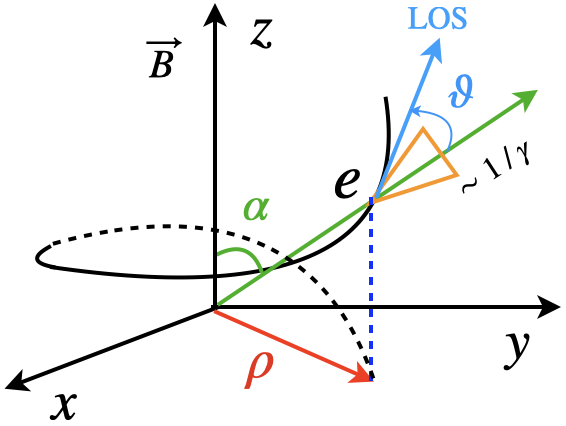}}
\end{tabular}
\caption{Left panel: A cartoon picture for the synchrotron maser model. The angle $\theta_j$ denotes the half opening angle of the FRB jet. Within  $\theta_j$ is called the on-beam case and there is no circular polarization expected ($\Pi_V=0$). Outside $\theta_j$ is called the off-beam case where $\Pi_V\neq0$. Right panel: An electron orbit with the pitch angle $\alpha$ between the electron momentum and the background magnetic field $\vec B$ (along the $z$-axis) and the Larmor radius $\rho$. The blue solid line is the LOS and $1/\gamma$ is the narrow beaming angle of synchrotron radiation.}
\label{fig:maser}
\end{center}
\end{figure*}

The GRB-like models invoke internal shocks or external shocks to accelerate particles, as a highly magnetized relativistic outflow collide internally or with a circumstellar medium. FRBs may be generated by a plasma maser process as charged particles gyrate coherently in an ordered magnetic field by forming a ring in the momentum space and radiating synchrotron/cyclotron photons coherently \citep{Lyubarsky2014,Beloborodov2017,Plotnikov&Sironi2019,Metzger2019,Beloborodov2020,Margalit2020}.

To calculate the polarization properties of emission in such a model, we first consider an electron in gyro-motion around a uniform magnetic field with a pitch angle $\alpha$, as shown in the right panel of Fig.\ref{fig:maser}. The perpendicular component velocity with respect to the $z$-axis is in the $x-y$ plane, i.e $\vec v=(v\cos\omega_Bt,v\sin\omega_Bt,0)$. The general motion equation of a single electron can be described by
\begin{equation}\label{motion equation}
\frac{d(\gamma m_e\vec v)}{dt}=e\left(\vec E+\frac{\vec v}{c}\times\vec B\right)+\vec F_{\rm rad},
\end{equation}
where $\vec F_{\rm rad}$ is the radiation reaction force.
{For simplicity, we consider that the electron only moves circularly around the magnetic field, i.e. $\alpha=\pi/2$.} This is roughly consistent with the physical picture because in this model the ordered magnetic field lines are expected to be parallel to the shock plane. We use the Cartesian coordinates with $\hat{z}$ in the direction of the background magnetic field ($\vec B=B_0\hat{z}$) and assume that the observer direction is in the $x-z$ plane for generality, i.e. $\vec n=(\sin\theta,0,\cos\theta)$. 
The position vector can be solved from Equation (\ref{motion equation}), giving
\begin{equation}
\Vec{r}=\frac{\gamma v_\perp}{\omega_B}\sin(\omega_Bt/\gamma)\hat{x}-\frac{\gamma v_\perp}{\omega_B}\cos(\omega_Bt/\gamma)\hat{y}.
\end{equation}
For the low $\gamma$ case, the electron emits cyclotron radiation.
Based on Parseval's theorem, the radiation spectrum of a periodic electron with no parallel velocity along B-field can be written as
\begin{equation}\label{radiation period}
\frac{dP}{d\Omega}=\frac{e^2s^2\omega_B^4}{8\pi^3c\gamma^4}\left\vert \int_0^Te^{is\omega_0(t'-\vec n\cdot\vec r)/c}[\vec n\times(\vec n\times\vec \beta)]dt'\right\vert^2,
\end{equation}
where $s$ is the harmonic number.
The integral term in Eq.(\ref{radiation period}) can be written as \citep{Landau1975}
\begin{equation}
\begin{aligned}
\int_0^Te^{is\omega_0(t-\vec n\cdot\vec r/c)}&[\vec n\times(\vec n\times\vec \beta)]dt=-\frac{\cos^2\theta}{\nu_B\sin\theta}J_s(s\beta\sin\theta)\hat{x}\\
&-\frac{i\beta}{\nu_B}J_s(s\beta\sin\theta)\hat{y}+\frac{\cos\theta}{\nu_B}J_s(s\beta\sin\theta)\hat{z}.
\end{aligned}
\end{equation}
The electric field of cyclotron emission at the base frequency ($s=1$) can be written as
\begin{equation}
\Vec{E}_{\rm cyc}\simeq-\frac{1}{2\nu_0}\beta\cos^2{\theta}\hat{x}-\frac{i\beta}{2\nu_0}\hat{y}+\frac{\beta}{2\nu_0}\cos\theta\sin\theta\hat{z}.
\end{equation}
One can see that for the case of $\theta=0$ (along the magnetic field), one has $E_z=0$ and $E_x=iE_y$, suggesting that the cyclotron wave is circularly polarized. For $\theta=\pi/2$ (along the LOS for a parallel shock as required for the synchrotron/cyclotron maser model), one has $E_z\neq0$ and $E_x=E_y=0$, suggesting that the wave is linear polarized. In general, one can consider an arbitrary LOS and define the amplitudes of the electric field components $A_{\parallel}$ and $A_{\perp}$,
with the magnetic field always along the $z$-axis. The direction of the LOS is chosen as ($\theta_v,\phi_v$). Therefore, the parallel (along the $x$-axis) and perpendicular (along the $y$-axis) components of electric field in the LOS frame can be written as
\begin{equation}
\begin{aligned}
A_{\parallel,\rm cyc}&=E_{{\rm cyc},i}\sin\phi_v-(E_{{\rm cyc},j}\cos\theta_v-E_{{\rm cyc},k}\sin\theta_v)\cos\phi_v,
\end{aligned}
\end{equation}
and
\begin{equation}
A_{\perp,\rm cyc}=E_{{\rm cyc},i}\cos\phi_v+(E_{{\rm cyc},j}\cos\theta_v-E_{{\rm cyc},k}\sin\theta_v)\sin\phi_v.
\end{equation}

\begin{figure*}
\begin{center}
\begin{tabular}{ll}
\resizebox{80mm}{!}{\includegraphics[]{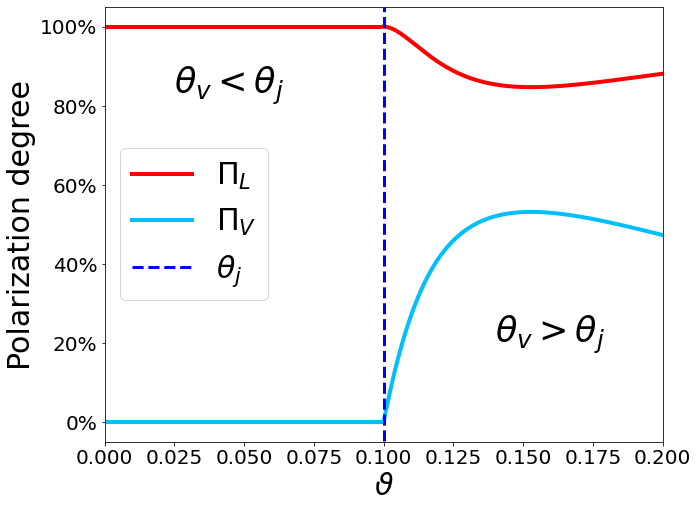}}&
\resizebox{80mm}{!}{\includegraphics[]{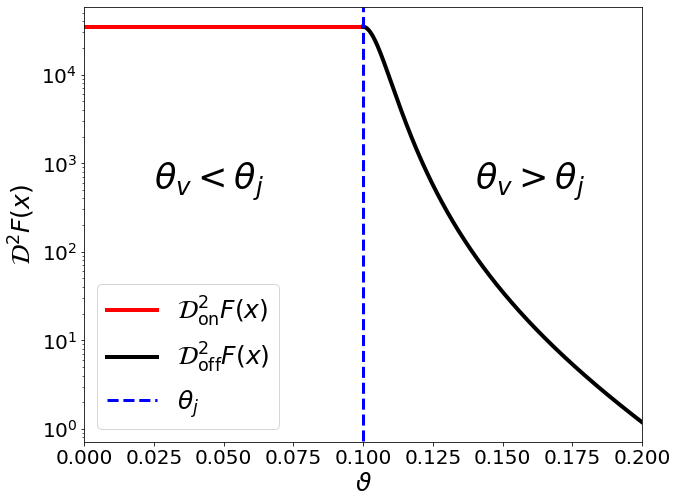}}
\end{tabular}
\caption{Left panel: Simulated circular (blue solid line) and linear polarization (red solid line) degree of one electron via synchrotron radiation as a function of angle $\vartheta$. Right panel: Doppler factor as a function of $\vartheta$ for two regions, i.e. $\theta_v<\theta_{j}$ and $\theta_v>\theta_{j}$. The vertical blue dashed line is the shock boundary ($\vartheta=0.1$). Following parameters are adopted: the bulk angle $\theta_j=0.1$, electrons Lorentz factor $\gamma=10$, bulk Lorentz factor $\Gamma=10^2$.}
\label{fig:syn_degree}
\end{center}
\end{figure*}

For the high $\gamma$ case, the electron emits synchrotron radiation. 
There are many electrons in the gyro-motion trajectories in the shocked region with a bulk motion Lorentz factor $\Gamma$. We consider the emission of one electron for simplicity, which is adequate to calculate polarization properties.  
We discuss synchrotron radiation of the electron in the comoving frame of the bulk motion.
{Let the angle between the line of sight and the electron trajectory plane is $\vartheta$ in the lab frame and $\vartheta'$ in the comoving frame.} One has $\sin\vartheta'={\cal D}\sin\vartheta$, or $\vartheta'={\cal D}\vartheta$ when both angles are small, where ${\cal D}=1/\Gamma(1-\beta\cos\vartheta)$ is the Doppler factor. 
Because the shock moving direction is perpendicular to the background magnetic field and the pitch angle is $\pi/2$, $\vartheta'$ is also the angle between the LOS and electron's trajectory plane in the bulk motion comoving frame. The amplitudes of the electric field can be written as
\begin{equation}
A_{\parallel,\rm syn}\simeq\frac{i2\rho'}{\sqrt{3}c}\left(\frac{1}{\gamma^2}+\vartheta'^2\right)K_{2/3}(\varsigma'),
\end{equation}
and
\begin{equation}
A_{\perp,\rm syn}\simeq\frac{2\rho'\vartheta'}{\sqrt{3}c}\left(\frac{1}{\gamma^2}+\vartheta'^2\right)^{1/2}K_{1/3}(\varsigma'),
\end{equation}
where $K_{\nu}(\varsigma')$ is the modified Bessel function of the second kind and $\varsigma'$ is given by \citep{Jackson1998}
\begin{equation}
\varsigma'=\frac{\omega'\rho'}{3c}\left(\frac{1}{\gamma^2}+\vartheta'^2\right)^{3/2}.
\end{equation}
The required magnetic field in the shock region can be calculated through the critical frequency of synchrotron radiation as
\begin{equation}
B_s \simeq \frac{2m_ec\omega_B}{3e\gamma^2},
\end{equation}
where $\omega_B$ is equal to the critical synchrotron radiation frequency.
The corresponding radius can be estimated as 
\begin{equation}
\rho'\simeq\rho=\frac{c}{\omega_B}=\frac{m_ec^2}{eB_s}.
\end{equation}
According to Eq.(\ref{degree}) and electric field amplitudes of synchrotron emission, the degree of linear and circular polarization at angle $\vartheta'$ in the comoving frame can be written as
\begin{equation}
\Pi_L=1-\frac{2(1/\gamma^2+\vartheta'^2)K_{2/3}^2(\varsigma')}{(1/\gamma^2+\vartheta'^2)K_{2/3}^2(\varsigma')+\vartheta^2K_{1/3}^2(\varsigma')},
\end{equation}
and
\begin{equation}
\Pi_V=\frac{2\vartheta'(1/\gamma^2+\vartheta'^2)^{1/2}K_{2/3}(\varsigma')K_{1/3}(\varsigma')}{(1/\gamma^2+\vartheta'^2)K_{2/3}^2(\varsigma')+\vartheta'^2K_{1/3}^2(\varsigma')}.
\end{equation}
We transform the expression of the circular polarization degree from the co-moving frame to the lab frame as
\begin{equation}
\Pi_V=\frac{2\vartheta{\cal D}(1/\gamma^2+\vartheta^2{\cal D}^2)^{1/2}K_{2/3}(\varsigma)K_{1/3}(\varsigma)}{(1/\gamma^2+\vartheta^2{\cal D}^2)K_{2/3}^2(\varsigma)+\vartheta^2{\cal D}^2K_{1/3}^2(\varsigma)}.
\end{equation}
where we have used $\omega=\omega'{\cal D}$ and $\varsigma={\omega{\cal D}\rho}\left({\gamma^{-2}}+\vartheta^2{\cal D}^2\right)^{3/2}\big/{3c}$. 
We define a bulk half opening angle $\theta_{j}$ for the FRB-emitting shock and rewrite the general degree of circular polarization as
\begin{equation}
\Pi_V=\left\{
\begin{aligned}
&\Pi_V(\vartheta =0), \ \ &\theta_v<\theta_{j},\\
&\Pi_V(\vartheta = \theta_v-\theta_{j}), \ &\theta_v>\theta_{j}.
\end{aligned}
\right.
\end{equation}

We present the degree of linear and circular polarization of synchrotron-maser FRB emission as a function of viewing angle $\vartheta$ in the lab frame in the left panel of Fig.\ref{fig:syn_degree}. 
The electron Lorentz factor in the co-moving frame is adopted as $\gamma=10$ and the bulk Lorentz factor is $\Gamma=100$. 
In order to produce the typical 1-GHz radio waves, the frequency in the co-moving frame of the bulk motion can be calculated as 
$\nu=\nu_B\simeq\nu_{\rm frb}/\Gamma\sim(10^7 \ {\rm Hz}) \ \nu_{\rm frb,9}\Gamma_2$ for $\theta_v<\theta_j$ case and 
$\nu=\nu_B=\nu_{\rm frb}/{\cal D}$ fir $\theta_v>\theta_j$ case.
One can see that within the bulk angle $\theta_{j}$, the radiation is completely linearly polarized since the LOS can always intersect with the trajectory of ``on-axis'' electrons. When $\theta_v>\theta_{j}$, one can start to observe circular polarization, with the maximum achievable circular polarization degree $\Pi_V\sim50\%$. However, the observed isotropic out side the jet cone decreases rapidly with $\theta$ when $\theta_v>\theta_{j}$ as $L={\cal D}^2L'$, where the Doppler factor $\cal D$ is defined as
\begin{equation}
    {\cal D}_{\rm on/off}(\vartheta)=\left\{
    \begin{aligned}
    &{\cal D}_{\rm on}(\vartheta\simeq0)=\frac{1}{\Gamma(1-\beta)} \simeq 2\Gamma, \ &\theta_v<\theta_{j},\\
    &{\cal D}_{\rm off}(\vartheta)=\frac{1}{\Gamma[1-\beta\cos(\theta-\theta_j)]}, \ &\theta_v>\theta_j.
    \end{aligned}
\right.
\end{equation}
We present the value of ${\cal D_{\rm on/off}}^2F(x)$ as a function of $\theta$ in the right panel of Fig.\ref{fig:syn_degree}, where $F(x)$ is the single electron's synchrotron spectrum function defined as
\begin{equation}
F(x)=x\int_x^{\infty}K_{5/3}(\varsigma)d\varsigma,
\end{equation}
where $x=\nu_{\rm frb}/\nu_B$ is the ratio between the emission frequency and the characteristic frequency.
The vertical blue dashed line is the jet angle which is adopted as $\theta_j=0.1$. One can see the isotropic luminosity (${\cal D}^2 F(x)$ as a proxy) remains constant when $\theta_v<\theta_{j}$ but decreases rapidly when $\theta_v>\theta_{j}$. 
This suggests that the off-axis emission from a synchrotron maser shock is likely barely observable unless the line of sight is slightly outside $\theta_j$. In this case, one cannot see high circular polarization. The observed high circular polarization (up to 75\% for FRB 20201124A, \citet{Xu2021}) therefore disfavors the synchrotron maser  as the FRB emission mechanism at least for some bursts.

\subsection{Propagation effect}

In this section, we discuss three propagation effects far outside of the magnetosphere of the FRB engine: synchrotron absorption by relativistic electrons in a synchrotron emitting nebula, cyclotron absorption in a cold, strongly magnetized medium, and Faraday conversion in a medium with magnetic field reversals. 

Ignoring the spontaneous emission of the medium because the FRB waves are very bright, we expand the radiation transfer equation (Eq.\ref{general}) as four differential equations to describe the evolution of Four Stokes parameters: 
\begin{equation}
\begin{aligned}
&\frac{dI}{ds}=-\eta I-\eta_Q Q-\eta_U U-\eta_V V\\
&\frac{dQ}{ds}=-\eta_Q I-\eta Q-\rho_V U+\rho_U V\\
&\frac{dU}{ds}=-\eta_U I+\rho_V Q-\eta U-\rho_Q V\\
&\frac{dV}{ds}=-\eta_V I-\rho_U Q+\rho_Q U-\eta V.
\end{aligned}
\end{equation}
With these, one can further derive the the differential equations describing the evolution of the circular, linear and total polarization degree \footnote{Equations (\ref{eq:dPi_V/ds}), (\ref{eq:dPi_L/ds}) and (\ref{eq:dPi_P/ds}) can be reduced to a simplified form in \citet{Xu2021} (the first arXiv version) when $\eta_U$ is chosen to be zero.}
\begin{equation}\label{eq:dPi_V/ds}
\frac{d\Pi_V}{ds}=\eta_V(\Pi_V^2-1)+(\eta_Q\Pi_V-\rho_U)\Pi_L\cos\Phi+(\eta_U\Pi_V+\rho_Q)\Pi_L\sin\Phi.
\end{equation}
\begin{equation}
\begin{aligned}
\frac{d\Pi_L}{ds}&=[(\Pi_V^2-1)\eta_Q+\rho_U\Pi_V]\cos\Phi+[(\Pi_L^2-1)\eta_U-\rho_Q\Pi_V]\\
&\times\sin\Phi+\eta_V\Pi_L\Pi_V.
\label{eq:dPi_L/ds}
\end{aligned}
\end{equation}
\begin{equation}\label{eq:dPi_P/ds}
\frac{d\Pi_P}{ds}=(\eta_Q\Pi_L\cos\Phi+\eta_U\Pi_L\sin\Phi+\eta_V\Pi_V)\left(\Pi_p-\frac{1}{\Pi_p}\right).
\end{equation}
as well as the evolution of the angle $\Phi=\tan^{-1} (U/Q)$ 
\begin{equation}\label{eq:dPhi/ds}
\frac{d\Phi}{ds}=\rho_V+\frac{\sin\Phi}{\Pi_L}(\eta_Q-\rho_U\Pi_V)-\frac{\cos\Phi}{\Pi_L}(\eta_U+\rho_Q\Pi_V).
\end{equation}

One can generally discuss the propagation effects from Eqs. (\ref{eq:dPi_V/ds}), (\ref{eq:dPi_L/ds}) and (\ref{eq:dPi_P/ds}). In particular, the three propagation effects that can modify the circular polarization state of the FRB waves can be understood from Eq.(\ref{eq:dPi_V/ds}).
\begin{itemize}
\item Synchrotron absorption: This corresponds to the case without Faraday conversion and rotation coefficients, i.e. $\rho_Q=\rho_U=\rho_V=0$, but with uneven absorption coefficients $\eta_Q$, $\eta_U$ and $\eta_V$. Since synchrotron radiation generate linear polarization, there is no absorption to circular polarization if the incoming emission has no circular polarization, i.e. $\eta_V=0$ if $V=0$ (see Eq.(\ref{eq:eta_V}) for a quantitative discussion later). Inspecting Eq.(\ref{eq:dPi_V/ds}), one can see $d \Pi_V/ds \neq 0$ only when $V \neq 0$. In other words, for a 100\% linearly polarized wave ($\Pi_L=100\%$), synchrotron absorption would not generate circular polarization.   
\item Cyclotron absorption: Because cyclotron radiation is circularly polarized, selective cyclotron absorption (which is the case for an electron-ion plasma) would generate circular polarization even if the incident wave is $\sim 100\%$ linearly polarized. Again assuming no Faraday rotation and conversion ($\rho_Q=\rho_U=\rho_V=0$), Eq.(\ref{eq:dPi_V/ds}) shows $d \Pi_V/ds \neq 0$ in general with a non-zero $\eta_V$. This means that for a 100\% linearly polarized wave, $|\Pi_V|$ will increase with distance along the path of propagation. For a 100\% circularly polarized wave, with circular absorption only ($\eta_V \neq 0$), the circular polarization degree $\Pi_V$ remains constant, even though the absolute intensity drops due to absorption.
\item Faraday conversion: Let us consider that there is no absorption 
i.e. $\eta_Q=\eta_U=\eta_V=0$. One can see the total polarization degree is constant, i.e. $d\Pi_P/ds=0$. When the magnetic field has a parallel component along the line of sight, Faraday rotation will occur with $\rho_V \neq 0$. When the magnetic field is perpendicular to the line of sight (i.e. field reversal), $\rho_Q$ and $\rho_U$ would be non-zero and Faraday conversion will happen, leading to conversion of linear polarization to  circular polarization and vice versa. We can re-write Eqs.(\ref{eq:dPi_V/ds}) and (\ref{eq:dPi_L/ds}) as
\begin{equation}\label{eq:FC_V}
\frac{d\Pi_V}{ds}=-\rho_U\Pi_L\cos\Phi+\rho_Q\Pi_L\sin\Phi.
\end{equation}
\begin{equation}
\frac{d\Pi_L}{ds}=\rho_U\Pi_V\cos\Phi-\rho_Q\Pi_V\sin\Phi.
\end{equation}
With $\rho_V=0$, Eq.(\ref{eq:dPhi/ds}) can be simplified as
\begin{equation}\label{eq:FC_Phi}
\frac{d\Phi}{ds}=-\frac{\Pi_V}{\Pi_L}(\rho_Q\cos\Phi+\rho_U\sin\Phi).
\end{equation}
It should be pointed out that the quantity $\rho_Q Q+\rho_U U+\rho_V V$ is invariant \citep{Melrose2010}. 

\end{itemize}

We discuss the three processes in detail in the following.

\subsubsection{Synchrotron Absorption}\label{sec:synchrotron-absorption}
We consider that the FRB sources are surrounded by a synchrotron-emitting nebula, in which there exist relativsitic electrons that can absorb FRB photons in a certain polarization modes and change the polarization state of the waves. 
This process is the opposite process of synchrotron radiation and can be described by several absorption coefficients in the general radiative transfer equation (Eq.\ref{general}).
The environment outside the magnetosphere is considered as an electron-ion plasma, thus the plasma is asymmetry since the gyration radius of electrons is much larger than that of ions. A pair plasma whose positive and negative species are symmetric with the same distribution of number density and Lorentz factor cannot produce circular polarization through synchrotron absorption. 
It should be pointed out that the Razin effect is negligible due to $\gamma\omega_p\ll\omega_{\rm frb}$ in the nebula.

The absorption coefficient for synchrotron radiation is given by \citep{Rybicki&Lightman1979}
\begin{equation}
\alpha_\nu=-\frac{1}{8\pi\nu^2m_e}\int_{\gamma_{\rm min}}^{\gamma_{\rm max}}d\gamma P(\gamma,\nu)\gamma^2\frac{\partial}{\partial \gamma}\left[\frac{N(\gamma)}{\gamma^2}\right],
\end{equation}
where $P(\gamma,\nu)$ is the specific synchrotron radiation power. 
We consider a relativistic electron gas with a power-law distribution in Lorentz factor, i.e.
$N(\gamma_e)d\gamma_e=C_{\gamma_{e}}\gamma_e^{-p}d\gamma_e$ with $\gamma_{\rm min}<\gamma_e<\gamma_{\rm max}$ and $p>1$. The total electron number density can be calculated as
\begin{equation}
n_{e}=\int_{\gamma_{\rm min}}^{\gamma_{\rm max}}C_{\gamma_e}\gamma_e^{-p}d\gamma_e=\frac{C_{\gamma_{e}}}{p-1}(\gamma_{\rm min}^{-p+1}-\gamma_{\rm max}^{-p+1}).
\end{equation}
For numerical purposes, we consider a specific case that the radius of the magnetar wind nebula is $r=10^{18}$ cm, the region length scale is $\Delta r=10^{17}$ cm and the magnetic field strength is $B=10^{-3}$ G. These parameters are relevant to the persistent radio source (PRS) associated with FRB 121102. This PRS has a sharp break at frequency $10$ GHz and a specific synchrotron emission luminosity $L_\nu=10^{29} \ {\rm erg \ s^{-1} \ Hz^{-1}}$. Thus the total number of electrons can be estimated as 
\begin{equation}
L_\nu\simeq \frac{N_{\rm tot}}{\kappa}\frac{\sqrt{3}e^3B\Gamma}{m_ec^2} \ \Rightarrow \ N_{\rm tot}\simeq4.3\times10^{53} \ \kappa L_{\nu,29}B_{-3}^{-1}\Gamma^{-1},
\end{equation}
where $\kappa > 1$ is a parameter to connect the total number density of electrons to the 10 GHz nebula luminosity. We have assumed that there is no bulk motion in the nebula.
The corresponding electron number density can be estimated as $n_e=N_{\rm tot}/(4\pi r^2 \Delta r)$.
Then the normalized coefficient can be calculated as
\begin{equation}
C_{\gamma_e}=\frac{n_{e}(p-1)}{\gamma_{\rm min}^{-p+1}-\gamma_{\rm max}^{-p+1}}.
\end{equation}
Plugging in the specific form of total power $P(\gamma,\nu)$ for synchrotron radiation, one can write the absorption coefficient as
\begin{equation}
\alpha_{\nu,e}=\frac{p+2}{8\pi m_e}C_{\gamma_{e}}\nu^{-2}\int_{\gamma_{\rm min}}^{\gamma_{\rm max}}\frac{\sqrt{3}e^2B_\perp}{m_{e}c^2}F(x)\gamma_{e}^{-(p+1)}d\gamma_{e},
\end{equation}
where 
\begin{equation}
    F(x)=x\int_{x}^{\infty}K_{5/3}(\xi)d\xi\sim\left\{
    \begin{aligned}
    &\frac{4\pi}{\sqrt{3}\Gamma(1/3)}\left(\frac{x}{2}\right)^{1/3}, &x\ll 1,\\
    &\left(\frac{\pi}{2}\right)^{1/2}x^{1/2}e^{-x}, &x\gg 1
    \end{aligned}
\right.
\end{equation}
describes the synchrotron spectrum of a single particle in a uniform magnetic field, $\Gamma(1/3)$ is the gamma function of argument $1/3$, $x=\omega/\omega_{\rm ch,frb}=\nu/\nu_{\rm ch,frb}$ and $\nu_{\rm ch,frb}=\omega_{\rm ch,frb}/(2\pi)=3\gamma_{\rm ch,frb}^2eB_\perp/(4\pi m_ec)$ is the characteristic synchrotron emission frequency. Assuming $\nu_{\rm ch,nebula}$ is in the 10-GHz band, the corresponding Lorentz factor can be estimated as $\gamma_{\rm ch,nebula}=\sqrt{4\pi m_ec\nu_c/(3eB)}\simeq1.5\times10^3 \ \nu_{\rm ch,nebula,10}^{1/2}B_{-3}^{-1/2}$. For FRBs $\nu_{\rm ch,frb}$ is in the 1-GHz band, the corresponding Lorentz factor can be estimated as $\gamma_{\rm ch,frb}=\sqrt{4\pi m_ec\nu_c/(3eB)}\simeq488 \ \nu_{\rm ch,frb,9}^{1/2}B_{-3}^{-1/2}$.
One can see that the outcome of the integration depends on whether the characteristic electron Lorentz factor $\gamma(\nu_{\rm ch,frb})$ lie in the range of $\gamma_{\rm min}$ to $\gamma_{\rm max}$ or outside (e.g. below $\gamma_{\rm min}$). 
There are two regimes:
\begin{itemize}
\item Case (i): When $\gamma_{\rm min}\ll\gamma({\nu_{\rm ch,frb}})\ll\gamma_{\rm max}$, the absorption coefficient can be integrated as (see Appendix \ref{C} for a derivation)
\begin{equation}
\begin{aligned}
&\alpha_{\nu,e}=\frac{\sqrt{3}e^3C_{\gamma_e}}{8\pi m_e^2c^2}\left(\frac{3e}{2\pi m_ec}\right)^{p/2} B_\perp^{\frac{p+2}{2}}\Gamma\left(\frac{3p+2}{12}\right)\Gamma\left(\frac{3p+22}{12}\right)\nu^{-\frac{p+4}{2}}\\
&\simeq[10^4(8.4\times10^6)^{\frac{p}{2}} \ {\rm cm^{-1}}]C_{\gamma_e} B_\perp^{\frac{p+2}{2}}\Gamma\left(\frac{3p+2}{12}\right)\Gamma\left(\frac{3p+22}{12}\right)\nu^{-\frac{p+4}{2}}.
\end{aligned}
\end{equation}
According to the PRS spectrum of FRB 121102 \citep{Chatterjee2017}, we assume the power-law index is $p=1.1$ and consider that the synchrotron self-absorption effect is important in the nebula region. The typical length scale of the nebula is estimated as the 10-yr supernova remnant radius $r =(3\times10^{17} \ {\rm cm}) \ v_{9}t_{8.5}$ with the velocity defined by $v\simeq(2E_0/M_{\rm ej})^{1/2}\simeq(1.4\times10^{9} \ {\rm cm \ s^{-1}}) \ E_{0,51}^{1/2}M_{\rm ej,33}^{-1/2}$, where $E_0$ is the released energy and $M_{\rm ej}$ is the ejecta mass. The normalized coefficient can be calculated as
\begin{equation}
C_{\gamma_e}=\frac{n_{e}(p-1)}{\gamma_{\rm min}^{-p+1}-\gamma_{\rm max}^{-p+1}}\simeq (7.1 \ {\rm cm^{-3}}) \ \kappa L_{\nu,29}B_{-3}^{-1}\Gamma^{-1}r_{17.48}^{-2}\Delta r_{16.48}^{-1},
\end{equation}
where $\gamma_{\min}=10$ and $\gamma_{\max}=10^3$ are adopted, $\kappa\simeq(\gamma_{\rm ch,nebula}/\gamma_{\rm min})^{p-1}\simeq1.7$. 
Then the optical depth for electrons can be estimated as
\begin{equation}
\tau_{e,\rm (i)}\simeq\alpha_{\nu,e} \Delta r\simeq
7.2\times10^{-3} \ C_{\gamma_e,0.58}B_{\perp,-3}^{1.55}\nu_{9}^{-2.55}\Delta r_{16.48}.
\end{equation}
\item Case (ii): When $\gamma({\nu_{\rm ch,frb}})\ll\gamma_{\rm min}\ll\gamma_{\rm max}$, one can replace the synchrotron spectrum by an asymptotic form $\propto x^{1/3}$ and integrate over Lorentz factor as
\begin{equation}
\begin{aligned}
\alpha_{\nu,e}&=\frac{1}{2^{4/3}\Gamma(1/3)}\frac{(p+2)}{(p+2/3)}\frac{e^3B_\perp C_{\gamma_e}}{m_e^2c^2}\left(\frac{4\pi m_ec}{3eB_\perp}\right)^{1/3}\gamma_{\rm min,e}^{-(p+2/3)}\nu^{-5/3}\\
&\simeq(136 \ {\rm cm^{-1}}) \ \frac{(p+2)}{(p+2/3)}C_{\gamma_e} B_\perp^{2/3}\gamma_{\rm min,e}^{-(p+2/3)}\nu^{-5/3}. 
\end{aligned}
\end{equation}
The normalized coefficient is
\begin{equation}
C_{\gamma_e}=\frac{n_{e}(p-1)}{\gamma_{\rm min}^{-p+1}-\gamma_{\rm max}^{-p+1}}\simeq (12.7 \ {\rm cm^{-3}}) \ \kappa L_{\nu,29}B_{-3}^{-1}\Gamma^{-1}r_{17.48}^{-2}\Delta r_{16.48}^{-1},
\end{equation}
where $\gamma_{\min}=10^3$ and $\gamma_{\max}=10^4$ are adopted, and $\kappa\simeq(\gamma_{\rm ch,nebula}/\gamma_{\rm min})^{(p-1)}\simeq1.0$. 
Then the optical depth for electrons can be estimated as
\begin{equation}
\tau_{e,\rm (ii)}\simeq\alpha_{\nu,e} \Delta r\simeq
4.2 \ C_{\gamma_e,1.1}B_{\perp,-3}^{2/3}\gamma_{\rm min,e,3}^{-53/30}\nu_9^{-5/3} \Delta r_{16.48}.
\end{equation}
\end{itemize}

\begin{figure}
	\includegraphics[width=\columnwidth]{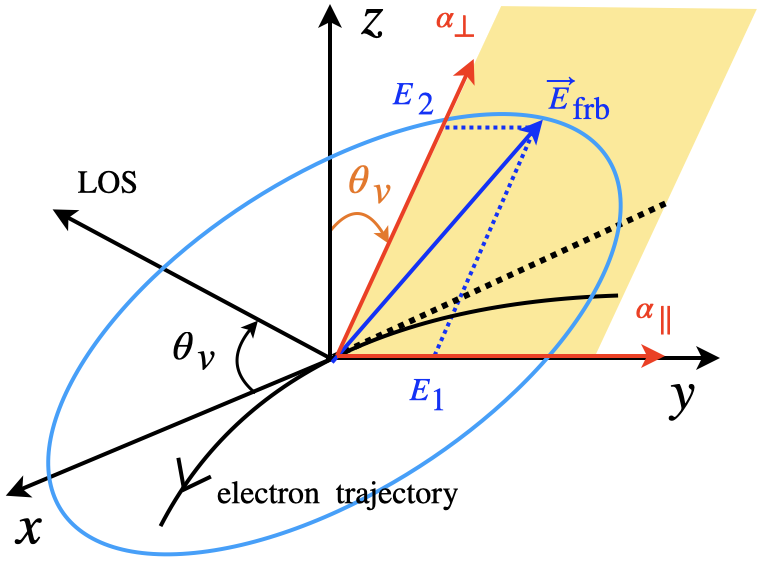}
    \caption{Geometry for instantaneous circular motion of electrons. The black curved line is the motion trajectory in the $x-y$ plane so the magnetic field is in the $-z$ direction. The blue arrow is the electric field of incident FRB waves ($\vec{E}_{\rm frb}$). The LOS is in the $x-z$ plane. The two red arrows ($E_{\rm frb,\parallel}$ is along $y$-axis and $E_{\rm frb,\perp}$ is in the $x-z$ plane) are the orthogonal components of $\vec E_{\rm frb}$. The blue ellipse describes that the incident FRB wave is elliptically polarized. $\alpha_\parallel$ is along the $y$-axis in the electron's trajectory $x-y$ plane and $\alpha_\perp$ direction is $\vec n\times\hat{y}$.}
    \label{fig:cartoon_synabs}
\end{figure}

\begin{figure*}
\begin{center}
\begin{tabular}{lll}
\resizebox{56mm}{!}{\includegraphics[]{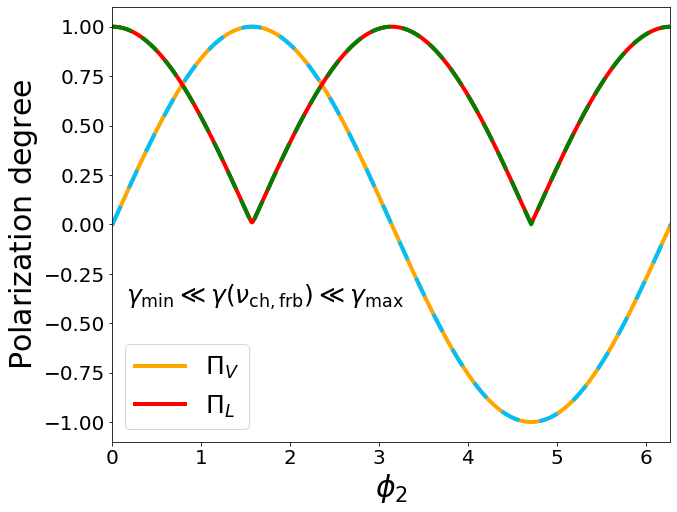}}&
\resizebox{56mm}{!}{\includegraphics[]{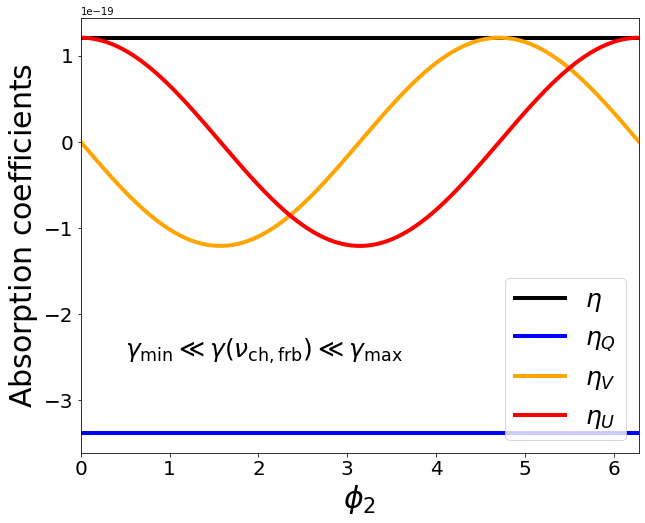}}&
\resizebox{56mm}{!}{\includegraphics[]{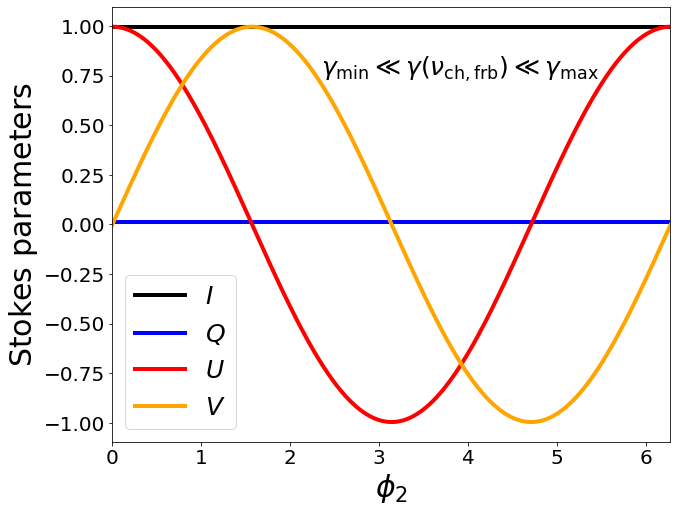}}\\
\resizebox{56mm}{!}{\includegraphics[]{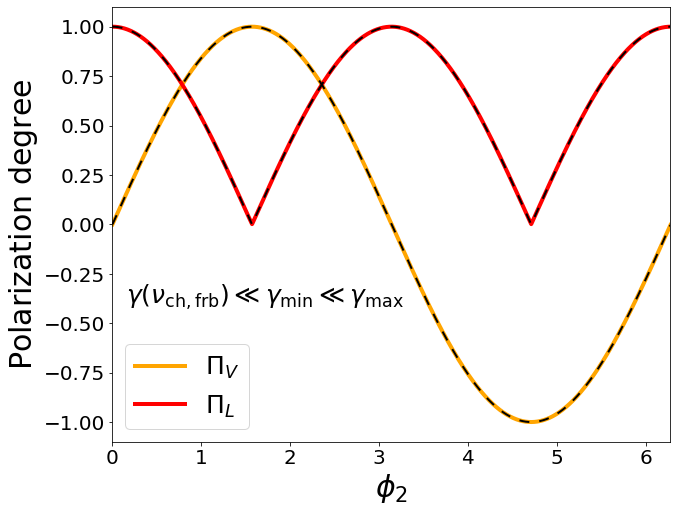}}&
\resizebox{56mm}{!}{\includegraphics[]{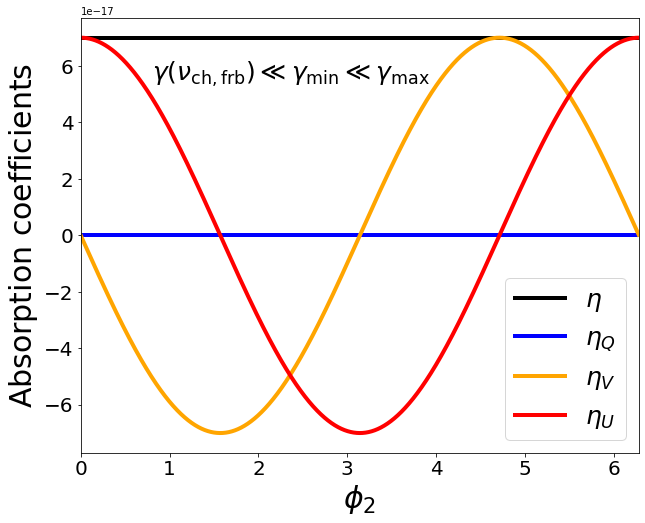}}&
\resizebox{56mm}{!}{\includegraphics[]{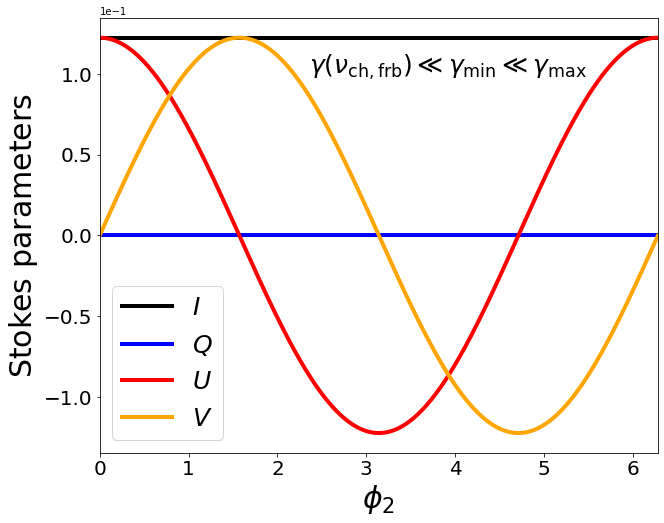}}
\end{tabular}
\caption{Linear/circular polarization degrees, absorption coefficients and Stokes parameters as a function of $\phi_2$ (from $0-2\pi$) for Case (i) with $\gamma_{\rm min}=10$ and $\gamma_{\rm max}=10^3$ (upper panels) and for Case (ii) with $\gamma_{\rm min}=10^3$ and $\gamma_{\rm max}=10^4$ (lower panels). Note $\phi_1$ is set to 0. Following parameters are adopted: nebula radius and absorption length scale $r=3\times10^{17}$ cm, absorption region $\Delta r=3\times10^{16}$ cm,  specific synchrotron luminosity of the nebular $L_{\rm \nu, nebula}=10^{29} \ \rm erg \ s^{-1} \ Hz^{-1}$, FRB typical frequency $\nu_{\rm frb}=10^9$ Hz, magnetic field $B=10^{-3}$ G, and the power-law index $p=1.1$, the ratio of two electric field amplitudes is $\epsilon_{1}/\epsilon_{2}=1$.}
\label{fig:syn_absorption}
\end{center}
\end{figure*}

In order to calculate the polarization state of an incident FRB wave, we should find the absorption coefficients of two orthogonal modes (see Fig. \ref{fig:syn_absorption}, the incident wave vector is perpendicular to the background magnetic field along the $y$-axis).
The synchrotron radiation powers of the two orthogonal modes are given by \citep{Rybicki&Lightman1979}
\begin{equation}
P_\parallel(\nu)=\frac{\sqrt{3}e^3B_\perp}{2 m_ec^2}[F(x)-G(x)],
\end{equation}
and
\begin{equation}
P_\perp(\nu)=\frac{\sqrt{3}e^3B_\perp}{2m_ec^2}[F(x)+G(x)],
\end{equation}
where
\begin{equation}
G(x)=xK_{2/3}(x)\sim\left\{
    \begin{aligned}
    &\Gamma\left(\frac{2}{3}\right)\left(\frac{x}{2}\right)^{1/3}, &x\ll 1,\\
    &x\sqrt{\frac{\pi}{2x}}e^{-x}, &x\gg 1.
    \end{aligned}
\right.
\end{equation}
We define the parallel and perpendicular components of absorption coefficients for electrons and  apply the power-law distribution of electrons for Case (i) as
\begin{equation}
\begin{aligned}
&\alpha_\parallel=-\frac{1}{8\pi\nu^2m_e}\int_{\gamma_{\rm min}}^{\gamma_{\rm max}}d\gamma P_\parallel(\nu)\gamma^2\frac{\partial}{\partial \gamma}\left[\frac{N(\gamma)}{\gamma^2}\right]\\
&=\frac{1}{2}\alpha_{\nu,e}-\frac{\sqrt{3}e^3B_\perp C_{\gamma_e}(p+2)}{32\pi\nu^2 m_e^2c^2}\int_{0}^{\infty} xK_{2/3}(x)\left(\frac{3eB_\perp}{4\pi m_e c\nu}\right)^{p/2}x^{\frac{p}{2}-1}dx,
\end{aligned}
\end{equation}
and
\begin{equation}
\begin{aligned}
&\alpha_\perp=-\frac{1}{8\pi\nu^2m_e}\int_{\gamma_{\rm min}}^{\gamma_{\rm max}}d\gamma P_\perp(\nu)\gamma^2\frac{\partial}{\partial \gamma}\left[\frac{N(\gamma)}{\gamma^2}\right]\\
&=\frac{1}{2}\alpha_{\nu,e}+\frac{\sqrt{3}e^3B_\perp C_{\gamma_e}(p+2)}{32\pi\nu^2 m_e^2c^2}\int_{0}^{\infty} xK_{2/3}(x)\left(\frac{3eB_\perp}{4\pi m_e c\nu}\right)^{p/2}x^{\frac{p}{2}-1}dx.
\end{aligned}
\end{equation}
We apply the integral formula
\begin{equation}
\int_{0}^{\infty}x^{\mu}G(x)dx=2^{\mu}\Gamma\left(\frac{\mu}{2}+\frac{4}{3}\right)\Gamma\left(\frac{\mu}{2}+\frac{2}{3}\right),
\end{equation}
with the variable $\mu=p/2-1$. The parallel and perpendicular absorption coefficients can be re-written as
\begin{equation}
\begin{aligned}
\alpha_{\parallel}=\frac{1}{2}\alpha_{\nu,e}-\alpha_{G(x)},
\end{aligned}
\end{equation}
and
\begin{equation}
\begin{aligned}
\alpha_{\perp}=\frac{1}{2}\alpha_{\nu,e}+\alpha_{G(x)},
\end{aligned}
\end{equation}
where the additional absorption coefficient of $\alpha_{G(x)}$ can be calculated as (see Appendix \ref{C} for a derivation)
\begin{equation}
\begin{aligned}
\alpha_{G(x)}&=\frac{\sqrt{3}e^3B_\perp C_{\gamma_e}}{64\pi m_e^2c^2(p+2)^{-1}}\left(\frac{3eB_\perp}{2\pi m_e c}\right)^{\frac{p}{2}}\Gamma\left(\frac{3p+10}{12}\right)\Gamma\left(\frac{3p+2}{12}\right)\nu^{-\frac{p+4}{2}}\\
&\simeq[5\times10^3(8.4\times10^6)^{\frac{p}{2}} \ {\rm cm^{-1}}](p+2)C_{\gamma_{e}}B_\perp^{\frac{p+2}{2}}\\
&\times\Gamma\left(\frac{3p+10}{12}\right)\Gamma\left(\frac{3p+2}{12}\right)\nu^{-\frac{p+4}{2}}.
\label{G_1}
\end{aligned}
\end{equation}

For Case (ii), the parallel and perpendicular components of the absorption coefficients for electrons can be written as
\begin{equation}
\alpha_{\parallel}=\frac{1}{2}\alpha_{\nu,e}-\frac{\sqrt{3}e^3B_\perp C_{\gamma_e}(p+2)}{16\pi\nu^2 m_e^2c^2}\int_{\gamma_{\rm min}}^{\gamma_{\rm max}} G(x)\gamma_e^{-(p+1)} d\gamma_e,
\end{equation}
and
\begin{equation}
\alpha_{\perp}=\frac{1}{2}\alpha_{\nu,e}+\frac{\sqrt{3}e^3B_\perp C_{\gamma_e}(p+2)}{16\pi\nu^2 m_e^2c^2}\int_{\gamma_{\rm min}}^{\gamma_{\rm max}} G(x)\gamma_e^{-(p+1)} d\gamma_e.
\end{equation}

Noticing that the emission frequency $\nu$ is in the $x\ll1$ regime, i.e. $G(x)\simeq\Gamma\left({2}/{3}\right)\left({x}/{2}\right)^{1/3}$, we replace $G(x)$ by its $\propto x^{1/3}$ asymptotic behaviour and integrate over $\gamma$ to obtain (see Appendix \ref{C} for a derivation)
\begin{equation}
\begin{aligned}
\alpha_{G(x)}&=\frac{\sqrt{3}e^3B_\perp C_{\gamma_e}(p+2)}{16\pi\nu^2 m_e^2c^2}\int_{\gamma_{\rm min}}^{\gamma_{\rm max}} G(x)\gamma_e^{-(p+1)} d\gamma_e\\
&=\frac{\sqrt{3}e^3B_\perp C_{\gamma_e}\Gamma(2/3)(p+2)}{16\pi\nu^2 m_e^2c^2} \left(\frac{2\pi m_ec\nu}{3eB_\perp}\right)^{1/3} \int_{\gamma_{\rm min}}^{\gamma_{\rm max}} \gamma_e^{-(p+\frac{5}{3})} d\gamma_e\\
&\simeq(34 \ {\rm cm}^{-1}) \ \frac{(p+2)}{(p+2/3)}C_{\gamma_e}B_\perp^{2/3}\gamma_{\rm min,e}^{-(p+2/3)}\nu^{-5/3}.
\label{G_2}
\end{aligned}
\end{equation}
We write the two components of the incident FRB wave electric field along the axis $\alpha_{\parallel}$ and $\alpha_{\perp}$ as (see Fig.\ref{fig:cartoon_synabs})
\begin{equation}
E_1=\varepsilon_1 e^{i\phi_1},
\end{equation}
and
\begin{equation}
E_2=\varepsilon_1 e^{i\phi_2},
\end{equation}
where $\varepsilon_1$ and $\varepsilon_2$ are the amplitudes, $\phi_1$ and $\phi_2$ are the phases of the two orthogonal electric fields.
Thus we can write the intensities of incident waves for parallel and perpendicular components as $I_{i,\parallel}=\varepsilon_1^2$ and $I_{i,\perp}=\varepsilon_2^2$, respectively. After passing the synchrotron self-absorption region, the two intensities of the waves can be written as $I_{s,\parallel}=I_{i,\parallel}e^{-\tau_{\parallel}}$ and $I_{s,\perp}=I_{i,\perp}e^{-\tau_{\perp}}$, where $\tau_{\parallel/\perp}$ is the optical depth. Thus the 
final electric fields of FRB waves and their complex conjugates after the absorption region can be written as
\begin{equation}
E_\parallel=\varepsilon_1 e^{-\tau_{\parallel}/2} e^{i\phi_1}, \ E_\parallel^*=\varepsilon_1 e^{-\tau_{\parallel}/2} e^{-i\phi_1},
\end{equation}
and
\begin{equation}
E_\perp=\varepsilon_2 e^{-\tau_{\perp}/2} e^{i\phi_2}, \ E_\perp^*=\varepsilon_2 e^{-\tau_{\perp}/2} e^{-i\phi_2}.
\end{equation}
The Stokes parameters can be calculated as
\begin{equation}
I=\frac{1}{2}(E_{\parallel}^*E_{\parallel}+E_{\perp}^*E_{\perp})=\frac{1}{2}(\varepsilon_1^2e^{-\tau_{\parallel}}+\varepsilon_2^2e^{-\tau_{\perp}}).
\end{equation}
\begin{equation}
Q=\frac{1}{2}(E_{\parallel}^*E_{\parallel}-E_{\perp}^*E_{\perp})=\frac{1}{2}(\varepsilon_1^2e^{-\tau_{\parallel}}-\varepsilon_2^2e^{-\tau_{\perp}}).
\end{equation}
\begin{equation}
U={\rm Re}(E_{\parallel}^*E_{\perp})=\varepsilon_1\varepsilon_2 e^{-(\tau_{\parallel}+\tau_{\perp})/2}\cos(\phi_2-\phi_1).
\end{equation}
\begin{equation}
V={\rm Im}(E_{\parallel}^*E_{\perp})=\varepsilon_1\varepsilon_2 e^{-(\tau_{\parallel}+\tau_{\perp})/2}\sin(\phi_2-\phi_1).
\end{equation}
Some general features can be pointed out: If there is no synchrotron absorption, then $\tau_\parallel=\tau_\perp=0$ and the polarization state remains constant. When the two modes have the same phase, i.e. $\phi_1=\phi_2$, and incident wave is completely linear polarized, i.e. $I^2=U^2+Q^2$. Then $V=0$ for the escaping wave and no circular polarization is generated. When the relative phase is $\phi_2-\phi_1=\pi/2$, the $V$-component reaches the maximum value.

\begin{figure*}
\begin{center}
\begin{tabular}{ll}
\resizebox{80mm}{!}{\includegraphics[]{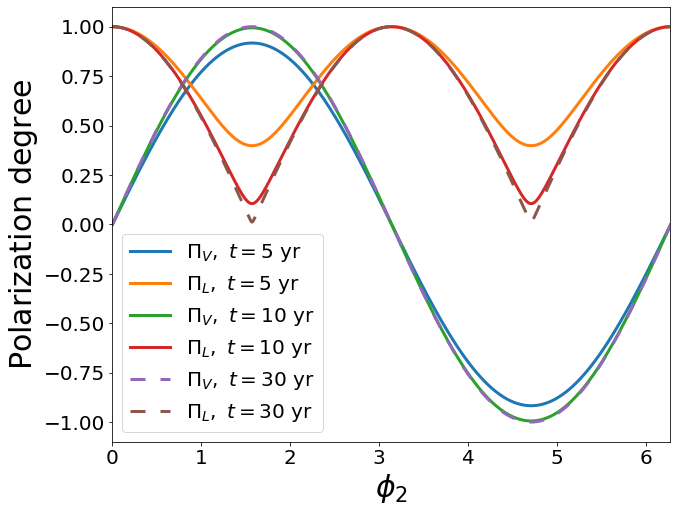}}&
\resizebox{80mm}{!}{\includegraphics[]{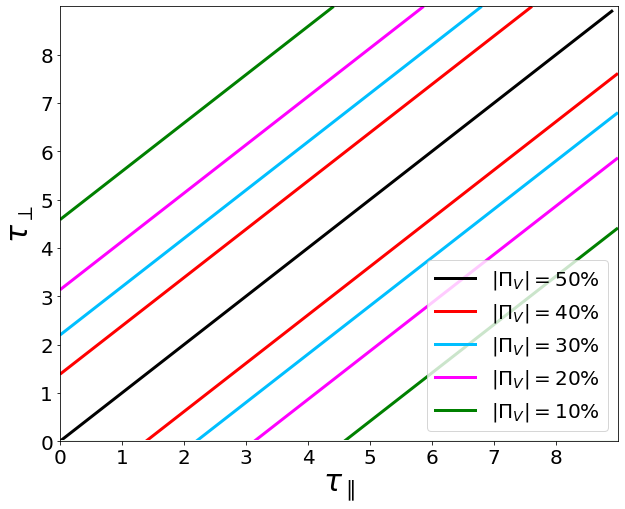}}
\end{tabular}
\caption{Left panel: Degrees of linear and circular polarization as a function of $\phi_2$ ($0-2\pi$) for Case (i) with $\gamma_{\rm min}=10$ and $\gamma_{\rm max}=10^3$, but assuming different supernova remnant ages. Blue and orange lines, green and red lines, purple and brown dashed lines are circular ($\Pi_{V}$) and linear ($\Pi_{L}$) polarization degree for a supernova remnant with $t=5, \ 10, \ 30 \ \rm yr$, respectively. Following parameters are adopted: the specific synchrotron luminosity $L_\nu=10^{29} \ \rm erg \ s^{-1} \ Hz^{-1}$, FRB typical frequency $\nu_{\rm frb}=10^9$ Hz, magnetic field $B=10^{-3}$ G, power-law index $p=2$, $\kappa=150$, and the ratio of the two electric field amplitudes $\epsilon_{1}/\epsilon_{2}=1$. Right panel: The circular polarization degree as a function of $\tau_\parallel$ and $\tau_\perp$. Following parameters are adopted: $\epsilon_1=\epsilon_2=1$, $\sin(\phi_2-\phi_1)=1$. One can see that $\Pi_V$ drops as either $\tau_\perp$ or $\tau_\parallel$ increase.}
\label{fig:syn_absorption_free}
\end{center}
\end{figure*}

We ignore the Faraday conversion and rotation coefficients, then the four absorption coefficients can be solved as
\begin{equation}
\eta=-\frac{{I}{I_s}-{Q}{Q_s}-{U}{U_s}-{V} {V_s}}{{I}^2-{Q}^2-{U}^2-{V}^2}=\frac{1}{2}(\alpha_\parallel+\alpha_\perp),
\label{eq:eta}
\end{equation}
\begin{equation}
\begin{aligned}
\eta_Q&=-\frac{{I}^2 {Q_s}-{I}{I_s} {Q}+{Q} {U}{U_s}+{Q}{V} {V_s}-{Q_s} {U}^2-{Q_s} {V}^2}{{I} \left({I}^2-{Q}^2-{U}^2-{V}^2\right)}\\
&=\frac{1}{2}(\alpha_\parallel-\alpha_\perp),
\end{aligned}
\end{equation}
\begin{equation}
\begin{aligned}
\eta_U&=-\frac{{I}^2 {U_s}-{I} {I_s} {U}-{Q}^2 {U_s}+{Q} {Q_s} {U}+{U} {V} {V_s}-{U_s} {V}^2}{{I} \left({I}^2-{Q}^2-{U}^2-{V}^2\right)}\\
&=\frac{\epsilon_1\epsilon_2{e^{-(\tau_\parallel+\tau_\perp)/2}}}{\epsilon_1^2e^{-\tau_\parallel}+\epsilon_2^2e^{-\tau_\perp}}(\alpha_\parallel+\alpha_\perp)\cos(\phi_2-\phi_1),
\end{aligned}
\end{equation}
\begin{equation}
\begin{aligned}
\eta_V&=-\frac{{I}^2 {V_s}-{I}{I_s} {V}-{Q}^2 {V_s}+{Q} {Q_s} {V}-{U}^2 {V_s}+{U} {U_s} {V}}{{I} \left({I}^2-{Q}^2-{U}^2-{V}^2\right)}\\
&=-\frac{\epsilon_1\epsilon_2{e^{-(\tau_\parallel+\tau_\perp)/2}}}{\epsilon_1^2e^{-\tau_\parallel}+\epsilon_2^2e^{-\tau_\perp}}(\alpha_\parallel+\alpha_\perp)\sin(\phi_2-\phi_1),
\label{eq:eta_V}
\end{aligned}
\end{equation}
where the derivative of the four Stokes parameters can be written as
\begin{equation}
I_s=\frac{dI}{ds}=\frac{1}{2}(-\alpha_\parallel\varepsilon_1^2e^{-\tau_{\parallel}}-\alpha_\perp\varepsilon_2^2e^{-\tau_{\perp}}).
\end{equation}
\begin{equation}
Q_s=\frac{dQ}{ds}=\frac{1}{2}(-\alpha_\parallel\varepsilon_1^2e^{-\tau_{\parallel}}+\alpha_\perp\varepsilon_2^2e^{-\tau_{\perp}}).
\end{equation}
\begin{equation}
U_s=\frac{dU}{ds}=[-(\alpha_\parallel+\alpha_\perp)/2]\varepsilon_1\varepsilon_2 e^{-(\tau_{\parallel}+\tau_{\perp})/2}\cos(\phi_2-\phi_1).
\end{equation}
\begin{equation}
V_s=\frac{dV}{ds}=[-(\alpha_\parallel+\alpha_\perp)/2]\varepsilon_1\varepsilon_2 e^{-(\tau_{\parallel}+\tau_{\perp})/2}\sin(\phi_2-\phi_1).
\end{equation}
One can see that $\eta$ is equal to the half value of $\alpha_\nu$. 
It should be pointed out that $I^2=Q^2+U^2+V^2$ is applied for the derivation, thus depolarization cannot happen via synchrotron self-absorption.
Notice that we have made a connection between $\eta$, $\eta_{Q,U,V}$ and  $\alpha_{\parallel,\perp}$ through Eqs.(\ref{eq:eta})-(\ref{eq:eta_V}).

\begin{itemize}
\item For Case (i) and $p=1.1$, the two-component optical depth can be estimated as
\begin{equation}
\tau_{\parallel,({\rm i})}=\alpha_{\parallel} \Delta r\simeq-6.5\times10^{-3} \ C_{\gamma_e,0.85}B_{\perp,-3}^{1.55}\nu_{9}^{-2.55}\Delta r_{16.48},
\end{equation}
and
\begin{equation}
\tau_{\perp,({\rm i})}=\alpha_{\perp} \Delta r\simeq0.01 \ C_{\gamma_e,0.85}B_{\perp,-3}^{1.55}\nu_{9}^{-2.55}r_{17.48}\Delta r_{16.48}.
\end{equation}
Note that the optical depth of parallel mode is negative, but the total optical depth is positive, thus there is no maser.
\item For Case (ii) and $p=1.1$, the two-component optical depth can be estimated as
\begin{equation}
\tau_{\parallel,({\rm ii})}=\alpha_{\parallel} \Delta r\simeq2.1 \ C_{\gamma_e,1.1}B_{\perp,-3}^{2/3}\gamma_{\rm min,e,3}^{-53/30}\nu_{9}^{-5/3}\Delta r_{16.48},
\end{equation}
and
\begin{equation}
\tau_{\perp,({\rm ii})}=\alpha_{\perp} \Delta r\simeq2.1 \ C_{\gamma_e,1.1}B_{\perp,-3}^{2/3}\gamma_{\rm min,e,3}^{-53/30}\nu_{9}^{-5/3}\Delta r_{16.48}.
\end{equation}
\end{itemize}

With the parameters to account for the PRS of FRB 121102, we present the numerical results of linear and circular polarization degree after synchrotron absorption as a function of $\phi_2$ ($\phi_1$ is set to 0) in Fig.\ref{fig:syn_absorption} for the two cases. We also present the four absorption coefficients and four Stokes parameters as a function of $\phi_2$. We have assumed that the two modes of the incident wave have the same wave amplitude\footnote{It should be pointed out that the amplitudes of the incident waves orthogonal modes can also determine the final polarization degree.}.
Case (i) is on the upper panel and Case (ii) is on the lower panel. Dashed lines denote the original linear and circular polarization degrees, respectively.
Red and orange lines denote the linear and circular polarization degrees after passing through the absorption region, respectively.  
Since we assumed $\phi_1=0$, one can see when $\phi_2=0,\pi$ or $2\pi$, the escaped wave has the same polarization degree as the incident wave, i.e. completely linearly polarized. One can see that for the adopted parameters the synchrotron absorption effect is insignificant to absorb the two modes of the incident wave. The linear and circular polarization degree profiles are nearly the same as those of the original incident wave.

In the above discussion, the calculations are carried out using the parameters to interpret the PRS of FRB 121102. Most other FRB sources do not have persistent radio emission (the only other case is FRB 190520B \citep{Niu2022}), likely due to a lower magnetic field and total number of electrons in an older nebula. This is consistent with the fact that both FRB 121102 and FRB 190520B have abnormally large RM whereas other FRBs have much smaller RM values \citep{YangLiZhang2020,YXZ}. Because the case FRB 121102 only gives a marginal absorption effect, one can conclude that for most FRBs, synchrotron absorption effect iis likely not important to modify the polarization properties of FRBs.

Nonetheless, one can explore the parameter space for the synchrotron nebula by taking the age age of supernova remnant as a free parameter. Left penal of Figure \ref{fig:syn_absorption_free} presents the linear/circular polarization degree as a function of $\phi_2$ for a range of age values. When the incident wave is completely linear polarized, i.e. $\phi_1=\phi_2=0, \ \pi \ {\rm or } \ 2\pi$, circular polarization components cannot be generated only through the synchrotron absorption since $V=0$ at the initial time and $\eta_V=0$ all the time, i.e. $d\Pi_V/ds=0$ (see Eq.(\ref{eq:dPi_V/ds})). However, the linear polarization component can be enhanced through the synchrotron absorption since  $L=\sqrt{Q^2+U^2}$ is influenced by both $Q$ and $U$. When the incident wave is completely circularly polarized, i.e. $\phi_2-\phi_1=\pi/2 \ {\rm or} \ 3\pi/2$, although $dU/ds=0$, $dQ/ds\neq0$. In such case, the absolute value circular polarization degree will decrease. 
For an elliptically polarized incident wave in general, synchrotron absorption tends to lower the circular polarization degree and increase the linear polarization degree.   
This can be readily seen in the right panel of Fig.\ref{fig:syn_absorption_free}, where we plot the circular polarization degree ($\Pi_V$) as a function of $\tau_\parallel$ and $\tau_\perp$.
Black dashed line corresponding to $\tau_\parallel=\tau_\perp$ with $\Pi_V=0.5$ (for 
$\sin(\phi_2-\phi_1)=1$). Once can see $\Pi_V$ always decreases monotonically as $\tau_\perp$ or $\tau_\parallel$ increase.

It should be pointed out that: (1) In order to have polarization-mode-dependent synchrotron absorption, an ordered magnetic field is required in the absorber. In reality, the synchrotron nebula surrounding the FRB engine may carry a random magnetic configuration. In such case, synchrotron absorption would globally decrease the flux below the absorption frequency without changing the polarization state of the incident waves. 
(2) If the FRB frequency is below the synchrotron self-absorption frequency of the nebula, electrons in the nebula would absorb FRB  photons. The electron spectrum could become harder due to synchrotron heating by the FRB emission \citep{Yang&Zhang16}. Such a process is not included in our analysis and needs further investigation.

\subsubsection{Cyclotron absorption}

For non-relativistic electrons, the cyclotron frequency can be estimated as $\omega_B=eB/(m_ec)\simeq(1.8\times10^7 \ {\rm rad \ s^{-1}}) \ B$. In order to satisfy the cyclotron absorption condition, i.e. $\omega_B=\omega_{\rm frb}=2\pi\nu_{\rm frb}$, the required magnetic field is
\begin{equation}
B=\frac{m_ec\omega_{\rm frb}}{e}\simeq(360 \ {\rm G}) \ \nu_{\rm frb,9}.
\end{equation}
Such a high magnetic field strength is unlikely to exist in the nebula around a single magnetar. A plausible physical scenario is to consider a binary system, so that a strong magnetic field could exist in the stellar wind from a companion star \citep[e.g.][]{Zhang18,IokaZhang20,Lyutikov20,Wada21,WFY2022}. The mass loss rate is $\dot M \sim 10^{-14}-10^{-10}M_\odot \ {\rm yr}^{-1}$ \citep{Wood2002} for a solar type star, and is $\dot M \sim 10^{-11}-10^{-8}M_\odot \ {\rm yr}^{-1}$ for a Be-star \citep{Snow1981,Poe1986}, for O-star is $\dot M=10^{-7}-10^{-5}M_\odot \ {\rm yr}^{-1}$ \citep{Puls1996,Muijres2012}.
Adopting a surface magnetic field strength $B_c=10^3$ G and a wind velocity  $v_w=10^3 \ {\rm km \ s^{-1}}$, one may define the Alfv\'en radius $R_A$ where the magnetic pressure balances the ram pressure
\begin{equation}\label{eq:Alfven radius}
\begin{aligned}
R_A&=\left(\frac{B_c^2R_c^6}{2\dot Mv_w}\right)^{1/4}\\
&\simeq\left\{
\begin{aligned}
&(5.1\times10^{11} \ {\rm cm}) \ B_{c,3}^{1/2}R_{c,10.8}^{3/2}\dot M_{15.8}^{-1/4}v_{w,8}^{-1/4},  &&{\rm Solar \ type \ star} \\
&(1.7\times10^{11} \ {\rm cm}) \ B_{c,3}^{1/2}R_{c,10.8}^{3/2}\dot M_{17.8}^{-1/4}v_{w,8}^{-1/4},  &&{\rm Be-star} \\
&(3.1\times10^{10} \ {\rm cm}) \ B_{c,3}^{1/2}R_{c,10.8}^{3/2}\dot M_{20.8}^{-1/4}v_{w,8}^{-1/4},  &&{\rm O-star}.
\end{aligned}
\right.
\end{aligned}
\end{equation}
The magnetic field strength at a distance $r$ from a highly magnetized companion star with $R_c=R_\odot$ can be estimated as
\begin{equation}\label{eq:B-field companion star}
\begin{aligned}
B&\simeq\left\{
\begin{aligned}
& B_c\left(\frac{r}{R_c}\right)^{-3}, \ &&R_c<r<R_A \ {\rm or} \ R_A<R_c<r, \\
& B_c\left(\frac{R_A}{R_c}\right)^{-3}\left(\frac{r}{R_A}\right)^{-1}, \ &&R_c<R_A<r.
\end{aligned}
\right.
\end{aligned}
\end{equation}
In order to satisfy the required magnetic field strength $B \simeq (360 \ {\rm G})~ \nu_{\rm frb,9}$, the corresponding radii for the three types of stars can be estimated as
\begin{equation}
\begin{aligned}
r&\simeq\left\{
\begin{aligned}
& (9.8\times10^{10} \ {\rm cm}) \ \nu_{\rm frb,9}^{-1/3}B_{c,3}^{1/3}R_{c,10.8}, \ &&{\rm Solar \ type \ star}  \\
& (9.8\times10^{10} \ {\rm cm}) \ \nu_{\rm frb,9}^{-1/3}B_{c,3}^{1/3}R_{c,10.8}, \ &&{\rm Be-star}\\
& (9.8\times10^{10} \ {\rm cm}) \ \nu_{\rm frb,9}^{-1} B_{c,3}^{1/3}R_{c,10.8}, \ &&{\rm O-star}.
\end{aligned}
\right.
\end{aligned}
\end{equation}
The electron density in the stellar wind of the three types of stars can be estimated as
\begin{equation}\label{eq:three types number density}
\begin{aligned}
n_e&=\frac{\dot M}{4\pi m_pr^2v_w}\\
&\simeq\left\{
\begin{aligned}
& (3.0\times10^{8} \ {\rm cm^{-3}}) \ \dot M_{15.8}r_{11}^{-2}v_{w,8}^{-1}, \ &&{\rm Solar \ type \ star} \\
& (3.0\times10^{10} \ {\rm cm^{-3}}) \ \dot M_{17.8}r_{11}^{-2}v_{w,8}^{-1}, \ &&{\rm Be-star} \\
& (3.0\times10^{13} \ {\rm cm^{-3}}) \ \dot M_{20.8}r_{11}^{-2}v_{w,8}^{-1}, \ &&{\rm O-star}.
\end{aligned}
\right.
\end{aligned}
\end{equation}
Similar to the resonant cyclotron absorption within the magnetosphere, we consider resonance cyclotron in the following treatment\footnote{The strengths of the magnetic field in both scenarios are comparable, and cyclotron absorption also proceeds between Landau levels. The main differences are two folds. First, the absorption region is now in the magnetosphere of the companion. Second, the absorption region is moving with a non-relativistic speed and the absorbing plasma is essentially cold.}
For resonant absorption, we consider 
Doppler broadening of the plasma with 
velocity $v\sim0.01 c$. The frequency shift due to Doppler motion can be estimated as $\Delta\nu=\nu-\nu_B\sim\beta\cos\theta\omega_B/2\pi\sim(10^{7} \ {\rm Hz}) \ \beta_{-2}B_{2.55}$, and we assumed the electron velocity is along the LOS, so that $\cos\theta=1$.
The cross section of the electrons cyclotron resonance absorption can be then calculated as
\begin{equation}
\sigma_{\rm cyc,c}=\frac{1}{2}\pi r_0c(1+\cos^2\theta)\phi(\nu-\nu_B)\simeq3.3\times10^{-21} \ {\rm cm^2}.
\end{equation}
We consider that the length scale of the cyclotron absorption region is $\Delta r\simeq0.1 r$.
Then the optical depth for electron  cyclotron absorption can be estimated as
\begin{equation}
\begin{aligned}
\tau_{\rm cyc,e}&\simeq n_e\sigma_{\rm cyc,e} \Delta r \\
&\simeq\left\{
\begin{aligned}
&0.01 \ \dot M_{15.8}r_{11}^{-2}v_{w,8}^{-1}\Delta r_{10},  &&{\rm Solar \ type \ star} \\
&1.0 \ \dot M_{17.8}r_{11}^{-2}v_{w,8}^{-1}\Delta r_{10},  &&{\rm Be-star} \\
&1.0\times10^3 \ \dot M_{20.8}r_{11}^{-2}v_{w,8}^{-1}\Delta r_{10},  &&{\rm O-star}.
\end{aligned}
\right.
\end{aligned}
\end{equation}
Then the circular polarization degree can be calculated as
\begin{equation}
\Pi_V=\frac{|e^{-\tau_{\rm cyc,p}}-e^{-\tau_{\rm cyc,e}}|}{e^{-\tau_{\rm cyc,p}}+e^{-\tau_{\rm cyc,e}}}\simeq\left\{
\begin{aligned}
&0.5\%,  &&{\rm Solar \ type \ star} \\
&46.3\%,  &&{\rm Be-star} \\
&100\%,  &&{\rm O-star}.
\end{aligned}
\right.
\end{equation}
for the adopted typical parameters. 
It should be pointed out that the absorption by the ion component is negligible, i.e. $\tau_{\rm cyc,p}=0$, so that net circular polarization will be produced. Thanks to the high electron number densities in their stellar winds, O stars with the highest mass loss rate could achieve a circular polarization degree as high as $100\%$ and Be stars can achieve a moderate value $46\%$.
The cyclotron absorption effect for solar-like stars, on the other hand, is negligible.

In general, in order to satisfy the required magnetic field strength ($B \simeq 360$ G) for cyclotron absorption, the absorption region is close to the magnetosphere of the companion star. At such a small radius, the binary would have an orbital period of $P_b \simeq 2\pi(GM)^{-1/2}r^{3/2}\simeq(0.2 \ {\rm days}) \ (M/M_{\odot})^{-1/2}r_{11}^{3/2}$, where $M$ is the total mass in the binary system. This scenario therefore predicts a periodic variation of the degree of cyclotron absorption, and hence, a periodic variation of the observed degree of circular polarization with a period of hours. Non-detection of such a pattern would disfavor such a scenario.

\subsubsection{Faraday rotation and conversion}\label{sec:conversion}
In this section, we discuss Faraday rotation and conversion far away from the FRB source. The ambient environment is believed to be an electron-ion plasma. Faraday rotation is the rotation of the polarization angle of a linearly polarized wave propagating through a magnetized plasma with a magnetic field component parallel to the line of sight. The Faraday rotation measure (RM) is defined as \citep{Rybicki&Lightman1979}
\begin{equation}
{\rm RM}=\frac{e^3}{2\pi m_ec^4}\int n_eB_\parallel ds
\end{equation}
for a non-relativistic, cold electron-ion plasma, and $B_\parallel$ is the magnetic field along the LOS. For a magnetic field configuration with a dominant $B_\parallel$ component, the eigenmodes are considered as R and L-modes, which can be derived by applying the quasi-parallel condition in section \ref{sec:polarization}. The dispersion relations can be written as\footnote{It is unlikely that a high degree of circular polarization of FRB waves can be generated because one of the R-mode or L-mode is in forbidden region for their dispersion relations, i.e. $n^2<0$. This is because for physical environments of magnetars or binary systems, two conditions for $n^2<0$, namely, the plasma frequency $\omega_p$ is quasi-equal to the cyclotron frequency $\omega_B$ and the incident FRB angular frequency is quasi-equal to $\omega_B$, cannot be satisfied.} (see Appendix \ref{B} for a derivation)
\begin{equation}
n_{\rm R}^2=1-\frac{\omega_p^2}{\omega(\omega-\omega_B\cos\theta)}\simeq1-\frac{\omega_p^2}{\omega(\omega-\omega_B)}
\end{equation}
and
\begin{equation}
n_{\rm L}^2=1-\frac{\omega_p^2}{\omega(\omega+\omega_B\cos\theta)}\simeq1-\frac{\omega_p^2}{\omega(\omega+\omega_B)},
\end{equation}
where the approximation $\theta\rightarrow 0$ is applied in the last step. The wave numbers of the two modes are $k_{\rm R,L}=\omega n_{\rm R,L}/c$. 
The difference in wave number between the two modes at $\theta=0$ can be written as 
\begin{equation}
\Delta k_{\rm RL}^{\theta=0}=\frac{\omega}{c}(n_{\rm R}-n_{\rm L})\simeq -\frac{\omega_p^2\omega_B}{c\omega^2}, 
\end{equation}
which varies with wavelength as $\omega^{-2}\sim \lambda^2$. The polarization angle varies as wave propagates and no circular component is generated via Faraday rotation.
It should be pointed out that RM is defined in the region where $\omega\gg\omega_B$ and $\omega\gg\omega_p$.

Faraday conversion via field reversal ($\vec k_{\rm frb}\perp\vec B$) is considered to be responsible for producing the circular polarization components of an initially linearly polarized FRB wave. A magnetic field reversal is defined as the region where the magnetic field component along the line of sight is nearly equal to zero, which connects the region where the field line points towards Earth and where it points away. In such a region, the wave vector is perpendicular to local magnetic field and the eigenmodes are considered as X and O-modes. The dispersion relations are presented in Eqs.(\ref{Xmodeion}) and (\ref{Omodeion}) for the two modes in an electron-ion plasma. The quasi-perpendicular condition in the region of $\omega\gg\omega_p$ and $\theta\rightarrow\pi/2$ can be simplified as \citep{Stix1992,Melrose2010}
\begin{equation}
|\cos\theta|\lesssim\frac{\omega_B}{2\omega}.
\end{equation}
The difference of the two modes wave vectors at $\theta=\pi/2$ can be written as 
\begin{equation}
\Delta k_{\rm XO}^{\theta=\pi/2}=\frac{\omega}{c}(n_{\rm X}-n_{\rm O}) \simeq -\frac{\omega_p^2\omega_B^2}{2c\omega^3},
\end{equation}
which varies as $\omega^{-3}\sim \lambda^3$ faster than rotation $\sim\lambda^2$.
The coefficients of Faraday rotation and conversion can be written as \citep[e.g.][]{Gruzinov&Levin2019}
\begin{equation}
\rho_V=\Delta k_{\rm RL}^{\theta=0}\hat{B}_z,
\end{equation}
and
\begin{equation}
\rho_Q+i\rho_U=\Delta k_{\rm XO}^{\theta=\pi/2}(\hat{B}_x+i\hat{B}_y)^2,
\end{equation}
where $\hat{B}_{x,y,z}=B_{x,y,z}/B$ are the normalized vector components of the magnetic field.
The relative phase of the two modes denoting the  Faraday conversion amplitude can be calculated as $\Delta\phi=\Delta k x = \Delta k \Delta r\cos\theta\simeq\omega_B \Delta r/(2\omega)$, where $\Delta r$ is the characteristic length of the magnetic field over which the angle changes in the field reversal region. 
Consider the relative phase $\Delta\phi=\pi/2$ as the optimistic condition to convert linear polarization to circular polarization, one can define a characteristic Faraday conversion frequency \citep{Cohen1960,Melrose2010}
\begin{equation}
\begin{aligned}
\omega_{\rm FC}&=\left(\frac{\omega_p^2\omega_B^3\Delta r}{2\pi c}\right)^{1/4}.
\end{aligned}
\end{equation}
The ratio between wave frequency and the characteristic Faraday conversion frequency $\omega/\omega_{\rm FC}$ determines the amplitude of Faraday conversion. When $\omega/\omega_{\rm FC}\gg1$, i.e. $\Delta\phi\ll1$ and the effect of Faraday conversion is weak since the relative phase is small and X and O-modes are nearly in the same phase. When $\omega/\omega_{\rm FC}\ll 1$, on the other hand, the conversion effect is very significant. The amount of change depends on $\Delta \phi$. 

\begin{figure}
	\includegraphics[width=\columnwidth]{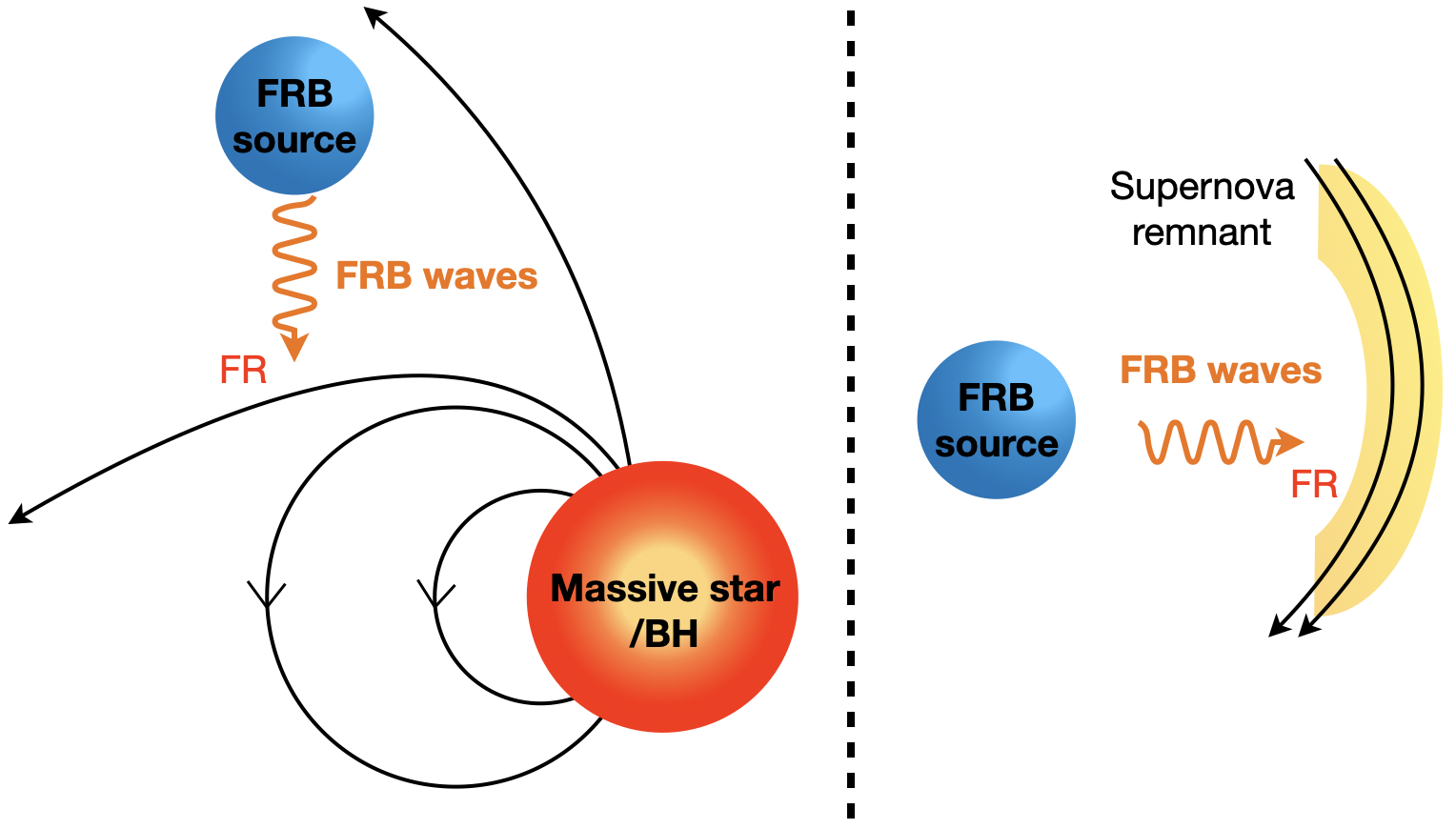}
    \caption{Cartoon pictures of three possible scenarios for field reversal (FR): In the left panel, the line of sight passes through the nearly perpendicular magnetic field of a massive star or black hole companion. In the right panel, the FRB source is surrounded by magnetar wind nebula or supernova remnant threaded by a perpendicular magnetic field. The orange wiggled arrows denote the FRB waves.}
    \label{fig:reversal}
\end{figure}

In the following, we discuss three astronomical scenarios of field reversal, which is the critical condition for Faraday conversion. 
These scenarios are illustrated in Fig.\ref{fig:reversal}, see also \cite{Dai21,YXZ}). We focus on the characteristic Faraday conversion frequency in the three scenarios to judge under what condition Faraday conversion is important in each scenario.

\begin{figure*}
\begin{center}
\begin{tabular}{lll}
\resizebox{58mm}{!}{\includegraphics[]{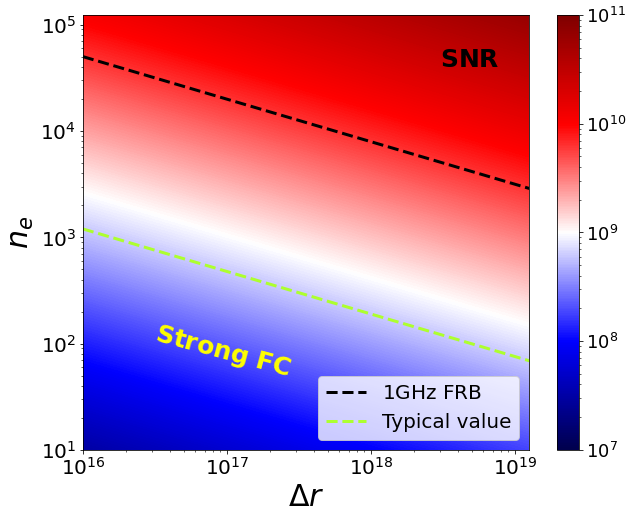}}&
\resizebox{58mm}{!}{\includegraphics[]{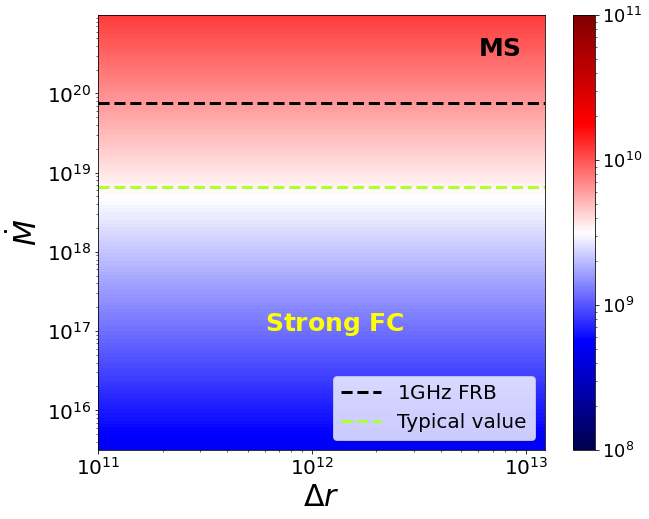}}&
\resizebox{58mm}{!}{\includegraphics[]{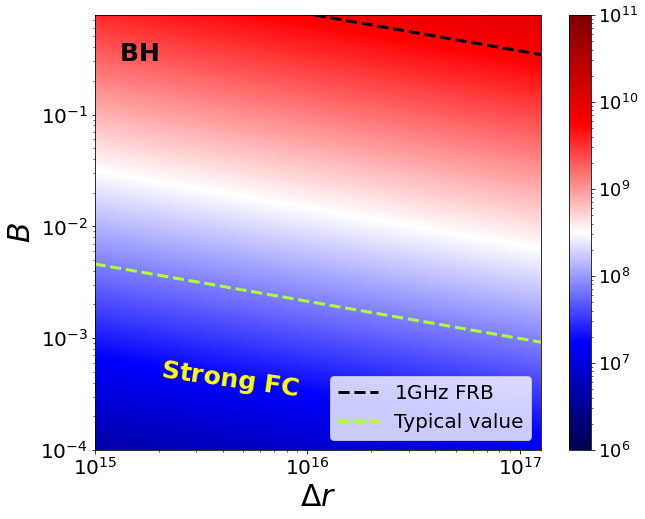}}\\
\resizebox{56mm}{!}{\includegraphics[]{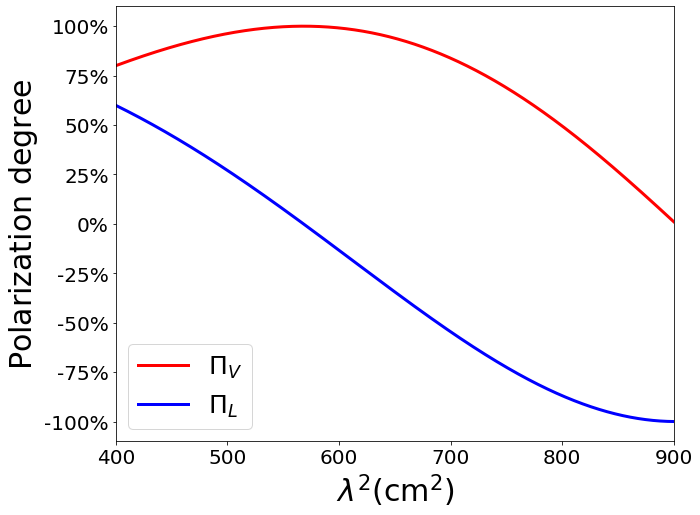}}&
\resizebox{56mm}{!}{\includegraphics[]{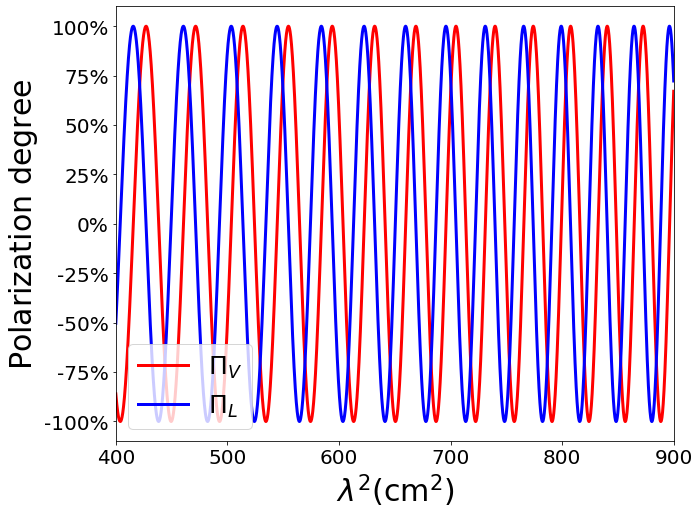}}&
\resizebox{56mm}{!}{\includegraphics[]{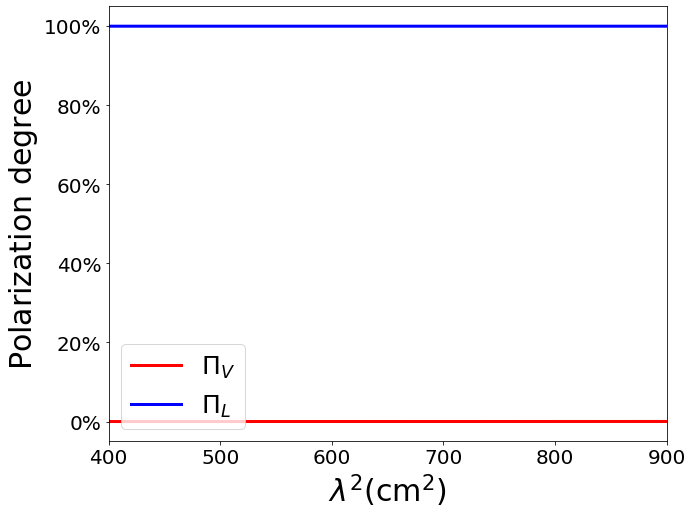}}
\end{tabular}
\caption{Degree of circular and linear polarization as a function of $\lambda^2$ in the lower panel.
Left column (field reversal by a supernova remnant): Characteristic Faraday conversion frequency as a function of typical length scale $\Delta r$ and electron number density $n_e$. Following parameters are adopted: $v_r=10^9 \ {\rm cm \ s^{-1}}$, $\nu_{\rm frb}=10^9$ Hz, $B=3.2\times10^{-2}$ G, $\epsilon_B=0.1$ and typical length scale $\Delta r=10^{17}$ cm. Center column (field reversal by a massive/giant companion star): Characteristic Faraday conversion frequency as a function of typical length scale $L$ and mass loss rate $\dot M$. Following parameters are adopted: $v_w=10^9 \ {\rm cm \ s^{-1}}$, typical length scale $\Delta r=10^{12}$ cm. Right column (field reversal by a massive black hole): Characteristic Faraday conversion frequency as a function of typical length scale $\Delta r$ and magnetic field strength $B$. Following parameters are adopted: $v_w=10^9 \ {\rm cm \ s^{-1}}$, $\epsilon_{\rm BH}=0.1$, $f=0.1$, $M_{\rm BH}=10^{38}$ g and typical length scale $\Delta r=10^{15}$ cm.}
\label{fig:Faraday conversion}
\end{center}
\end{figure*}
\begin{itemize}
\item Field reversal by a supernova remnant. A supernova remnant (SNR) is the blastwave due to interaction between a supernova ejecta and the ambient medium. We consider a simplified model that a spherically symmetric SNR expands into a uniform density medium, with a quasi-toroidal magnetic field threading the blastwave. 
We consider three phases of the blastwave evolution \citep[e.g.][]{Draine}: (1) The free-expansion phase, during which the SNR ejecta moves at nearly constant velocity which can be estimated as $v\simeq(2E_0/M_{\rm ej})^{1/2}\simeq(1.4\times10^{9} \ {\rm cm \ s^{-1}}) \ E_{0,51}^{1/2}M_{\rm ej,33}^{-1/2}$,
so that the SNR outer radius  
can be estimated as $r\simeq vt=(10^{16} \ {\rm cm}) \ v_{9}t_{7}$, where $t$ is the expansion time normalized to $10^7$ s (of the order of a year). 
The SNR mass is dominated by the ejecta mass and the mass contribution from the swept ambient materials can be neglected. The free-expansion phase ends at the radius  
$R_{\rm blast}=[3M_{\rm ej}/(4\pi m_pn_0)]^{1/3}\simeq(5.2\times10^{18} \ {\rm cm}) \ M_{\rm ej,33}^{1/3}n_0^{1/3}$, where $n_0$ is the blastwave swept number density. 
(2) The Sedov–Taylor phase: it starts
when the ejecta begins to slow down after the swept medium mass becomes comparable to the mass of the ejecta. 
During the phase the SNR mass is dominated by the swept ambient medium and the internal structure of the SNR can be described by a self-similar solution. (3) The snowplow phase, during which the blastwave undergoes fast radiative cooling so that the momentum of the blastwave is conserved.  
We consider an SNR blastwave at $\sim 1$ pc in the free-expansion phase. With the thin shell approximation, the number density in the shell can be estimated as
\begin{equation}
n_e\simeq\frac{M}{4\pi\mu_m m_pr^2\Delta r}\simeq(4.8\times10^2 \ {\rm cm^{-3}}) \ M_{33}r_{18}^{-2}\Delta r_{17}^{-1}\mu_m^{-1},
\end{equation}
where $\mu_{m}$ is the mean molecular weight.
The plasma frequency in the thin shell can be calculated as
\begin{equation}
\omega_{p,\rm shell}=\sqrt{\frac{4\pi e^2n_e}{m_e}}\simeq(1.2\times10^6 \ {\rm rad \ s^{-1}}) \ M_{33}^{1/2}r_{18}^{-1}\Delta r_{17}^{-1/2}\mu_m^{-1/2}.
\end{equation}
We consider that the magnetic field is in the reverse shock
and assume that the fraction of the shock energy that goes to magnetic fields is $\epsilon_{B}=0.1$. 
This gives \citep{Piro18}
\begin{equation}
\frac{B^2}{8\pi}\simeq\frac{1}{2}\epsilon_{B}\rho v_r^2,
\end{equation}
where $v_r$ is the velocity of the reverse shock. 
The magnetic field can be calculated as 
\begin{equation}
\begin{aligned}
B&\simeq\sqrt{4\pi\epsilon_{B}m_p n_e}v_r\\
&\simeq(3.2\times10^{-2} \ {\rm G}) \ \epsilon_{B,-1}^{1/2}M_{33}^{1/2}r_{18}^{-1}\Delta r_{17}^{-1/2}\mu_m^{-1/2}v_{r,9}.
\end{aligned}
\end{equation}
The corresponding characteristic Faraday conversion frequency can be then estimated as 
\begin{equation}
\omega_{\rm FC,SNR}\simeq(6.1\times10^{8} \ {\rm rad \ s^{-1}}) \ \epsilon_B^{3/8} M_{33}^{5/8}r_{18}^{-5/4}\Delta r_{17}^{-3/8}\mu_m^{-5/8}v_{r,9}^{3/4},
\end{equation}
where the typical length scale of the blastwave $\Delta r$ is chosen to be $10^{17}$ cm. One can see that it is close to the FRB emission frequency for the assumed parameters.
\item Field reversal by a massive/giant companion star. The FRB source might be in a binary system and the field reversal might occur under some special geometric configurations (Fig.\ref{fig:reversal}).
The electron density in the stellar wind of the three types of companion stars can be estimated as (see Eq.(\ref{eq:three types number density}))
\begin{equation}
\begin{aligned}
n_e&=\frac{\dot M}{4\pi m_pr^2v_w}\\
&\simeq\left\{
\begin{aligned}
& (3.0\times10^{4} \ {\rm cm^{-3}}) \ \dot M_{15.8}r_{13}^{-2}v_{w,8}^{-1}, \ &&{\rm Solar \ type \ star} \\
& (3.0\times10^{6} \ {\rm cm^{-3}}) \ \dot M_{17.8}r_{13}^{-2}v_{w,8}^{-1}, \ &&{\rm Be-star} \\
& (3.0\times10^{9} \ {\rm cm^{-3}}) \ \dot M_{20.8}r_{13}^{-2}v_{w,8}^{-1}, \ &&{\rm O-star}.
\end{aligned}
\right.
\end{aligned}
\end{equation}
The plasma frequency can be calculated as
\begin{equation}
\begin{aligned}
\omega_p&=\sqrt{\frac{4\pi e^2n_w}{m_e}}\\
&\simeq\left\{
\begin{aligned}
&(9.8\times10^6 \ {\rm rad \ s^{-1}}) \ \dot M_{15.8}^{1/2}r_{13}^{-1}v_{w,8}^{-1/2}, \ &&{\rm Solar \ type \ star}  \\
&(9.8\times10^7 \ {\rm rad \ s^{-1}}) \ \dot M_{17.8}^{1/2}r_{13}^{-1}v_{w,8}^{-1/2}, \ &&{\rm Be-star}  \\
&(3.1\times10^{9} \ {\rm rad \ s^{-1}}) \ \dot M_{20.8}^{1/2}r_{13}^{-1}v_{w,8}^{-1/2}, \ &&{\rm O-star}.
\end{aligned}
\right.
\end{aligned}
\end{equation}
According to Eq.(\ref{eq:Alfven radius}), we can calculate the magnetic field strength at a distance $r=10^{13}$ cm from the massive companion star as (see Eq.(\ref{eq:B-field companion star}))
\begin{equation}
\begin{aligned}
B&\simeq\left\{
\begin{aligned}
&(7.9\times10^{-2} \ {\rm G}) \ B_{c,3}R_{c,10.8}^3R_{A,11.8}^{-2}r_{13}^{-1}, \ &&{\rm Solar \ type \ star}  \\
&(7.8\times10^{-1} \ {\rm G}) \ B_{c,3}R_{c,10.8}^3R_{A,11.3}^{-2}r_{13}^{-1}, \ &&{\rm Be-star}  \\
&(3.4\times10^{-4} \ {\rm G}) \ B_{c,3}R_{c,10.8}^3r_{13}^{-3}, \ &&{\rm O-star}.
\end{aligned}
\right.
\end{aligned}
\end{equation}
The characteristic Faraday conversion frequency of the three types of stars can be estimated as
\begin{equation}
\begin{aligned}
&\omega_{\rm FC,MS}\\
&\simeq\left\{
\begin{aligned}
&(1.9\times10^{8} \ {\rm rad \ s^{-1}}) \ B_{c,3}^{3/4}R_{c,10.8}^{9/4}R_{A,11.8}^{-3/2}\dot M_{15.8}^{1/4}r_{13}^{-5/4}v_{w,8}^{-1/4}\Delta r_{12}^{1/4},  \\
&{\rm Solar \ type \ star}  \\
&(3.4\times10^9 \ {\rm rad \ s^{-1}}) \ B_{c,3}^{3/4}R_{c,10.8}^{9/4}R_{A,11.3}^{-3/2}\dot M_{17.8}^{1/4}r_{13}^{-5/4}v_{w,8}^{-1/4}\Delta r_{12}^{1/4},  \\
&{\rm Be-star}  \\
&(5.8\times10^7 \ {\rm rad \ s^{-1}}) \ B_{c,3}^{3/4}R_{c,10.8}^{9/4}\dot M_{20.8}^{1/4}r_{13}^{-11/4}v_{w,8}^{-1/4} \Delta r_{12}^{1/4},  \\
&{\rm O-star}.
\end{aligned}
\right.
\end{aligned}
\end{equation}
where the typical length scale $\Delta r$ is chosen to be $10^{12}$ cm. One can see for adopted parameters, O-stars and solar type stars have weak Faraday conversion while Be-stars have strong Faraday conversion.

\item Field reversal by a massive black hole. We take the radio loud magnetar PSR J1745-2900 residing $\sim0.12$ pc from Sgr $A^*$ as an example \citep{Eatough2013}.
The wind from a massive black hole is attributed to the outflow from an accretion disk. We consider a total  luminosity of the order of the Eddington luminosity $L_{\rm Edd}$ and assume a radiative efficiency of the black hole accretion disk of the order of $\epsilon_{\rm BH}=0.1$, i.e.
\begin{equation}
\epsilon_{\rm BH} \dot M_{\rm acc} c^2=L_{\rm Edd}=\frac{4\pi cGMm_p}{\sigma_{\rm T}},
\end{equation}
We assume the that the wind mass loss rate is proportional to the accretion rate with a parameter $f=0.1 f_{-1}$, so that the mass loss rate $\dot M$ can be estimated as
\begin{equation}
\dot M =f \dot M_{\rm acc}=\frac{4\pi GM_{\rm BH}m_p f}{\epsilon_{\rm BH}c\sigma_{\rm T}}\simeq(7.0\times10^{21} \ {\rm g \ s^{-1}}) \ M_{\rm BH,38}\epsilon_{\rm BH,-1}^{-1}f_{-1}.
\end{equation}
We consider the magnetic field strength $\sim 10^{-3}$ G at a distance $r\sim0.01 \ {\rm pc}\sim10^{16}$ cm from the black hole and the wind density can be estimated as
\begin{equation}
\begin{aligned}
n_e&=\frac{\dot M}{4\pi \mu_m m_pv_wr^2}\\
&\simeq(3.3\times10^3 \ {\rm cm^{-3}}) \ M_{\rm BH,38}\epsilon_{\rm BH,-1}^{-1}v_{w,9}^{-1}r_{16}^{-2}\mu_m^{-1}f_{-1}.
\end{aligned}
\end{equation}
The plasma frequency can be calculated as
\begin{equation}
\begin{aligned}
\omega_p&=\sqrt{\frac{4\pi e^2n_e}{m_e}}\\
&\simeq(3.3\times10^6 \ {\rm rad \ s^{-1}}) \ M_{\rm BH,38}^{1/2}\epsilon_{\rm BH,-1}^{-1/2}v_{9}^{-1/2}r_{16}^{-1}\mu_m^{-1/2}f_{-1}^{1/2},
\end{aligned}
\end{equation}
and the characteristic Faraday conversion frequency can be estimated as
\begin{equation}
\begin{aligned}
\omega_{\rm FC,BH}&\simeq(2.4\times10^7 \ {\rm rad \ s^{-1}}) \ M_{\rm BH,38}^{1/4}\epsilon_{\rm BH,-1}^{-1/4}v_{9}^{-1/4}r_{16}^{-1/2}B_{-3}^{3/4}\Delta r_{15}^{1/4}\\
&\times\mu_m^{-1/4}f_{-1}^{1/4},
\end{aligned}
\end{equation}
where the typical length scale $\Delta r$ is chosen to be $10^{15}$ cm. One can see that for the adopted parameters one has
$\omega_{\rm frb}\gg \omega_{\rm FC,BH}$, so that the Faraday conversion effect can be negligible.

\end{itemize}

We present the characteristic Faraday conversion frequency for the three scenarios in the upper panel of Fig.\ref{fig:Faraday conversion}. The black and green dashed lines in the three scenarios denote $1$-GHz FRB and the typical value of Faraday conversion angular frequency. The upper left panel is the field reversal scenario for a supernova remnant. One can see $\omega_{\rm FC,SNR}<\omega_{\rm FRB}$ for typical values.
The upper central panel is the field reversal scenario for a Be companion star. One can see that typically $\omega_{\rm FC,MS}\sim \omega_{\rm frb}$, i.e. the oscillation effect between linear and circular polarization modes is strong. The typical Faraday conversion frequencies of a Solar type star or an O-star are much smaller than the typical FRB frequency $\sim 1$-GHz, so we ignore those two cases.
The upper right panel is the field reversal scenario for a massive black hole. One has $\omega_{\rm FRB}\gg\omega_{\rm FC,MS}$, i.e. the Faraday effect is weak and the X-mode and O-mode nearly have the same phase. In the three scenarios, we solve the differential equations (\ref{eq:FC_V})-(\ref{eq:FC_Phi}) and present the circular and linear polarization degrees as a function of $\lambda^2$ in the bottom panels in Fig.\ref{fig:Faraday conversion}, with the same field reversal scenarios presented in the columns. One can see that the field reversal for the Be-star case is the most possible scenario for  Faraday conversion, and black hole scenario has the weakest Faraday conversion effect. For the supernova remnant case, the Faraday conversion effect could exist and be significant for the case of a large electron number density environment. 
It should be pointed out that Faraday conversion cannot change the total polarization degree, i.e. $\Pi_P=\sqrt{\Pi_V^2+\Pi_L^2}$ remains unchanged all the time.

\section{Conclusions and discussions}\label{sec:conclusions}
In this paper, we have investigated a variety of intrinsic radiation mechanisms and propagation effects inside (curvature radiation, inverse Compton scattering and cyclotron resonance absorption) and outside (cyclotron/synchrotron radiation, synchrotron absorption, cyclotron  absorption and Faraday conversion) the  magnetosphere of an FRB source (presumably a magnetar) and study their effects on the observed polarization properties of FRBs. We pay special attention to the mechanisms that might give rise to circular polarization. 
We draw the following conclusions:
\begin{itemize}
\item In general, the intrinsic radiation mechanisms tend to generate nearly 100\% linear polarization. Circular polarization can be generated from an off-axis configuration, especially for magnetospheric mechanisms. Propagation effects both within and outside the magnetosphere can contribute to circular polarization under certain physical conditions, even though some conditions are quite stringent.  
\item The first magnetospheric intrinsic radiation mechanism we consider is coherent curvature radiation of charged bunches. We find that it is a possible mechanism to produce both linear and circular polarization from FRBs. An on-axis observer will detect 100\% linear polarized waves for the on-axis case since there is only one mode existing in the plane perpendicular to LOS.
For a point-source bunch, an off-axis observer starts to see circular polarization even if the line of sight is within the relativistic beam
i.e. $\theta<1/\gamma$.
The high circular polarization stays at $\theta>1/\gamma$ where the emission flux is much degraded.
Observationally, since only a small fraction of bursts are detected with circular polarization and most bursts have high linear polarization degree, the geometry of charged bunches cannot be point-source like. Rather, the bunch cross section needs to be much larger acrossing a bundle of open magnetic field lines. High linear polarization is sustained as long as the LOS is within the opening angle of the bundle. Only when the LOS falls outside the bundle when significant circular polarization is observed. This conclusion is consistent with \cite{Wang2022b}.
\item Another magnetospheric intrinsic  radiation mechanism is coherent inverse Compton scattering by charged bunches.
The orthogonal modes of coherent ICS radiation by relativistic single particle and charged bunch are investigated. For ICS radiation of a
single charge particle (point-like bunch), the linear polarization degree is always 100\% since the electric field of the scattered wave is determined by that of the incident low frequency waves.
Circular polarization can be produced when a geometric bunch is considered because of the asymmetry introduced into the system. More specifically, following conditions are satisfied for an asymmetric system:  1. If the LOS is not aligned with the symmetric axis of bunch, both the incident waves and outgoing waves have phase differences within the bunch; 2. For a bunch involving curved field lines, different particles move along slightly different field lines so that particles at different locations in the bunch have different polarization angles in emission. 
In general, circular polarization can be generated in an off-axis geometry even within the $1/\gamma$ cone (similar to the curvature radiation case). In order to produce the right ratio of linearly polarized bursts and highly circularly polarized bursts, the ICS bunch also needs to have a large cross section invoking a wide bundle of magnetic field lines. 
\item One important propagation effect within the magnetosphere is cyclotron resonance absorption. It can basically generate circular polarization by selectively absorbing the one circular polarization mode (R-mode or L-mode) in the incident FRB waves. One condition is that the wave vector is quasi-parallel to the local magnetic field line, which is the case for emission invoking open field line regions. Another condition is that electrons and positrons should have different Lorentz factors ($\gamma_-\neq\gamma_+$), so that the cyclotron resonance absorption optical depths of the two modes are different.
Cyclotron resonance absorption effect could be responsible for generating high circular polarization of FRBs for an extremely asymmetric distribution of the lepton Lorentz factors. The effect could be negligible when $\gamma_-\simeq\gamma_+$, and the escape waves should be highly linear polarized.
\item Outside the magnetosphere, the intrinsic radiation mechanism we consider is the synchrotron maser mechanism invoking relativistic shocks with ordered magnetic fields. Because the mechanism requires that the field lines are highly ordered and lie in the shock plane, this mechanism can only produce nearly 100\% linear polarization when viewed on beam, i.e. $\theta_v<\theta_j$.
The circular polarization can be produced in an off-beam geometry, i.e. $\theta_v>\theta_j$. However, because the high Lorentz factors of the shock and the emitting particles, the isotropic luminosity (${\cal D}^2F(x)$ as a proxy) decreases rapidly when $\theta_v>\theta_j$. Highly circularly polarized bright FRBs are therefore impossible to be generated with this mechanism. We conclude that highly circularly polarized FRBs cannot be generated via the synchotron maser radiation mechanism.
\item Outside the magnetosphere, the polarization state of the FRB waves can be modified through the following three propagation effects: selected synchrotron absorption, selected cyclotron absorption, and Faraday conversion. The application of each of these mechanisms requires some stringent conditions. 
\item Synchtrotron absorption: If the FRB source is surrounded by a synchrotron-emitting nebula (e.g. a magnetar wind nebula), FRB waves may undergo absorption by the electrons in the nebula. 
The incident FRB modes can be decomposed into two orthogonal modes and we find the parallel and perpendicular modes have different absorption coefficients ($\alpha_\parallel$ and $\alpha_\perp$). If the incident waves are 100\% linearly polarized, i.e. $V=0$, we find that circular polarization cannot be generated because $d\Pi_V/ds=0$. More generally, for elliptically polarized incident waves, synchrotron absorption effect tends to reduce circular polarization more so that the circular polarization degree keeps falling. In our calculation, we adopted the parameters that can account for the persistent radio source of FRB 121102 but found that the synchrotron absorption effect is not important. 
Since most other FRBs do not have PRS as bright as that of FRB 121102, we draw the conclusion that synchrotron absorption effect is in general not important for FRBs unless the FRB source is surrounded by a denser and brighter synchrotron nebula (e.g. for a magnetar of an even younger age). 
\item Cyclotron absorption: If the FRB waves go through a cold, dense, and highly magnetized medium along the line of sight, one circular polarization mode may be absorbed through electron cyclotron absorption at a cyclotron resonance. The required magnetic field strength $B \simeq 360$ G, which is usually unachievable. One plausible scenario may be that the FRB source has a very nearby magnetized companion.
We discuss three possible types of companion stars with different mass loss rates, i.e. a solar type star, a Be-star and an O-star. We find that the O-star has the highest mass loss rate, which could in principle completely absorb one mode and leave the escaped waves to have a circular polarization degree of $\sim100\%$. The Be-star can allow the escaped waves to reach a moderate degree of circular polarization, but the effect of a solar type star is negligible. 
\item Faraday conversion arises from different propagation speeds of the X-mode and O-mode waves in the quasi-perpendicular configuration, and therefore requires a magnetic field reversal in an astronomical environment. 
We discuss three possible astronomical scenarios that might give rise to field reversal: (i) in a supernova remnant; (2) in the wind of a companion star; and (3) in the wind of an accreting black hole. Comparing the FRB frequency with the Faraday conversion circular frequency $\omega_{\rm FC}$ under all three scenarios, we find that the most plausible scenario is to have a companion star, most favorably of Be-type. The supernova remnant case can also induce moderate Faraday conversion for typical parameters, but the black hole companion case is the least plausible case for Faraday conversion to occur.

\end{itemize}

It is worth noting that some FRBs show total polarization degree much lower than 100\%, i.e.  $Q^2+U^2+V^2<I^2$. The intrinsic radiation mechanisms discussed in this paper generally produce a $\sim 100\%$ total polarization degree. Some propagation effects, e.g. Faraday conversion or mode-dependent absorption, conserve the total polarization degree. Thus, it is highly interesting to consider whether any mechanism might depolarise the FRB emission. In general, this requires incoherent superposition of multiple coherent radiation units. 
There are two possibilities. First, for intrinsic radiation mechanisms, one may have the situation that within the instantaneously observed radiation beam there are multiple incoherent radiation units. This is possible for the magnetospheric coherent radiation mechanisms by bunches (both for curvature radiation and ICS). For the synchrotron maser model, this may be realized by invoking field lines that are not strictly parallel to each other or leptons with a wide energy distribution so that they do not form a perfect ring in the momentum space. However, these deviations from the perfect synchrotron maser conditions also pay the price of lowering the coherent degree so that such models predict a negative correlation between the brightness of the bursts and their polarization degrees. There is no observational evidence for such a behavior so far. Second, even if throughout the paper we only discuss the propagation effects involving one ray of emission reaching the observer, in reality it is possible that the observed emission originates from multiple rays due to scattering by a plasma screen along the line of sight. For an inhomogeneously magnetized electron-ion plasma environment, there will be a scatter in the rotation measure from different rays, causing depolarization at lower frequencies. Such a feature has been claimed to interpret the frequency-dependent polarization observed in some FRBs \citep{Feng2022b,McKinven2023}. For detailed studies of multi-path effects, see \cite{Beniamini2021} and \cite{Yang2022}.

\section*{Acknowledgements}
We thank Kejia Lee, Dongzi Li, Wenbin Lu, Myles Sherman, Chen Wang, Yihan Wang and Yuanpei Yang for helpful discussion and an anonymous referee for helpful comments. This work is supported by the Nevada Center for Astrophysics and a Top Tier Doctoral Graduate Research Assistantship (TTDGRA) at the University of Nevada, Las Vegas.

\section*{Data Availability}
There is no data generated from this theoretical work. The code developed to perform the calculations in this paper is available upon request.







\appendix

\section{Dispersion relation of electromagnetic wave modes in a cold plasma}\label{B}
In this section, we present a brief derivation of the dispersion relation of electromagnetic wave modes in a cold plasma. Within the magnetosphere and in the wind of a magnetar, the plasma mainly consists of electron-positron pairs. Far from the magnetosphere, on the other hand, the plasma is composed of electrons and ions. Therefore, we discuss both cases in the following discussion. In general, the wave equation can be written as
\begin{equation}
\vec n\times(\vec n\times\vec E)+\stackrel{\leftrightarrow}{\epsilon}\cdot\vec E=0,
\end{equation}
where $\vec n=\vec kc/\omega$ is the refractive index and $\stackrel{\leftrightarrow}{\epsilon}$ is the dielectric tensor. Generally, we denote $\theta$ as the angle between the wave vector and magnetic field ($\vec B=B_0\hat{z}$), thus the wave vector is $\vec k=(|\vec k|\sin\theta,0,|\vec k|\cos\theta)$. Making use of the refractive index and dielectric tensor of plasma, the Maxwell response tensor is defined as
\begin{equation}
\stackrel{\leftrightarrow} M=\stackrel{\leftrightarrow}\epsilon-n^2(\stackrel{\leftrightarrow}{I}-{\stackrel{\leftrightarrow}{k}}\stackrel{\leftrightarrow}{{k}}).
\end{equation}
The dielectric tensor can be calculated by the Ohm's law in the Fourier space, i.e. $\vec\sigma(\vec{k},\omega)=\vec{j}(\vec{k},\omega)/\vec{E}(\vec{k},\omega)$, as
\begin{equation}
\stackrel{\leftrightarrow}{\epsilon}=\left( 
  \begin{array}{ccc}  
    S & -iD & 0\\
    iD & S & 0\\
    0 & 0 & P\\
  \end{array}
\right),
\end{equation}
where
\begin{equation}\label{eqS}
S=\frac{1}{2}(R+L)=1-\frac{\omega_p^2(\omega^2+\Omega_i\Omega_e)}{(\omega^2-\Omega_i^2)(\omega^2-\Omega_e^2)},
\end{equation}
\begin{equation}\label{eqD}
D=\frac{1}{2}(R-L)=\frac{\omega_p^2\omega(\Omega_e+\Omega_i)}{(\omega^2-\Omega_i^2)(\omega^2-\Omega_e^2)},
\end{equation}
\begin{equation}
R=S+D=1-\frac{\omega_p^2}{(\omega+\Omega_i)(\omega+\Omega_e)},
\end{equation}
\begin{equation}
L=S-D=1-\frac{\omega_p^2}{(\omega-\Omega_i)(\omega-\Omega_e)},
\end{equation}
\begin{equation}\label{eqP}
P=1-\frac{\omega_p^2}{\omega^2}.
\end{equation}
Here $\Omega_e=-eB/(m_ec)=-\omega_B$ is the electron gyration frequency, $\Omega_i$ is the ion gyration frequency, and $\Omega_i=\omega_B$ for a pair plasma. The parameters  $R, L, P$ denote the the "right", "left", and "plasma" modes, respectively, and $S$ and $D$ denote "sum" and "difference", respectively. The wave equation can be written in terms of dielectric tensor components as
\begin{equation}\label{Eq:generalfunction}
\left( 
  \begin{array}{ccc}  
    S-n^2\cos\theta^2 & -iD & n^2\cos\theta\sin\theta\\
    iD & S-n^2 & 0\\
    n^2\cos\theta\sin\theta & 0 & P-n^2\sin\theta^2\\
  \end{array}
\right)\left( 
  \begin{array}{ccc}  
    E_x\\
    E_y\\
    E_z\\
  \end{array}
\right)=0,
\end{equation}
where the left matrix is the Maxwell response tensor and we obtain the dispersion relation by letting the determinant of the matrix be zero
\begin{equation}
{\rm det}\left[\stackrel{\leftrightarrow}{\epsilon}-n^2\left(\stackrel{\leftrightarrow}{I}-{\stackrel{\leftrightarrow}{k}}{\stackrel{\leftrightarrow}{k}}\right)\right]=0.
\end{equation}
The general dispersion relation of electromagnetic waves in a cold magnetized plasma can be written as
\begin{equation}\label{general dispersion equation}
An^4-Bn^2+C=0,
\end{equation}
where
\begin{equation}
A=S\sin\theta^2+P\cos{\theta}^2,
\end{equation}
\begin{equation}
B=RL\sin{\theta}^2+PS(1+\cos\theta^2),
\end{equation}
\begin{equation}
C=P(S^2-D^2)=PRL.
\end{equation}
The general solution of Eq.(\ref{general dispersion equation}) can be calculated as
\begin{equation}
n^2=\frac{B\pm\sqrt{B^2-4AC}}{2A}=1-\frac{2(A-B+C)}{2A-B\pm\sqrt{B^2-4AC}}.
\end{equation}
We discuss the dispersion relations in electron-ion plasma and electron-positron (pair) plasma below separately. For convenience, we first discuss $\vec k\parallel\vec B$ and $\vec k\perp\vec B$ and then put it forward to a general case.
\begin{itemize}
\item When $\vec k\parallel\vec B$, i.e. $\theta=0$, then the dielectric tensor Eq.(\ref{Eq:generalfunction}) can be written as
\begin{equation}
\left( 
  \begin{array}{ccc}  
    S-n^2 & -iD & 0\\
    iD & S-n^2 & 0\\
    0 & 0 & P\\
  \end{array}
\right)\left( 
  \begin{array}{ccc}  
    E_x\\
    E_y\\
    E_z\\
  \end{array}
\right)=0,
\end{equation}
and the coefficients can be simplified as
\begin{equation}
A=P, \ B=2PS, \ C=PRL,
\end{equation}
one can see $E_z$ is independent of $E_x$ and $E_y$ and $P$ is required to be zero, i.e. $\omega^2=\omega_p^2$ denotes the plasma mode. In such a case, both $E_x$ and $E_y$ are perpendicular to the LOS and we can obtain the dispersion relations for R-mode and L-mode as
\begin{equation}
n_R^2=R=\left\{
    \begin{aligned}
    &1-\frac{\omega_p^2}{\omega^2-\omega_B^2}, &{\rm pair},\\
    &1-\frac{\omega_p^2}{\omega(\omega-\omega_B)}, &{\rm ion},
    \end{aligned}
\right.
\end{equation}
and
\begin{equation}
n_L^2=L=\left\{
    \begin{aligned}
    &1-\frac{\omega_p^2}{\omega^2-\omega_B^2}, &{\rm pair},\\
    &1-\frac{\omega_p^2}{\omega(\omega+\omega_B)}, &{\rm ion},
    \end{aligned}
\right.
\end{equation}
respectively. 
It should be pointed out that the dispersion relations can be modified by replacing $\omega_B$ to $\omega_B\cos\theta$ for quasi-parallel propagation, i.e. $\theta\ll1$.

\item When $\vec k\perp\vec B$, i.e. $\theta=\pi/2$, then the dielectric tensor (Eq. \ref{Eq:generalfunction}) can be written as
\begin{equation}
\left( 
  \begin{array}{ccc}  
    S & -iD & 0\\
    iD & S-n^2 & 0\\
    0 & 0 & P-n^2\\
  \end{array}
\right)\left( 
  \begin{array}{ccc}  
    E_x\\
    E_y\\
    E_z\\
  \end{array}
\right)=0,
\end{equation}
and the coefficients can be simplified as
\begin{equation}
A=S, \ B=RL+PS, \ C=PRL,
\end{equation}
one can see $E_z$ can be directly solved as $n^2=P$ and the electric field is parallel to the LOS, i.e. $\vec E_z$ is perpendicular to the $(\Vec{k},\Vec{B})$ plane. Thus the dispersion relation for O-mode is
\begin{equation}
n_O^2=P=1-\frac{\omega_p^2}{\omega^2}.
\end{equation}
Both $E_x$ and $E_y$ are perpendicular to the $(\Vec{k},\Vec{B})$ plane, i.e. X-mode. The dispersion relation can be calculated as
\begin{equation}
\begin{aligned}
n_X^2=\frac{RL}{S}&=\frac{\left[1-\frac{\omega_p^2}{(\omega+\Omega_i)(\omega+\Omega_e)}\right]\left[1-\frac{\omega_p^2}{(\omega-\Omega_i)(\omega-\Omega_e)}\right]}{\left[1-\frac{\omega_p^2(\omega^2+\Omega_i\Omega_e)}{(\omega^2-\Omega_i^2)(\omega^2-\Omega_e^2)}\right]}\\
&\simeq\left\{
    \begin{aligned}
    &1-\frac{\omega_p^2}{\omega^2-\omega_B^2}, &{\rm pair},\\
    &\frac{(\omega^2-\omega_p^2)^2-\omega^2\omega_B^2}{\omega^2(\omega^2-\omega_p^2-\omega_B^2)}, &{\rm ion},
    \end{aligned}
\right.
\end{aligned}
\end{equation}
It should be pointed out that the dispersion relation of X-mode can be modified by replacing $\omega_B$ by $\omega_B\sin\theta$ for quasi-perpendicular case, i.e. $\theta\rightarrow \pi/2$.

\end{itemize}

\section{Scattered electric field in a strong magnetic field}\label{A}

\begin{figure}
	\includegraphics[width=\columnwidth]{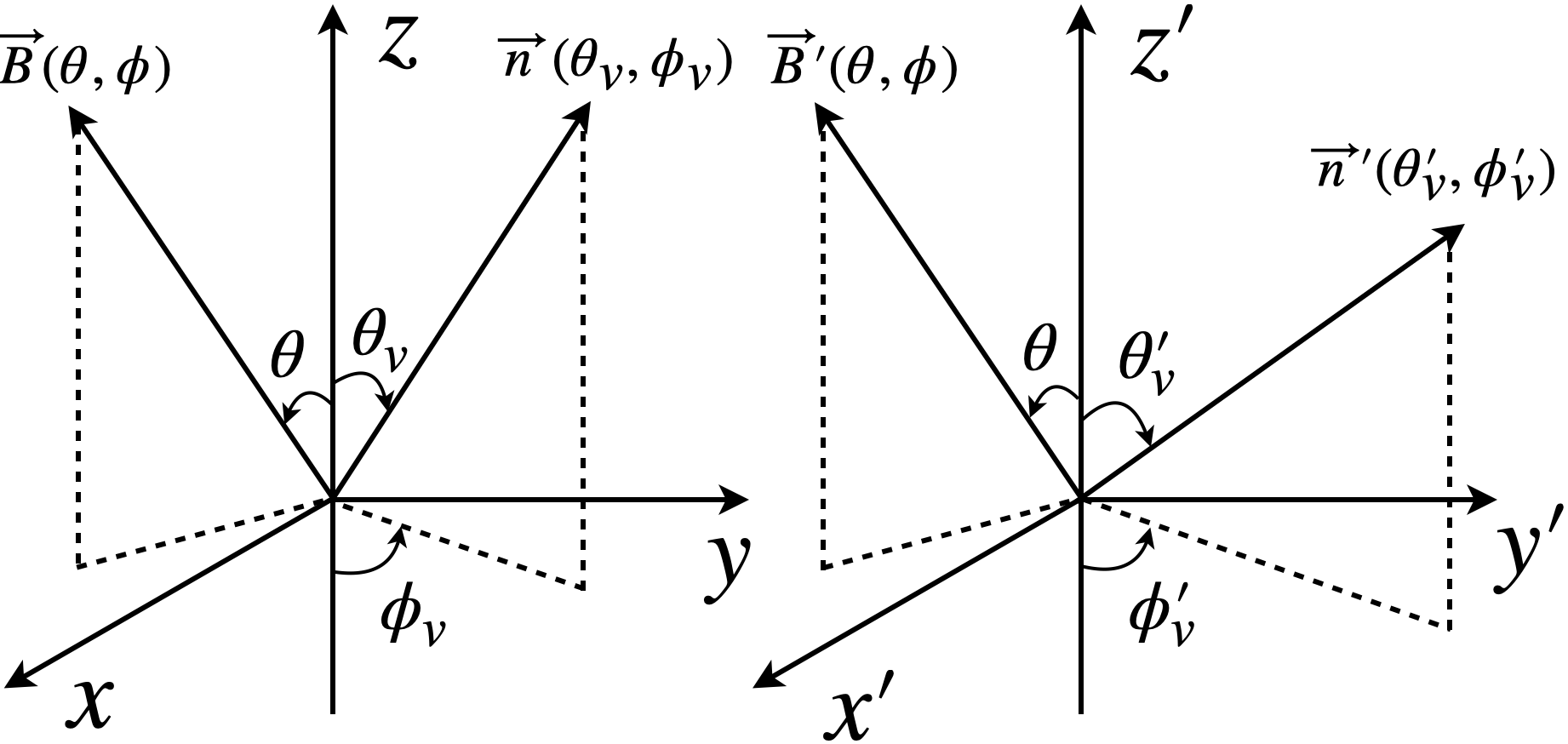}
    \caption{Geometry for inverse Compton scattering. In the lab frame (left-hand panel), we consider the LOS is along the $\vec{n}$ and electron moves along an arbitrary magnetic field $\vec{B}$. Some calculations in this appendix are done in the co-moving frame of electrons (right-hand panel), where all quantities are denoted with a prime ($'$). All parameters are written in $xyz$-frame.}
    \label{fig:secondrelation}
\end{figure}

In this appendix, we consider an incident electromagnetic wave scattered by a relativistic charged particle in a strong magnetic field and derive the electric field of the scattered wave in the lab frame. The radiation electric field produced by a moving 
electron 
is given by \citep{Rybicki&Lightman1979,Jackson1998}
\begin{equation}
\begin{aligned}
\vec E_{\rm rad}(\vec r,t)&=\frac{e}{c}\left[\frac{\hat{n}\times\{(\hat{n}-\vec \beta)\times\Dot{\Vec{\beta}}\}}{(1-\hat{n}\cdot\Vec{\beta})R}\right]_{\rm rec}\simeq\frac{e}{c}\left[\frac{\vec{n}\times(\vec{n}\times\Dot{\vec{{\beta}}})}{R}\right]_{\rm rec},
\end{aligned}
\end{equation}
where $\vec{n}$ is the unit vector in the observer direction, 
$R$ is the distance from a scattering point to the observer. We treat the problem in the comoving frame of the relativistically moving electron, in which the electron is initially at rest but later under acceleration in the (redshifted) electromagnetic fields of the incident low-frequency waves. For our problem, the low-frequency waves do not necessarily have large amplitudes \citep{Zhang22}, so the electron motion can be treated as non-relativistic, so that $\vec \beta\rightarrow0$. 
Here the subscript “ret” means that all quantities are evaluated at a retarded time. In the first step, all calculations are carried out in the rest frame of electron and the parameters are denoted by prime ($'$). We assume an initially rest electron at the origin and the background arbitrary strong magnetic field is $\vec{B}'=\vec B=(B_0\sin\theta\cos\phi,B_0\sin\theta\sin\phi,B_0\cos\theta)$.
The electrons can only move along the strong magnetic field line. The electron motion equation in the co-moving frame can be written as
\begin{equation}
m\frac{d^2\vec{r}'}{dt'^2}=e\Vec{E}'+\frac{e}{c}\left(\frac{d\vec{r}'}{dt'}\times\Vec{B}'\right),
\end{equation}
We assume that the incident electric field behave sinusoidally in the lab frame as
\begin{equation}
\vec E=E_0{\rm exp}(-i\omega_{i}t)\hat{e}_i,
\end{equation}
where $E_0$ and $\omega_{i}$ are the amplitude and the angular frequency of the incident wave, respectively. $\hat{e}_i=(\cos\theta_{i},0,\sin\theta_i)$ is the unit polarization vector of the incident wave in the $x-z$ plane, i.e. the three components of the incident electric field can be written as $E_{i,x}=E_0\cos\theta_i$, $E_{i,y}=0$ and $E_{i,z}=E_0\sin\theta_i$. We project the three components onto the parallel and perpendicular directions of the electron motion as
\begin{equation}
\vec E_{i,\parallel,B}=|E_{i,\parallel,B}|\sin\theta\cos\phi\hat{x}+|E_{i,\parallel,B}|\sin\theta\sin\phi\hat{y}+|E_{i,\parallel,B}|\cos\theta\hat{z},
\end{equation}
where the corresponding projection magnitude is 
\begin{equation}
|E_{i,\parallel,B}|=E_0\cos\phi\cos\theta_{i}\sin\theta+E_0\cos\theta\sin\theta_i,
\end{equation}
and
\begin{equation}
\begin{aligned}
\vec E_{i,\perp,B}&=\vec E_i-\vec E_{i,\parallel,B}\\
&=(E_0\cos\theta_i-|E_{i,\parallel,B}|\sin\theta\cos\phi)\hat{x}-|E_{i,\parallel,B}|\sin\theta\sin\phi\hat{y}\\
&+(E_0\sin\theta_i-|E_{i,\parallel,B}|\cos\theta)\hat{z}.
\end{aligned}
\end{equation}
In order to find the incident electric field in the electron's co-moving frame, we perform the relativistic transformation of the field and obtain
\begin{equation}
\begin{aligned}
\vec E_{i,\parallel,B}'
&=|E_{i,\parallel,B}|\sin\theta\cos\phi\hat{x}+|E_{i,\parallel,B}|\sin\theta\sin\phi\hat{y}+|E_{i,\parallel,B}|\cos\theta\hat{z},
\end{aligned}
\end{equation}
and
\begin{equation}
\begin{aligned}
\vec E_{i,\perp,B}'&=\gamma\left(\vec E_{i,\perp,B}+\frac{\vec{v}}{c}\times\vec{B}_w\right)\\
&=\gamma \vec E_{i,\perp,B}-(E_0\cos\theta\cos^2\theta_i+E_0\cos\theta\sin^2\theta_i)\hat{x}\\
&+(E_0\cos\phi\cos^2\theta_i\sin\theta+E_0\cos\phi\sin\theta\sin^2\theta_i)\hat{z},
\end{aligned}
\end{equation}
where the second term $\vec v\times\Vec{B}_w/c\simeq\hat{B}\times(\hat k_i\times\vec E_i)$, $\hat{B}=(\sin\theta\cos\phi,\sin\theta\sin\phi,\cos\theta)$, the incident electric filed $\vec E_i=(E_0\cos\theta_i,0,E_0\sin\theta_i)$, and the unit vector of the incident wave vector is $\hat k_i=(-\sin\theta_i,0,\cos\theta_i)$.

Thus we can write the three incident electric field components in the $xyz$-frame as
\begin{equation}
\begin{aligned}
E_{i,x}'&=|E_{i,\parallel,B}|\sin\theta\cos\phi+\gamma(E_0\cos\theta_i-|E_{i,\parallel,B}|\sin\theta\cos\phi)\\
&-E_0\cos\theta\cos^2\theta_i-E_0\cos\theta\sin^2\theta_i.
\end{aligned}
\end{equation}
\begin{equation}
E_{i,y}'=|E_{i,\parallel,B}|\sin\theta\sin\phi-\gamma|E_{i,\parallel,B}|\sin\theta\sin\phi
\end{equation}
\begin{equation}
\begin{aligned}
E_{i,z}'&=|E_{i,\parallel,B}|\cos\theta+\gamma(E_0\sin\theta_i-|E_{i,\parallel,B}|\cos\theta)\\
&+E_0\cos\phi\cos^2\theta_i\sin\theta+E_0\cos\phi\sin\theta\sin^2\theta_i.
\end{aligned}
\end{equation}
The position vector in the co-moving frame can be written as
\begin{equation}
\vec r'=\vec r_0'{\rm exp}(-i\omega_{i}'t').
\end{equation}
The time derivative $\partial/\partial t'$ can be written as $-i\omega_{i}'$ and the motion equation can be written as
\begin{equation}
m_e\omega_{i}'^2\vec r_0'+e\vec{E}_i'-i\frac{e\omega_{i}'}{c}\vec r_0'\times\vec B'=0.
\end{equation}
We assume $\vec{r}_0'=(r_x',r_y',r_z')$ and the electron motion equation can be solved by considering $B_0\gg1$ ($\omega_B'/\omega_i'\gg1$ is applied) and the Lorentz force can be ignored, i.e. electrons can only move along the local magnetic field line in the co-moving frame, so that
\begin{equation}
\begin{aligned}\label{Eq:r_x'}
r_x'
&=-\frac{e}{m_e\omega_i'^2}(E_{i,x}'\cos^2 \phi\sin^2\theta+E_{i,y}\sin\phi\cos\phi\sin^2\theta\\
&+E_{i,z}'\cos\phi\sin\theta\cos\theta).
\end{aligned}
\end{equation}
\begin{equation}
\begin{aligned}\label{Eq:r_y'}
{r}_{y}'
&=-\frac{e}{m_e\omega_i'^2}(E_{i,x}'\sin\phi\cos\phi\sin^2\theta+E_{i,y}'\sin^2\phi\sin^2\theta\\
&+E_{i,z}'\sin\phi\sin\theta\cos\theta).
\end{aligned}
\end{equation}
\begin{equation}
\begin{aligned}\label{Eq:r_z'}
{r}_{z}'
&=-\frac{e}{m_e\omega_i'^2}(E_{i,x}'\cos\phi\sin\theta\cos\theta+E_{i,y}'\sin\phi\sin\theta\cos\theta\\
&+E_{i,z}'\cos^2\theta).
\end{aligned}
\end{equation}
With $\omega_i'=\gamma(1-\beta\cos\theta_{iB})\omega_{i}$ and $\cos\theta_{iB}=\hat{B}\cdot\hat{z}_i$,one can see $r_x'$, $r_y'$ and $r_z'$ are independent with time.

The unit vector of the LOS in the comoving frame can be written as $\vec{n}'=(\sin\theta_v'\cos\phi_v',\sin\theta_v'\sin\phi_v',\cos\theta_v')$,
thus the the electric field of the scattered wave can be calculated as
\begin{equation}
\begin{aligned}
\vec E_s^{'}&=\frac{e}{c}\left[\frac{\vec{n}'\times(\vec{n}'\times\Dot{\vec{{\beta}'}})}{R'}\right]_{\rm rec}\\
&=-\frac{e\omega_i'^2}{c^2D'}\left[{\vec{n}'\times(\vec{n}'\times{\vec{{r_0}}'})}\right]_{\rm rec}{\rm exp}(-i\omega_i' t'),
\end{aligned}
\end{equation}
where $r_c=e^2/(m_ec^2)$ is the classical electron radius, $D'$ is the distance from a scattering point to an observer. The three components of the  electric field of the scattered wave can be written as
\begin{equation}
\begin{aligned}
E_{s,x}'=&-\frac{e\omega_i'^2}{c^2D'}(-{r_x'}\sin^2{\phi_v'} \sin ^2{\theta_v'}+{r_y'}\sin{\phi_v'}\cos{\phi_v'}\sin^2{\theta_v'}\\
&+{r_z'}\cos{\phi_v'}\sin{\theta_v'} \cos {\theta_v'}-{r_x'}\cos^2{\theta_v'}).
\end{aligned}
\end{equation}
\begin{equation}
\begin{aligned}
E_{s,y}'=&-\frac{e\omega_i'^2}{c^2D'}({r_x'} \sin{\phi_v'} \cos {\phi_v'} \sin ^2{\theta_v'}-{r_y'} \cos^2{\phi_v'}\sin^2{\theta_v'}\\
&+{r_z'}\sin {\phi_v'}\sin{\theta_v'}\cos{\theta_v'}-{r_y'}\cos^2{\theta_v'}).
\end{aligned}
\end{equation}
\begin{equation}
\begin{aligned}
E_{s,z}'=&-\frac{e\omega_i'^2}{c^2D'}({r_x'} \cos{\phi_v'} \sin {\theta_v'} \cos{\theta_v'}+{r_y'} \sin{\phi_v'} \sin{\theta_v'}\cos{\theta_v'}\\
&-{r_z'} \sin^2{\phi_v'} \sin ^2{\theta_v'}-{r_z'} \cos^2{\phi_v'} \sin^2{\theta_v'}).
\end{aligned}
\end{equation}

Now we consider the configuration of a charged bunch in the right panel of Fig.\ref{fig:ICSfig}.
The amplitudes of the electric fields of the scattered waves by electrons in different magnetic field lines in the co-moving frame are different. 
Another different parameter is the wave phase. Even along the same magnetic field line, the electrons may stay at different altitudes, i.e. different longitudinal coordinates in the bunch, which will influence the phase term. Thus the wave electric field in the rest frame of an electron that moves along an arbitrary magnetic field $(\theta',\phi')$ can be expressed in the $xyz$ coordinates as
\begin{equation}
\Vec{E}_s'=(E_{s,x}'\hat{x}+E_{s,y}'\hat{y}+E_{s,z}'\hat{z}){\rm exp}[i({k}_{s}{R_s}+{k}_{i}{R_i}-\omega_{s}t)],
\end{equation}
where the two factors in the phase term can be calculated as
\begin{equation}
\begin{aligned}
R_s=|A_1x_0+B_1y_0+C_1z_0+D|,
\end{aligned}
\end{equation}
and
\begin{equation}
R_i=|A_2x_0+B_2y_0+C_2z_0|,
\end{equation}
where $A_1=\sin\theta_v\cos\phi_v$, $B_1=\sin\theta_v\sin\phi_v$, $C_1=\cos\theta_v$, 
$D=-A_1H\tan\theta_c\cos\phi_v-B_1H\tan\theta_c\sin\phi_v-C_1H$, $H=d\cos\theta_c$, $\theta_c=1/\gamma$,
$A_2=-\sin\theta_i$, $B_2=0$, $C_2=\cos\theta_i$, $x_0=r\sin\theta\cos\phi$, $y_0=r\sin\theta\sin\phi$ and $z_0=r\cos\theta$. $k_{s/i}=2\pi\nu_{s/i}/c$ is the wave vector value, where $\nu_s=10^9$ Hz and $\nu_i=10^4$ Hz for scattered and incident wave frequency, respectively.
It should be pointed out that the wave phase term is a Lorentz invariant.

One can see that the three electric field components of the scattered wave, $E_{s,x}'$, $E_{s,y}'$ and $E_{s,z}'$ are written in the $xyz$ coordinate frame and be calculated in an arbitrary magnetic field line at $(\theta,\phi)$. 
In the first step, we project the three components into a new coordinate system defined by an arbitrary magnetic field frame in the bunch with $z'\parallel\vec{B}'$ (the $z$ axis is defined along the symmetric axis of the bunch). The rotation matrix can be operated as
\begin{equation}
\begin{aligned}
\left( 
  \begin{array}{ccc}  
    E_{s,x,B'}'\\
    E_{s,y,B'}'\\
    E_{s,z,B'}'\\
  \end{array}
\right)=&\left( 
  \begin{array}{ccc}  
    1 & 0 & 0\\
    0 & \cos\theta & -\sin\theta\\
    0 & \sin\theta & \cos\theta\\
  \end{array}
\right)\\
&\cdot\left(
  \begin{array}{ccc}  
    \cos\phi & \sin\phi & 0\\
    -\sin\phi & \cos\phi & 0\\
    0 & 0 & 1\\
  \end{array}
\right)\left(
  \begin{array}{ccc}  
    E_{s,x}'\\
    E_{s,y}'\\
    E_{s,z}'\\
  \end{array}
\right).
\end{aligned}
\end{equation}
The three components in the above matrix calculation can be expanded as
\begin{equation}
E_{s,x,B'}'=E_{s,x}'\cos\phi+E_{s,y}'\sin\phi.
\end{equation}
\begin{equation}
E_{s,y,B'}'=-E_{s,x}'\sin\phi\cos\theta+E_{s,y}'\cos\phi\cos\theta-E_{s,z}'\sin\theta.
\end{equation}
\begin{equation}
E_{s,z,B'}'=-E_{s,x}'\sin\phi\sin\theta+E_{s,y}'\cos\phi\sin\theta+E_{s,z}'\cos\theta.
\end{equation}
In the second step, for a relativistic electron with $\gamma\gg1$, we can transform the electric field in the $z'$-axis frame, i.e. in the comoving frame, to the lab frame through the transformation law of electromagnetic field. The two components can be calculated in the lab frame as
\begin{equation}
\Vec{E}_{\parallel,B}={E}_{\parallel,B,k}\hat{z}_B=E_{s,z,B'}'{\rm exp}[i({k}_{s}{R_s}+{k}_{i}{R_i})]\hat{z}_{B},
\end{equation}
and
\begin{equation}
\begin{aligned}
&\Vec{E}_{\perp,B}={E}_{\perp,B,i}\hat{x}_B+{E}_{\perp,B,j}\hat{y}_B={\gamma}\left({\Vec{E}_{\perp}'}-\frac{\Vec{v}}{c}\times\Vec{B}_w'\right)\\
&=\left({\gamma}{E_{s,x,B'}'}\hat{x}_{B}+{\gamma}{E_{s,y,B'}'}\hat{y}_{B}\right){\rm exp}[i({k}_{s}{R_s}+{k}_{i}{R_i})]\\
\end{aligned}
\end{equation}
where the three unit vectors $\hat{x}_B$, $\hat{y}_B$ and $\hat{
k}_B$ are defined in the arbitrary magnetic field frame (the frame defined by $z'$), one can see $v'\times\vec B_w'/c\simeq0$ since $v'/c\sim0$ in the electrons comoving frame. 
The three electric field components in the arbitrary magnetic field frame can be re-written as
\begin{equation}
\begin{aligned}
&{E}_{\perp,B,i}=(\gamma E_{s,x}'\cos\phi+\gamma E_{s,y}'\sin\phi){\rm exp}[i({k}_{s}{R_s}+{k}_{i}{R_i})].
\end{aligned}
\end{equation}
\begin{equation}
\begin{aligned}
{E}_{\perp,B,j}&=\gamma(-E_{s,x}'\sin\phi\cos\theta+E_{s,y}'\cos\phi\cos\theta\\
&-E_{s,z}'\sin\theta){\rm exp}[i({k}_{s}{R_s}+{k}_{i}{R_i})].
\end{aligned}
\end{equation}
\begin{equation}
{E}_{\parallel,B,k}=E_{s,z,B'}'{\rm exp}[i({k}_{s}{R_s}+{k}_{i}{R_i})].
\end{equation}
We drop the factor related to time in the phase term which cannot affect the polarization state of the scattered wave. We project $\Vec{E}_{\parallel,B}$ and $\Vec{E}_{\perp,B}$ back onto the $xyz$-frame and the matrix calculation can be re-written as
\begin{equation}
\begin{aligned}
\left( 
  \begin{array}{ccc}  
    E_{\perp,i}\\
    E_{\perp,j}\\
    E_{\parallel,k}\\
  \end{array}
\right)=&\left( 
  \begin{array}{ccc}  
    1 & 0 & 0\\
    0 & \cos\theta & \sin\theta\\
    0 & -\sin\theta & \cos\theta\\
  \end{array}
\right)\\
&\cdot\left( 
  \begin{array}{ccc}  
    \cos\phi & -\sin\phi & 0\\
    \sin\phi & \cos\phi & 0\\
    0 & 0 & 1\\
  \end{array}
\right)\left(
  \begin{array}{ccc}  
    E_{\perp,B,i}\\
    E_{\perp,B,j}\\
    E_{\parallel,B,k}\\
  \end{array}
\right),
\end{aligned}
\end{equation}
where $E_{\perp,B,i}$, $E_{\perp,B,j}$ and $E_{\perp,B,k}$ are the components of $\Vec{E}_{\perp,B}$ and $\Vec{E}_{\parallel,B}$, respectively.
The three components in the above matrix calculation can be expanded as
\begin{equation}
E_{\perp,i}=E_{\perp,B,i}\cos\phi-E_{\perp,B,j}\sin\phi.
\end{equation}
\begin{equation}
E_{\perp,j}=E_{\perp,B,i}\sin\phi\cos\theta+E_{\perp,B,j}\cos\phi\cos\theta+E_{\parallel,B,k}\sin\theta.
\end{equation}
\begin{equation}
E_{\parallel,k}=-E_{\perp,B,i}\sin\phi\sin\theta-E_{\perp,B,j}\cos\phi\sin\theta+E_{\parallel,B,k}\cos\theta.
\end{equation}
We decompose the three components into the LOS coordinate system $(\theta_v,\phi_v)$ and obtain three new components labeled as "parallel"
(which is along new $x$-axis), "perpendicular" (which is along new $y$-axis) and the LOS (which is along new  $z$-axis)
\begin{equation}
{A}_{\parallel}=E_{\perp,i}\cos\phi_v+E_{\perp,j}\sin\phi_v,
\end{equation}
\begin{equation}
{A}_{\perp}=-E_{\perp,i}\cos\theta_v\sin\phi_v+E_{\perp,j}\cos\theta_v\cos\phi_v-E_{\parallel,k}\sin\theta_v,
\end{equation}
\begin{equation}
{A}_{\rm LOS}=-E_{\perp,i}\sin\theta_v\sin\phi_v+E_{\perp,j}\sin\theta_v\cos\phi_v+E_{\parallel,k}\cos\theta_v,
\end{equation}
The total scattered electric field perpendicular to the line of sight can be calculated as
\begin{equation}
\begin{aligned}
{A}_{\rm tot,\parallel}=\int_0^{\theta_c}d\theta\int_0^{2\pi}d\phi\int_{r_{\rm min}}^{r_{\rm max}}r^2\sin\theta dr[A_\parallel],
\end{aligned}
\end{equation}
and
\begin{equation}
\begin{aligned}
{A}_{\rm tot,\perp}=\int_0^{\theta_c}d\theta\int_0^{2\pi}d\phi\int_{r_{\rm min}}^{r_{\rm max}}r^2\sin\theta dr[A_\perp].
\end{aligned}
\end{equation}
where $r_{\rm min}=(d-\lambda)\cos\theta_c/\cos\theta$ and $r_{\rm max}=d\cos\theta_c/\cos\theta$.

In order to find the transformation of angle $\theta_v$ and $\phi_v$ from co-moving frame to lab frame, we define the angle $\theta_{s}$ between LOS ($\theta_v,\phi_v$) and arbitrary magnetic field ($\theta,\phi$) as
\begin{equation}\label{Eq:firstrelation}
\cos\theta_s=\cos\theta_v\cos\theta+\sin\theta_v\sin\theta\cos(\phi_v-\phi)
\end{equation}
and
\begin{equation}
\cos\theta_s'=\cos\theta_v'\cos\theta+\sin\theta_v'\sin\theta\cos(\phi_v-\phi).
\end{equation}
It should be pointed out that $\phi_v'=\phi_v$.
The angle $\theta_{s}$ satisfies
\begin{equation}\label{Eq:secondrelation}
\cos\theta_{s}'=\frac{\cos\theta_{s}-\beta}{1-\beta\cos\theta_{s}}.
\end{equation}
The angles $\theta_v'$ and $\phi_v'$ can be numerically solved by calculating Eqs.(\ref{Eq:firstrelation}) and (\ref{Eq:secondrelation}) together.

\section{Synchrotron absorption coefficient}\label{C}
In this Appendix, we present a derivation of synchrotron self-absorption coefficients of electrons for the total intensity ($\alpha_{\nu,e}$) which are the classical results in the GRB field \citep{Wu2003,Zhangbook} and for the parallel ($\alpha_{\parallel}$) and perpendicular ($\alpha_{\perp}$) components (which is newly derived here) in different regimes for a power law distribution of electrons. 
\begin{itemize}
\item Absorption of total intensity: The absorption coefficient of synchrotron radiation can be written as
\begin{equation}
\begin{aligned}
\alpha_{\nu,e}&=\frac{p+2}{8\pi m_e}C_\gamma \nu^{-2}\int_{\gamma_{\rm min}}^{\gamma_{\rm max}}\frac{\sqrt{3}e^3B_\perp}{m_ec^2}F(x)\gamma^{-(p+1)}d\gamma,
\end{aligned}
\end{equation}
where
\begin{equation}
x=\frac{\nu}{\nu_{\rm ch}}=\frac{4\pi m_e c\nu}{3eB_\perp \gamma^2}.
\end{equation}

For Case (i), i.e. $\gamma_{\rm min}\ll\gamma(\nu_{\rm  ch,frb})\ll\gamma_{\rm max}$,
one can see when $\gamma\gg1$ we have $x_{\rm min}\ll 1 \simeq0$, when $\gamma\sim1$ we have $x_{\rm max}\gg0\simeq\infty$. Thus we can replace $p/2-1$ by $\mu$ and apply the integral formula 
\begin{equation}
\int_0^{\infty}x^{\mu}F(x)dx=\frac{2^{\mu+1}}{\mu+2}\Gamma\left(\frac{\mu}{2}+\frac{7}{3}\right)\Gamma\left(\frac{\mu}{2}+\frac{2}{3}\right)
\end{equation}
to obtain
\begin{equation}
\begin{aligned}
&\alpha_{\nu,e}=\frac{\sqrt{3}e^3C_\gamma}{8\pi m_e^2c^2}
\left(\frac{3e}{2\pi m_ec}\right)^{p/2} B_\perp^{\frac{p+2}{2}}\Gamma\left(\frac{3p+2}{12}\right)\Gamma\left(\frac{3p+22}{12}\right)\nu_{\rm ch}^{-\frac{p+4}{2}}\\
&\simeq[10^4(8.4\times10^6)^{\frac{p}{2}} \ {\rm cm^{-1}}]C_{\gamma_e} B_\perp^{\frac{p+2}{2}}\Gamma\left(\frac{3p+2}{12}\right)\Gamma\left(\frac{3p+22}{12}\right)\nu^{-\frac{p+4}{2}}.
\end{aligned}
\end{equation}

For Case (ii), i.e. $\gamma(\nu_{\rm ch,frb})\ll\gamma_{\rm min}\ll\gamma_{\rm max}$, we can replace $F(x)$ by the asymptotic expression $\propto x^{1/3}$ when $x\ll 1$. Then we can directly integrate over Lorentz factor to obtain
\begin{equation}
\begin{aligned}
\alpha_{\nu,e}&=\frac{1}{2^{4/3}\Gamma(1/3)}\frac{(p+2)}{(p+2/3)}\frac{e^3B_\perp C_{\gamma_e}}{m_e^2c^2}\left(\frac{4\pi m_ec}{3eB_\perp}\right)^{1/3}\gamma_{\rm min,e}^{-(p+2/3)}\nu^{-5/3}\\
&\simeq(136 \ {\rm cm^{-1}}) \ \frac{(p+2)}{(p+2/3)}C_{\gamma_e} B_\perp^{2/3}\gamma_{\rm min,e}^{-(p+2/3)}\nu^{-5/3}. 
\end{aligned}
\end{equation}
\item Absorption of parallel and perpendicular components:  The synchrotron radiation power of a single electron have two orthogonal components (with respect to the magnetic field direction in the plane perpendicular to the LOS) which are given by \citep{Rybicki&Lightman1979}
\begin{equation}
P_\parallel(\nu)=\frac{\sqrt{3}e^3B_\perp}{2 m_ec^2}[F(x)-G(x)],
\end{equation}
and
\begin{equation}
P_\perp(\nu)=\frac{\sqrt{3}e^3B_\perp}{2m_ec^2}[F(x)+G(x)],
\end{equation}
where $F(x)=x\int_x^\infty K_{5/3}(\xi)d\xi$ and $G(x=)xK_{2/3}(x)$.

For Case (i), the parallel and perpendicular absorption coefficients can be calculated as
\begin{equation}
\begin{aligned}
&\alpha_\parallel=-\frac{1}{8\pi\nu^2m_e}\int_{\gamma_{\rm min}}^{\gamma_{\rm max}}d\gamma P_\parallel(\nu)\gamma^2\frac{\partial}{\partial \gamma}\left[\frac{N(\gamma)}{\gamma^2}\right]\\
&=\frac{1}{2}\alpha_{\nu,e}-\frac{\sqrt{3}e^3B_\perp C_{\gamma_e}(p+2)}{16\pi\nu^2 m_e^2c^2}\int_{\gamma_{\rm min}}^{\gamma_{\rm max}} xK_{2/3}(x)\gamma_e^{-(p+1)} d\gamma_e\\
&=\frac{1}{2}\alpha_{\nu,e}-\frac{\sqrt{3}e^3B_\perp C_{\gamma_e}(p+2)}{32\pi\nu^2 m_e^2c^2}\int_{0}^{\infty} xK_{2/3}(x)\left(\frac{3eB_\perp}{4\pi m_e c\nu}\right)^{p/2}x^{\frac{p}{2}-1}dx,
\end{aligned}
\end{equation}
and
\begin{equation}
\begin{aligned}
&\alpha_\perp=-\frac{1}{8\pi\nu^2m_e}\int_{\gamma_{\rm min}}^{\gamma_{\rm max}}d\gamma P_\perp(\nu)\gamma^2\frac{\partial}{\partial \gamma}\left[\frac{N(\gamma)}{\gamma^2}\right]\\
&=\frac{1}{2}\alpha_{\nu,e}+\frac{\sqrt{3}e^3B_\perp C_{\gamma_e}(p+2)}{16\pi\nu^2 m_e^2c^2}\int_{\gamma_{\rm min}}^{\gamma_{\rm max}} xK_{2/3}(x)\gamma_e^{-(p+1)} d\gamma_e\\
&=\frac{1}{2}\alpha_{\nu,e}+\frac{\sqrt{3}e^3B_\perp C_{\gamma_e}(p+2)}{32\pi\nu^2 m_e^2c^2}\int_{0}^{\infty} xK_{2/3}(x)\left(\frac{3eB_\perp}{4\pi m_e c\nu}\right)^{p/2}x^{\frac{p}{2}-1}dx.
\end{aligned}
\end{equation}
We define $\alpha_{G(x)}$ (the second term in both $\alpha_{\parallel}$ and $\alpha_{\perp}$), which can be calculated as
\begin{equation}
\begin{aligned}
\alpha_{G(x)}&=\frac{\sqrt{3}e^3B_\perp C_{\gamma_e}(p+2)}{32\pi\nu^2 m_e^2c^2}\int_{0}^{\infty} xK_{2/3}(x)\left(\frac{3eB_\perp}{4\pi m_e c\nu}\right)^{p/2}x^{\frac{p}{2}-1}dx\\
&=\frac{\sqrt{3}e^3B_\perp C_{\gamma_e}}{64\pi m_e^2c^2(p+2)^{-1}}\left(\frac{3eB_\perp}{2\pi m_e c}\right)^{\frac{p}{2}}\Gamma\left(\frac{3p+10}{12}\right)\Gamma\left(\frac{3p+2}{12}\right)\nu^{-\frac{p+4}{2}}\\
&\simeq[5\times10^3(8.4\times10^6)^{\frac{p}{2}} \ {\rm cm^{-1}}](p+2)C_{\gamma_{e}}B_\perp^{\frac{p+2}{2}}\\
&\times\Gamma\left(\frac{3p+10}{12}\right)\Gamma\left(\frac{3p+2}{12}\right)\nu^{-\frac{p+4}{2}},
\end{aligned}
\end{equation}
which is consistent with Eq.(\ref{G_1}). 

For Case (ii), we have
\begin{equation}
\alpha_{\parallel}=\frac{1}{2}\alpha_{\nu,e}-\frac{\sqrt{3}e^3B_\perp C_{\gamma_e}(p+2)}{16\pi\nu^2 m_e^2c^2}\int_{\gamma_{\rm min}}^{\gamma_{\rm max}} G(x)\gamma_e^{-(p+1)} d\gamma_e,
\end{equation}
and
\begin{equation}
\alpha_{\perp}=\frac{1}{2}\alpha_{\nu,e}+\frac{\sqrt{3}e^3B_\perp C_{\gamma_e}(p+2)}{16\pi\nu^2 m_e^2c^2}\int_{\gamma_{\rm min}}^{\gamma_{\rm max}} G(x)\gamma_e^{-(p+1)} d\gamma_e.
\end{equation}
We define $\alpha_{G(x)}$ (the second term in both $\alpha_{\parallel}$ and $\alpha_{\perp}$), we calculate it by assuming $\gamma_{\rm max}\gg1$ and obtain
\begin{equation}
\begin{aligned}
\alpha_{G(x)}&=\frac{\sqrt{3}e^3B_\perp C_{\gamma_e}(p+2)}{16\pi\nu^2 m_e^2c^2}\int_{\gamma_{\rm min}}^{\gamma_{\rm max}} G(x)\gamma_e^{-(p+1)} d\gamma_e\\
&=\frac{\sqrt{3}e^3B_\perp C_{\gamma_e}\Gamma(2/3)(p+2)}{16\pi\nu^2 m_e^2c^2} \left(\frac{2\pi m_ec\nu}{3eB_\perp}\right)^{1/3} \int_{\gamma_{\rm min}}^{\gamma_{\rm max}} \gamma_e^{-(p+\frac{5}{3})} d\gamma_e\\
&=\frac{\sqrt{3}e^3B_\perp C_{\gamma_e}\Gamma(2/3)(p+2)}{16\pi m_e^2c^2(p+2/3)} \left(\frac{2\pi m_ec}{3eB_\perp}\right)^{1/3}\gamma_{\rm min}^{-(p+2/3)}\nu^{-5/3}\\
&\simeq(34 \ {\rm cm}^{-1}) \ \frac{(p+2)}{(p+2/3)}C_{\gamma_e}B_\perp^{2/3}\gamma_{\rm min,e}^{-(p+2/3)}\nu^{-5/3}.
\end{aligned}
\end{equation}
which is consistent with Eq.(\ref{G_2}).

\end{itemize}


\bsp	
\label{lastpage}
\end{document}